\documentclass[prd,aps,nofootinbib,floatfix,10 pt]{revtex4}
\usepackage{amssymb}
\usepackage{amsmath,graphicx,color,epsfig}
\usepackage{slashed}
\begin{document}

\title{Impact of $Z^{\prime}$ and UED parameters on different asymmetries in $B_{s} \to \phi \ell^{+} \ell^{-}$ decays}
\author{Ishtiaq Ahmed$^{1,2}$}
\email{ishtiaq@ncp.edu.pk}
\author{M. Jamil Aslam$^{1,2}$}
\email{jamil@phys.qau.edu.pk}
\author{M. Ali Paracha$^{3,4}$}
\email{ali@ncp.edu.pk}
\affiliation{$^{1}$National Centre for Physics,\\
Quaid-i-Azam University Campus, Islamabad 45320, Pakistan}
\affiliation{$^{2}$Department of Physics,\\
Quaid-i-Azam University, Islamabad 45320, Pakistan}
\affiliation{$^{3}$Centre for Advanced Mathematics and Physics, \\
National University of Science and Technology, Islamabad, Pakistan}
\affiliation{$^{4}$Laborat\`orio de Física Te\`orica e
Computacional, Universidade Cruzeiro do Sul, 01506-000 S\~ao Paulo,
Brazil}

\date{\today}
\begin{abstract}

 A comprehensive study of the impact of new-physics on different observables for
$B_{s}\to \phi \ell^{+}\ell^{-}$ has been done. We examine the new physics models such as $Z^{\prime}$
and UED models, where the effects of new physics coming through
the modification of Wilson coefficients. We have analyzed these effects through the theoretical prediction of the branching ratio,
the forward-backward asymmetry, lepton polarization asymmetries and the helicity fractions of the final
state meson. These all observables will definitely be measured in present and future colliders with great precision.
We also pointed out that hadronic uncertainties in various physical observables are small which make them
an ideal probe to establish new physics. Therefore, the measurements of these observables for the same decay
 would permit the detection of physics beyond the Standard Model and will also help us to distinguish
 between different new physics scenarios.
\end{abstract}

\maketitle

\section{Introduction}

Studies of flavor-changing neutral current (FCNC) decays have played a
pivotal role in formulating the theoretical description of particle
physics known as the Standard Model (SM). In the SM, at tree level, all the neutral currents
conserve flavor so that FCNC decays do not occur at lowest order and are
induced by the Glahsow-Iliopoulos-Maiani (GIM) amplitude \cite{1} at the
loop level, which make their effective strength small. In addition to this
loop suppression these are also suppressed in the SM due to their dependence
on the weak mixing angles of the Cabbibo-Kobayashi-Maskawa (CKM) matrix $%
V_{CKM}$ \cite{2,3}. Therefore, these two circumstances make the FCNC decays
relatively rare and hence important to provide stringent tests of SM in the
flavor sector.

Although many measurements of observables in the $B$ meson systems agree
with the SM. However, there are
several observables whose measured values differ from the
predictions of the SM such as (i) the values of $B_{d}^{0}-\bar{B_{d}^{0}}$
mixing phase $\sin \left( 2\beta \right) $ obtained from different penguin
dominated $b\rightarrow s$ channels tend to be systematically smaller than
that obtained from $B_{d}^{0}\rightarrow J/\psi K_{s}$ \cite{4a, 4b, 4c},
(ii) the $B_{s}^{0}-\bar{B_{s}^{0}}$ mixing phase by the CDF and D$0$
collaboration deviates from the SM prediction \cite{5a, 5b}, (iii) in $%
B\rightarrow K\pi $ decays, it is difficult to account for all the
experimental measurements within the SM \cite{6}, (iv) the isospin asymmetry
between neutral and the charged decay modes of the $\bar{B}\rightarrow \bar{%
K^{\ast }}\ell ^{+}\ell ^{-}$ decay also deviate from the SM \cite{7}. These
disagreements are typically at the level $2\sigma $ which are not
statistically significant, but still these are fertile ground to test SM and check the NP, as they appear in the
$b\rightarrow s$ transitions. In this context there have been numerous
papers examining the possible new physics (NP) FCNC scenarios through the
various $b\rightarrow s$ processes \cite{8}.

On the experimental side, the Large Hadron Collider (LHC) is already up and running where CMS, ATLAS
and LHCb have started taking data while the Belle II is on its
way. We are already witnessing that SM is still standing tall at least in
the data taken till to date and the recent discovery of Higgs like boson in
the mass range of $126$GeV has left enough air to breath for the SM. It is
therefore an ideal time to test the predictions of SM and try to identify
the nature of physics that is beyond it.

Moreover, in the SM the zero crossing of the leptons forward-backward
asymmetry $(A_{FB}(q^{2}))$ in $B\rightarrow K^{\ast
}l^{+}l^{-}$ is at a well determined position which is free from the
hadronic uncertainties at the leading order (LO) in strong coupling $\alpha
_{s}$ \cite{9,10,11}. On the other hand the LHCb has announced the results of $%
A_{FB} $ , the fraction of longitudinal polarization $F_{L}$ and the
differential branching ratio $dB/dq^{2}$, as a function of the dimuon
invariant mass for the decay $\bar{B}\rightarrow \bar{K}^{\ast }\mu ^{+}\mu
^{-}$ using $0.37$fb$^{-1}$ of data taken in the year 2011 \cite{CERN}.
These results of $A_{FB}(\bar{B}\rightarrow \bar{K}^{\ast }\mu ^{+}\mu ^{-})$
are close to the SM predictions with slight error bars, thus they have
overwritten the earlier measurements by Babar and Belle which measure this
asymmetry with opposite sign with better statistics \cite{9a,10a,11a,CERN}.
The collaboration plans to continue to the study of the channel $\bar{B}%
\rightarrow \bar{K}^{\ast }\mu ^{+}\mu ^{-}$ in finer detail, with
more angular variables, and expected to achieve
high sensitivity to any small deviation from the SM \cite{CERN}.

In order to incorporate the experimental predictions of different physical
observables in $\bar{B} \to \bar{K}^{*} \mu^{+} \mu^{-}$ this decay has been
studied in SM and in number of different NP scenarios \cite%
{17,18,20,21,22,23,24,25,26,27,28,29,30, 31,32,33,34,35,36,37,38} where NP
effects display themselves through modification in the Wilson coefficients
as well as through the new operators. Beside these models the general
analysis of $\bar{B} \to \bar{K}^{*} \mu^{+} \mu^{-}$ decay has also been
performed which allow us to include all possible NP operators such as
vector-axial vector ($VA$), scalar-pseudoscalar ($SP$) and tensor-axial
tensor ($TE$) \cite{39}.

Apart from the ordinary $B$ meson decays an interesting avenue for the NP
is opened by the $B_{s}$ meson decays, where $B_{s}^{0}-\bar{%
B_{s}^{0}}$ mixing is the exciting feature, which in the SM, is originated
from the box topologies and hence is strongly suppressed. In the presence of
NP, new particles could give rise to additional box topologies or even
these decays can occurs at the tree level. In this regard, the key channel to address this possibility is
the $B_{s}^{0}\rightarrow J/\Psi \phi $ where the pertinent feature is that
its final state contains two vector mesons and thereby require the time
dependent angular analysis of the $J/\Psi \rightarrow \mu ^{+}\mu ^{-}$ and $%
\phi \rightarrow K^{+}K^{-}$ decay products \cite{RF1}. In addition, over the last couple
of years, measurements of CP violating asymmetries in "tagged" analysis
(distinguishing between initially present $B_{s}^{0}$ or $\bar{B^{0}}_{s}$
mesons) of the $B_{s}^{0}\rightarrow J/\Psi \phi $ channel at the Tevatron
indicate possible NP effects in $B_{s}^{0}-\bar{B_{s}^{0}}$ mixing \cite%
{RF2,RF3,RF4}. These results are complemented by the measurement of the
anomalous like-sign dimuon charge asymmetry at D0, which was found to differ
by $3.9\sigma $ from the SM prediction \cite{RF5}. However, in the last summer the LHCb collaboration
has also reported, the results
that disfavor large NP effects \cite{RF6}. Therefore, the more data is needed to clarify the potential and status of NP.

Following the same footprints, the exclusive $B_{s}\rightarrow \phi \ell
^{+}\ell ^{-}$ become also attractive since at quark level these decays are
also induced by $b\rightarrow s\ell ^{+}\ell ^{-}$, and could be measured at
the running Tevatron, LHC and future super-B factories. Recently, the CDF
collaboration had observed $B_{s}\rightarrow \phi \mu ^{+}\mu ^{-}$ with the
branching ratio \cite{39}
\begin{equation}
Br(B_{s}\rightarrow \phi \mu ^{+}\mu ^{-})=[1.44\pm 0.33(\text{stat.})\pm
0.46(\text{syst.})]\times 10^{-6}.  \label{measurement}
\end{equation}%
On theoretical side this exclusive process is well studied in the literature
\cite{40, 41, 42, 43, 44} with varying degrees of theoretical rigor and
emphasis. In order to study different physical observables such as, dilepton
invariant mass spectrum, the forward-backward asymmetry, helicity fractions
of final state meson and different lepton polarization asymmetries, the crucial
ingredients are the form factors which needed to be calculated using a
non-perturbative QCD methods and therefore form the bulk of theoretical
uncertainties. Form factors parameterizing $B_{s}\rightarrow
\phi \mu ^{+}\mu ^{-}$ have already been calculated in different models,
such as light cone sum rules (LCSRs) \cite{45, 46, 47}, perturbative QCD
approach \cite{48}, relativistic constituent quark model \cite{49},
constituent quark model \cite{50} and light front quark model \cite{51}.
Among them, the LCSRs deal with form factors at small momentum region and is
complementary to the lattice QCD approach and consistent with
perturbative QCD as well as heavy quark limit, therefore, we will adopt the form
factors calculated by this approach in our forthcoming analysis of $%
B_{s}\rightarrow \phi \ell ^{+}\ell ^{-}$ decays in the SM and two different NP
models, namely, $Z^{\prime}$ and Universal Extra Dimension (UED).
The form factors calculated from LCSRs are restricted to the low $q^2$ region,
where at the high $q^2$ region, ongoing efforts aim at the first unquenched prediction
from lattice \cite{latt}.

It is well known that the NP play its role in the rare $B$ meson decays in
two different ways, one is through the modification in the Wilson
coefficients corresponding to the SM operators and the other is due to the
appearance of new operators in the effective Hamiltonian, which are absent
in the SM. The UED and $Z^{\prime }$ models' belong to the category,
commonly known as Minimal Flavor Violating models, which do not change the
operator basis of the SM and hence their contribution is absorbed in the
Wilson coefficients. In the present study, we perform the analysis
of branching ratio, the forward-backward asymmetry $(A_{FB})$, the helicity
fraction of final state $\phi $ meson $(f_{L,T})$, the lepton
polarization asymmetries (both single and double)
in the $B_{s}\rightarrow \phi
\ell ^{+}\ell ^{-}$ decay in aforementioned NP scenarios.

The outline of the paper is as follows: In Sec. 2 we briefly
discuss the different NP scenarios and introduce the effective Hamiltonian
formalism for semileptonic rare $B_{s}$ decays. Section 3 contains the
definitions and parameterizations of $B_{s}\rightarrow \phi $ matrix
elements and summarizes the form factor calculated in LCSRs. In Sec. 4, we
display the mathematical expressions for the branching ratio, the forward-backward
asymmetry, the helicity fractions of $\phi $ meson and the different
lepton polarization asymmetries. Sec. 5 contains our numerical results for
the above mentioned physical observables both in the SM and in different NP
scenarios where we show the influence of the NP parameters on the various
asymmetries outlined above. A brief summary and some
concluding remarks are also given at the end of this section.

\section{Theoretical Framework}

To calculate the decay amplitude
of $B_{s} \to \phi \ell^{+} \ell^{-}$ decays requires some theoretical steps.
Among them the most important and relevant are:

\begin{itemize}
\item the separation of short distance effects (encoded in Wilson coefficients)
 from the long-distance QCD effects (encoded in the matrix elements) in the effective Hamiltonian;

\item the calculation of matrix elements of local quark bilinear operators
of the type $\langle\phi|J|B_{s}\rangle$ in terms of form factors.
\end{itemize}

As the effective Hamiltonian will be changed in different models therefore
at first step we will describe the effective Hamiltonian in the aforementioned models and
the discussion about the form factors will be postponed to next section.

\subsection{Standard Model (SM)}

At quark level the decay $B_{s}\rightarrow \phi \ell ^{+}\ell ^{-}$ is
governed by the transition $b\rightarrow \ s\ell ^{+}\ell ^{-}$ for which
the effective Hamiltonian can be written as
\begin{equation}
H_{eff}=-\frac{4G_{F}}{\sqrt{2}}V_{tb}^{\ast }V_{ts}{\sum\limits_{i=1}^{10}}%
C_{i}({\mu })O_{i}({\mu }),  \label{effective hamiltonian 1}
\end{equation}%
where $O_{i}({\mu })$ $(i=1,\ldots ,6)$ are the four-quark operators, $i=7,8$
are dipole operators and $i=9,10$ are the semileptonic operators. The $C_{i}(%
{\mu })$ are the corresponding Wilson\ coefficients at the energy scale ${%
\mu }$. The terms that corresponds to the running of $u$-quark in the loop,
i.e. $V_{ub}^{\ast }V_{us}$ can be safely ignored because $\frac{%
V_{ub}^{\ast }V_{us}}{V_{tb}^{\ast }V_{ts}}\prec 2\times 10^{-2}$. The
operators responsible for $B_{s}\rightarrow \phi \ell ^{+}\ell ^{-}$ are $%
O_{7}$, $O_{9}$ and $O_{10}$ and their form is given by
\begin{eqnarray}
O_{7} &=&\frac{e^{2}}{16\pi ^{2}}m_{b}\left( \bar{s}\sigma _{\mu \nu
}P_{R}b\right) F^{\mu \nu },\,  \notag \\
O_{9} &=&\frac{e^{2}}{16\pi ^{2}}(\bar{s}\gamma _{\mu }P_{L}b)(\bar{l}\gamma
^{\mu }l),\,  \label{op-form} \\
O_{10} &=&\frac{e^{2}}{16\pi ^{2}}(\bar{s}\gamma _{\mu }P_{L}b)(\bar{l}%
\gamma ^{\mu }\gamma _{5}l),  \notag
\end{eqnarray}%
with $P_{L,R}=\left( 1\pm \gamma _{5}\right) /2$. The Wilson coefficients $%
C_{i}$ can be calculated perturbatively and the explicit expressions of
these in the SM at next-to-leading order (NLO) and at next-to-next leading
logarithm (NNLL) are given in \cite{52, 53, 54, 55, 56, 57, 58, 59, 60, 61,
62, 63, 64, 65, 66, 67}. Since the $Z$ -boson is absent in the effective
theory, therefore the operator $O_{10}$ can not be induced by the insertion
of four-quark operators. Hence, the Wilson coefficient $C_{10}$ does not
renormalize under QCD corrections and so it is independent on the energy
scale.

The\ Wilson coefficient $C_{9}^{SM}(\mu )$ \ which is commonly written as $%
C_{9}^{eff}(\mu )$ corresponds to the semileptonic operator $O_{9}$. It can
be decomposed into three parts
\begin{equation}
C_{9}^{SM}=C_{9}^{eff}(\mu )=C_{9}(\mu )+Y_{SD}(z,s^{\prime
})+Y_{LD}(z,s^{\prime }),  \label{c9coefficient}
\end{equation}%
where the parameters $z$ and $s^{\prime }$ are defined as $%
z=m_{c}/m_{b},\,\,\,s^{\prime }=q^{2}/m_{b}^{2}$. The short distance
function $Y_{SD}(z,s^{\prime })$ describes the perturbative part which
include the indirect contributions from the matrix element of four-quark
operators $\sum_{i=1}^{6}\langle l^{+}l^{-}s|O_{i}|b\rangle $ and this lies
sufficiently far away from the $c\bar{c}$ resonance regions. The manifest
expressions for $Y_{SD}(z,s^{\prime })$ can be written as \cite{53, 54}
\begin{eqnarray}
Y_{SD}(z,s^{\prime }) &=&h(z,s^{\prime })(3C_{1}(\mu )+C_{2}(\mu
)+3C_{3}(\mu )+C_{4}(\mu )+3C_{5}(\mu )+C_{6}(\mu ))  \notag \\
&&-\frac{1}{2}h(1,s^{\prime })(4C_{3}(\mu )+4C_{4}(\mu )+3C_{5}(\mu
)+C_{6}(\mu ))  \notag \\
&&-\frac{1}{2}h(0,s^{\prime })(C_{3}(\mu )+3C_{4}(\mu ))+{\frac{2}{9}}%
(3C_{3}(\mu )+C_{4}(\mu )+3C_{5}(\mu )+C_{6}(\mu )),  \label{short-distance}
\end{eqnarray}%
with
\begin{eqnarray}
h(z,s^{\prime }) &=&-{\frac{8}{9}}\mathrm{ln}z+{\frac{8}{27}}+{\frac{4}{9}}x-%
{\frac{2}{9}}(2+x)|1-x|^{1/2}\left\{
\begin{array}{l}
\ln \left\vert \frac{\sqrt{1-x}+1}{\sqrt{1-x}-1}\right\vert -i\pi \quad
\mathrm{for}{{\ }x\equiv 4z^{2}/s^{\prime }<1} \\
2\arctan \frac{1}{\sqrt{x-1}}\qquad \mathrm{for}{{\ }x\equiv
4z^{2}/s^{\prime }>1}%
\end{array}%
\right. ,  \notag \\
h(0,s^{\prime }) &=&{\frac{8}{27}}-{\frac{8}{9}}\mathrm{ln}{\frac{m_{b}}{\mu
}}-{\frac{4}{9}}\mathrm{ln}s^{\prime }+{\frac{4}{9}}i\pi \,\,.  \label{hzs}
\end{eqnarray}%
The long-distance contributions $Y_{LD}(z,s^{\prime })$ from four-quark
operators near the $c\bar{c}$ resonance cannot be calculated from first
principles of QCD and are usually parameterized in the form of a
phenomenological Breit-Wigner formula making use of the vacuum saturation
approximation and quark-hadron duality. In the present study we ignore this
part because this lies far away from the region of interest.

The non-factorizable effects \cite{68, 69, 70} from the charm loop can bring
about further corrections to the radiative $b\rightarrow s\gamma $
transition, which can be absorbed into the effective Wilson coefficient $%
C_{7}^{eff}$. Specifically, the Wilson coefficient $C_{7}^{eff}$ is given by
\begin{equation}
C_{7}^{SM}=C_{7}^{eff}(\mu )=C_{7}(\mu )+C_{b\rightarrow s\gamma }(\mu ),
\label{c7coefficient}
\end{equation}%
with
\begin{eqnarray}
C_{b\rightarrow s\gamma }(\mu ) &=&i\alpha _{s}\bigg[{\frac{2}{9}}\eta
^{14/23}(G_{1}(x_{t})-0.1687)-0.03C_{2}(\mu )\bigg], \\
G_{1}(x_{t}) &=&{\frac{x_{t}(x_{t}^{2}-5x_{t}-2)}{8(x_{t}-1)^{3}}}+{\frac{%
3x_{t}^{2}\mathrm{ln}^{2}x_{t}}{4(x_{t}-1)^{4}}},
\end{eqnarray}%
where $\eta =\alpha _{s}(m_{W})/\alpha _{s}(\mu )$, $%
x_{t}=m_{t}^{2}/m_{W}^{2}$, $C_{b\rightarrow s\gamma }$ is the absorptive
part for the $b\rightarrow sc\bar{c}\rightarrow s\gamma $ rescattering and
we have dropped out the tiny contributions proportional to CKM sector $%
V_{ub}V_{us}^{\ast }$.

In terms of the above Hamiltonian, the free quark decay amplitude for $%
b\rightarrow s\ell ^{+}\ell ^{-}$ in the SM can be derived as:
\begin{eqnarray}
\mathcal{M}(b &\rightarrow &s\ell ^{+}\ell ^{-})=-\frac{G_{F}\alpha }{\sqrt{2%
}\pi }V_{tb}V_{ts}^{\ast }\bigg\{C_{9}^{SM}(\bar{s}\gamma _{\mu }P_{L}b)(%
\bar{l}\gamma ^{\mu }l)+C_{10}^{SM}(\bar{s}\gamma _{\mu }P_{L}b)(\bar{l}%
\gamma ^{\mu }\gamma _{5}l)  \notag \\
&&-2m_{b}C_{7}^{SM}(\bar{s}i\sigma _{\mu \nu }\frac{q^{\nu }}{q^{2}}P_{R}b)(%
\bar{l}\gamma ^{\mu }l)\bigg\},  \label{quark-amplitudeSM}
\end{eqnarray}%
where $q^{2}=(p_{l^{+}}+p_{l^{-}})^{2}$ is the square of momentum transfer.

\subsection{Universal Extra Dimension Model}

Among different new physics models cooked during last 20 years or so, a
special role is played by one with universal extra dimensions (UED). In
this model all SM\ fields are allowed to propagate in all available
dimensions. The economy of UED\ model is that there is only one additional
parameter to that of SM which is the radius $R$ of the compactified extra
dimension. Now above the compactification scale $1/R$ a given UED model
becomes a higher dimensional field theory whose equivalent description in
four dimensions consists of SM\ fields and the towers of Kaluza-Klein (KK)
modes having no partner in the SM. A simplest model of this type was
proposed by Appelquist,Cheng and Dobrescu (ACD) \cite{ACD1}. In this model,
all the masses of the KK particles and their interactions with SM\ particles
and also among themselves are described in terms of the inverse of
compactification radius $R$ and the parameters of the SM \cite{ACD2, ACD3}.

The ACD model belongs to the class of Minimal Flavor Violating
models where the effects beyond SM are only encoded in the Wilson
coefficients of the effective Hamiltonian without changing the operator basis
of SM. Wilson coefficients contributing
in the calculation of $b\rightarrow s\ell ^{+}\ell ^{-}$ i.e. $C_{7}$, $%
C_{9} $ and $C_{10}$ get modified due to the KK excitation inducing a
dependence on the compactification radius $R$. As the value of the
compactification radius $R$ becomes smaller or in other words the value of $%
1/R$, becomes larger we can recover the Standard\ Model phenomenology
because the massive KK states started to decouple. As a general expression,
the Wilson coefficients are represented by periodic functions $F\left(
x_{t},1/R\right) $ generalizing their SM analogues $F_{0}\left( x_{t}\right)
:$%
\begin{equation}
F\left( x_{t},1/R\right) =F_{0}\left( x_{t}\right)
+\sum\limits_{n=1}^{\infty }F\left( x_{t},x_{n}\right)  \label{UEDeq}
\end{equation}%
with $x_{t}=\frac{m_{t}^{2}}{M_{w}^{2}}$, $x_{n}=\frac{m_{n}^{2}}{M_{w}^{2}}$
and $m_{n}=\frac{n}{R}$. The remarkable feature of above equation is that
the summation over the KK contribution is finite at the leading order (LO)
in all the cases as a consequence of generalized GIM mechanism \cite{ACD2,
ACD3}. As $R\rightarrow 0$, $F\left( x_{t},1/R\right) \rightarrow
F_{0}\left( x_{t}\right) $ which is the SM result. Now if we take $1/R$ to
be few hundred GeV, the values of the Wilson coefficients differ
considerable from their corresponding SM values, where the most pronounced
effects comes in the $C_{7}$. It is therefore expected that the various
physical observables differ significantly from the SM results for certain
range of compactification radius $R$.

Thus the effective Hamiltonian for $b\rightarrow s\ell ^{+}\ell ^{-}$
transition in UED model is given by%
\begin{eqnarray}
\mathcal{H}_{eff}^{UED}(b &\rightarrow &s\ell ^{+}\ell ^{-})=-\frac{%
G_{F}\alpha }{\sqrt{2}\pi }V_{tb}V_{ts}^{\ast }\bigg\{C_{9}^{UED}(\bar{s}%
\gamma _{\mu }P_{L}b)(\bar{l}\gamma ^{\mu }l)+C_{10}^{UED}(\bar{s}\gamma
_{\mu }P_{L}b)(\bar{l}\gamma ^{\mu }\gamma _{5}l)  \notag \\
&&-2m_{b}C_{7}^{UED}(\bar{s}i\sigma _{\mu \nu }\frac{q^{\nu }}{q^{2}}P_{R}b)(%
\bar{l}\gamma ^{\mu }l)\bigg\},  \label{HUED}
\end{eqnarray}%
where the explicit expressions of various Wilson coefficients are given in
Refs. \cite{ACD2, ACD3}. Using this Hamiltonian the free quark decay
amplitude becomes%
\begin{eqnarray}
\mathcal{M}(b &\rightarrow &s\ell ^{+}\ell ^{-})=-\frac{G_{F}\alpha }{\sqrt{2%
}\pi }V_{tb}V_{ts}^{\ast }\bigg\{C_{9}^{UED}(\bar{s}\gamma _{\mu }P_{L}b)(%
\bar{l}\gamma ^{\mu }l)+C_{10}^{UED}(\bar{s}\gamma _{\mu }P_{L}b)(\bar{l}%
\gamma ^{\mu }\gamma _{5}l)  \notag \\
&&-2m_{b}C_{7}^{UED}(\bar{s}i\sigma _{\mu \nu }\frac{q^{\nu }}{q^{2}}P_{R}b)(%
\bar{l}\gamma ^{\mu }l)\bigg\}.  \label{FAUED}
\end{eqnarray}

\subsection{Family Non-universal $Z^{\prime }$ Model}

A family non-universal $Z^{\prime }$ boson could be derived naturally in
many extensions of the SM and the one easiest way to get it is to include an
additional $U^{\prime }(1)$ gauge symmetry. This has been formulated in
detail by Langacker and Pl\"{u}macher \cite{76}. Now in a family
non-universal $Z^{\prime }$ model, FCNC transitions $b\rightarrow s\ell
^{+}\ell ^{-}$ could be induced at tree level because of the non-diagonal chiral
coupling matrix. Taken to be for granted, the couplings of right handed
quark flavors with $Z^{\prime }$ boson are diagonal and ignoring $%
Z-Z^{\prime }$ mixing, the effective Hamiltonian for $b\rightarrow s\ell
^{+}\ell ^{-}$ can be written as \cite{77}
\begin{equation}
\mathcal{H}_{eff}^{Z^{\prime }}(b\rightarrow s\ell ^{+}\ell ^{-})=-\frac{%
2G_{F}}{\sqrt{2}}V_{tb}^{\ast }V_{ts}B_{sb}[\frac{4\pi }{\alpha V_{tb}^{\ast
}V_{ts}}S_{\ell \ell }^{L}\bar{\ell}\gamma ^{\mu }\left( 1-\gamma
^{5}\right) \ell -\frac{4\pi S_{\ell \ell }^{R}}{\alpha V_{tb}^{\ast }V_{ts}}%
\bar{\ell}\gamma ^{\mu }\left( 1+\gamma ^{5}\right) \ell ]\bar{s}\gamma
_{\mu }\left( 1-\gamma ^{5}\right) b+h.c.  \label{10}
\end{equation}%
where $S_{\ell \ell }^{L}$ and $S_{\ell \ell }^{R}$ represents the coupling
of $Z^{\prime }$ boson with the left and right handed leptons, respectively
and $B_{sb}$ corresponds to the off diagonal left handed coupling of quarks
with new $Z^{\prime }$ boson in a case when the weak phase $\phi _{sb}$ is
neglected. In a situation when the weak phase is introduced in the off
diagonal coupling then this coupling reads as $B_{sb}=|B_{sb}|e^{-i\phi
_{sb}}$. One can reformulate the effective Hamiltonian given in Eq.(\ref{10}%
) as
\begin{equation*}
H_{eff}^{Z^{\prime }}(b\rightarrow s\ell ^{+}\ell ^{-})=-\frac{4G_{F}}{\sqrt{%
2}}V_{tb}^{\ast }V_{ts}\left[ \bar{C}_{9}^{Z^{\prime }}O_{9}+\bar{C}%
_{10}^{Z^{\prime }}O_{10}\right] +h.c.
\end{equation*}%
where%
\begin{eqnarray}
\bar{C}_{9}^{Z^{\prime }} &=&\frac{4\pi e^{-i\phi _{sb}}}{\alpha
V_{tb}^{\ast }V_{ts}}\Re[B_{sb}]S_{LL}  \notag \\
\bar{C}_{10}^{Z^{\prime }} &=&\frac{4\pi e^{-i\phi _{sb}}}{\alpha
V_{tb}^{\ast }V_{ts}}\Re[B_{sb}]D_{LL}  \label{C910bar}
\end{eqnarray}%
with%
\begin{eqnarray*}
S_{LL} &=&S_{\ell \ell }^{L}+S_{\ell \ell }^{R} \\
D_{LL} &=&S_{\ell \ell }^{L}-S_{\ell \ell }^{R}
\end{eqnarray*}

Hence the contribution of $Z^{\prime }\,$\ boson leads to the modification
the Wilson coefficients $C_{9}$ and $C_{10}$ which now take the form%
\begin{eqnarray*}
C_{9}^{Z^{\prime }} &=&C_{9}^{SM}+\bar{C}_{9}^{Z^{\prime }} \\
C_{10}^{Z^{\prime }} &=&C_{10}^{SM}+\bar{C}_{10}^{Z^{\prime }}
\end{eqnarray*}%
while the Wilson coefficient $C_{7}$ remains unchanged.

\section{Matrix Elements and Form Factors in Light Cone Sum Rules}

In order to calculate the decay amplitudes for $B_{s}\rightarrow \phi \ell
^{+}\ell ^{-}$ at hadron level, we have to sandwich the free quark
amplitudes between the initial and final meson states. Consequently, the
following four hadronic matrix elements
\begin{eqnarray}
\langle \phi (k,\varepsilon )|\bar{s}\gamma _{\mu }b|B_{s}(p)\rangle
&,&\,\,\,\langle \phi (k,\varepsilon )|\bar{s}\gamma _{\mu }\gamma
_{5}b|B_{s}(p)\rangle ,  \notag \\
\langle \phi (k,\varepsilon )|\bar{s}\sigma _{\mu \nu }b|B_{s}(p)\rangle
&,&\,\,\,\langle \phi (k,\varepsilon )|\bar{s}\sigma _{\mu \nu }\gamma
_{5}b|B_{s}(p)\rangle ,  \label{nonvanishing-melements}
\end{eqnarray}%
need to be computed. The above matrix elements can be parameterized in terms
of the form factors as
\begin{eqnarray}
\langle \phi (k,\varepsilon )|\bar{s}\gamma _{\mu }b|B_{s}(p)\rangle
&=&\varepsilon _{\mu \nu \rho \sigma }\varepsilon ^{\ast \nu }p^{\rho
}k^{\sigma }\frac{2V\left( q^{2}\right) }{M_{B_{s}}+M_{\phi }},\,\,
\label{vectormatrix element} \\
\,\langle \phi (k,\varepsilon )|\bar{s}\gamma _{\mu }\gamma
_{5}b|B_{s}(p)\rangle &=&i\varepsilon _{\mu }^{\ast }\left(
M_{B_{s}}+M_{\phi }\right) A_{1}\left( q^{2}\right) -i(p+k)_{\mu
}(\varepsilon ^{\ast }\cdot q)\frac{A_{2}\left( q^{2}\right) }{%
M_{B_{s}}+M_{\phi }}  \notag \\
&&-iq_{\mu }(\varepsilon ^{\ast }\cdot q)\frac{2M_{\phi }}{q^{2}}\left[
A_{3}\left( q^{2}\right) -A_{0}\left( q^{2}\right) \right] ,\,\,
\label{axialvector-melement} \\
\langle \phi \left( k,\varepsilon \right) |\bar{s}\sigma _{\mu \nu }q^{\nu
}b|B_{s}\left( p\right) \rangle &=&i\varepsilon _{\mu \nu \rho \sigma
}\varepsilon ^{\ast \nu }p^{\rho }k^{\sigma }2T_{1}\left( q^{2}\right) ,\,\,
\label{tensor matrix element 1} \\
\langle \phi \left( k,\varepsilon \right) |\bar{s}\sigma _{\mu \nu }\gamma
_{5}q^{\nu }b|B_{s}\left( p\right) \rangle &=&T_{2}\left( q^{2}\right) \left[
\varepsilon _{\mu }^{\ast }\left( M_{B_{s}}^{2}-M_{\phi }^{2}\right)
-(p+k)_{\mu }(\varepsilon ^{\ast }\cdot q)\right]  \notag \\
&&+T_{3}\left( q^{2}\right) (\varepsilon ^{\ast }\cdot q)\left[ q_{\mu }-%
\frac{q^{2}}{M_{B_{s}}^{2}-M_{\phi }^{2}}(p+k)_{\mu }\right] ,
\label{tensor matrix element 2}
\end{eqnarray}%
where all the form factors $A_{i}$ and $T_{i}$ are functions of the square
of momentum transfer $q^{2}=\left( p-k\right) ^{2}$ and $\varepsilon ^{\ast
\nu }$ is the polarization of the final state vector meson $\left( \phi
\right) $ .
The form factors $A_{i}$ and $T_{i}$ appearing in the above equations are
not independent but can be related to each other with the help of equation
of motion. The various relationship between the form factors are \cite{44}
\begin{eqnarray}
A_{3}\left( q^{2}\right) &=&\frac{M_{B_{s}}+M_{\phi }}{2M_{\phi }}%
A_{1}\left( q^{2}\right) -\frac{M_{B_{s}}-M_{\phi }}{2M_{\phi }}A_{2}\left(
q^{2}\right)  \notag \\
A_{3}\left( 0\right) &=&A_{0}\left( 0\right) \text{, }T_{1}\left( 0\right)
=T_{2}\left( 0\right) .  \label{form-factor-relation}
\end{eqnarray}%
The form factors for $B_{s}\rightarrow \phi $ transition are the
non-perturbative quantities and are needed to be calculated using different
approaches (both perturbative and non-perturbative) like Lattice QCD, QCD
sum rules, Light Cone sum rules, etc. Here, we will consider the form
factors calculated by using the Light Cone Sum Rules approach by Ball and
Braun \cite{45}. The form factors $V,A_{0}$ and $T_{1}$ are parameterized by

\begin{equation}
F(q^{2})=\frac{r_{1}}{1-q^{2}/m_{R}^{2}}+\frac{r_{2}}{1-q^{2}/m_{fit}^{2}}
\label{form-factor-sq}
\end{equation}%
While the form factors $A_{2}$ and $\tilde{T}_{3}$ are parameterized as
follows,
\begin{equation}
F(q^{2})=\frac{r_{1}}{1-q^{2}/m^{2}}+\frac{r_{2}}{(1-q^{2}/m^{2})^{2}}
\label{form-factor-sq1}
\end{equation}%
The fit formula for $A_{1}$ and $T_{2}$ is
\begin{equation}
F(q^{2})=\frac{r_{2}}{1-q^{2}/m_{fit}^{2}}  \label{form-factor-sq2}
\end{equation}%
The form factor $T_{3}$ can be obtained through the relation
\begin{equation*}
T_{3}(q^{2})=\frac{M_{B_{s}}^{2}-M_{\phi }^{2}}{q^{2}}[\tilde{T_{3}}%
(q^{2}-T_{2}(q^{2})],
\end{equation*}%
where the values of different parameters are summarized in Table I.
\begin{table}[tbh]
\caption{Fit parameters for $B_{s}\rightarrow \protect\phi $ transition form
factors. $F(0)$ denotes the value of form factors at $q^{2}=0$ Eq. (\protect
\ref{form-factor-sq})\protect\cite{45}. The theoretical uncertainty estimated
is around $15\%$.}
\label{di-fit B_s to phi)}%
\begin{tabular}{ccccccc}
\hline\hline
& $F(q^{2})$ & $\hspace{1 cm} F(0)$ & $\hspace{1 cm} r_{1}$ & $\hspace{1 cm}
m_{R}^{2}$ & $\hspace{1 cm} r_{2}$ & $\hspace{1 cm} m_{fit}^{2}$ \\ \hline
& $A_{1}(q^{2})$ & $\hspace{1 cm} 0.311$ & $\hspace{1 cm}-$ & $\hspace{1 cm}%
- $ & $\hspace{1 cm} 0.308$ & $\hspace{1 cm} 36.54$ \\ \hline
& $A_{2}(q^{2})$ & $\hspace{1 cm} 0.234$ & $\hspace{1 cm} -0.054$ & $\hspace{%
1 cm} -$ & $\hspace{1 cm} 0.288$ & $\hspace{1 cm} 48.94$ \\ \hline
& $A_{0}(q^{2})$ & $\hspace{1 cm} 0.474$ & $\hspace{1 cm} 3.310$ & $\hspace{%
1 cm} 5.28^{2}$ & $\hspace{1 cm} -2.835$ & $\hspace{1 cm} 31.57$ \\ \hline
& $V(q^{2})$ & $\hspace{1 cm} 0.434$ & $\hspace{1 cm} 1.484$ & $\hspace{1 cm}
5.32^{2}$ & $\hspace{1 cm} -1.049$ & $\hspace{1 cm} 39.52$ \\ \hline
& $T_{1}(q^{2})$ & $\hspace{1 cm} 0.349$ & $\hspace{1 cm} 1.303$ & $\hspace{%
1 cm} 5.32^{2}$ & $\hspace{1 cm} -0.954$ & $\hspace{1 cm} 38.28$ \\ \hline
& $T_{2}(q^{2})$ & $\hspace{1 cm} 0.349$ & $\hspace{1 cm} -$ & $\hspace{1 cm}
-$ & $\hspace{1 cm} 0.349$ & $\hspace{1 cm} 37.21$ \\ \hline
& $\tilde{T}_{3}(q^{2})$ & $\hspace{1 cm} 0.349$ & $\hspace{1 cm} 0.027$ & $%
\hspace{1 cm} -$ & $\hspace{1 cm} 0.321$ & $\hspace{1 cm} 45.56$ \\
\hline\hline
&  &  &  &  &  &
\end{tabular}%
\end{table}

From Eqs. (\ref{vectormatrix element} - \ref{tensor matrix element 2})
it is straightforward to find the matrix elements for $B_{s}\rightarrow \phi
\ell ^{+}\ell ^{-}$ as follows:
\begin{equation}
\mathcal{M=}-\frac{G_{F}\alpha }{2\sqrt{2}\pi M_{B_{s}}^{2}}%
V_{tb}V_{ts}^{\ast }\left[ \mathcal{T}_{\mu }^{1}\left( \bar{l}\gamma ^{\mu
}l\right) +\mathcal{T}_{\mu }^{2}\left( \bar{l}\gamma ^{\mu }\gamma
_{5}l\right) +\mathcal{T}^{3}\left( \bar{l}l\right) \right]
\label{matrix-elements}
\end{equation}%
where
\begin{eqnarray}
\mathcal{T}_{\mu }^{1} &=&f_{1}(q^{2})\epsilon _{\mu \nu \alpha \beta
}\varepsilon ^{\ast \nu }p^{\alpha }k^{\beta }-if_{2}(q^{2})\varepsilon
_{\mu }^{\ast }+if_{3}(q^{2})(\varepsilon ^{\ast }\cdot p)P_{\mu }
\label{60} \\
\mathcal{T}_{\mu }^{2} &=&f_{4}(q^{2})\epsilon _{\mu \nu \alpha \beta
}\varepsilon ^{\ast \nu }p^{\alpha }k^{\beta }-if_{5}(q^{2})\varepsilon
_{\mu }^{\ast }+if_{6}(q^{2})(\varepsilon ^{\ast }\cdot p)P_{\mu
}+if_{7}(q^{2})(\varepsilon ^{\ast }\cdot p)P_{\mu }  \label{61}
\end{eqnarray}%
with $P_{\mu }=p_{\mu }+k_{\mu }$. The auxiliary functions appearing in
above equation can be written as
\begin{align}
f_{1}(q^{2})=& 4\widetilde{C}_{7}(\frac{m_{b}+m_{s}}{q^{2}})T_{1}(q^{2})+%
\widetilde{C}_{9}\frac{2{V}(q^{2})}{M_{B_{s}}+M_{\phi }}  \label{621a} \\
f_{2}(q^{2})=& 2\widetilde{C}_{7}(\frac{m_{b}-m_{s}}{q^{2}})T_{2}(q^{2})\left(
M_{B_{s}}^{2}-M_{\phi }^{2}\right)   \notag \\
& +\widehat{C}_{9}A_{1}(q^{2})\left( M_{B_{s}}+M_{\phi }\right)
\label{621b} \\
f_{3}(q^{2})=& 4\widetilde{C}_{7}(\frac{m_{b}-m_{s}}{q^{2}})\left(
T_{2}(q^{2})+q^{2}\frac{T_{3}(q^{2})}{\left( M_{B_{s}}^{2}-M_{\phi
}^{2}\right) }\right)   \notag \\
& +\widehat{C}_{9}\frac{A_{+}(q^{2})}{M_{B_{s}}+M_{\phi }}  \label{621c} \\
f_{4}(q^{2})=& \widetilde{C}_{10}\frac{2V(q^{2})}{M_{B_{s}}+M_{\phi }}
\label{621d} \\
f_{5}(q^{2})=& 2\widetilde{C}_{10}A_{0}(q^{2})\left( M_{B_{s}}+M_{\phi
}\right)   \label{621e} \\
f_{6}(q^{2})=& 2\widetilde{C}_{10}\frac{A_{1}(q^{2})}{M_{B_{s}}+M_{\phi }}
\label{621f} \\
f_{7}(q^{2})=& 4\widetilde{C}_{10}\frac{A_{2}(q^{2})}{M_{B_{s}}+M_{\phi }}
\label{621g}
\end{align}%
Here the Wilson coefficients $\widetilde{C}_{i}$ will
be different for different models and these are gathered in Table II.
\begin{table}[tbh]
\caption{Wilson coefficients corresponding to the models under discussion here.}%
\label{WCs}\centering
\begin{tabular}{cccc}
\hline\hline
& $\hspace{1cm}$SM & UED & $\hspace{1cm}Z^{\prime }$model \\ \hline
$\widetilde{C}_{7}$ & $\hspace{1cm}C_{7}^{SM}$ & $C_{7}^{UED}$ & $\hspace{1cm%
}C_{7}^{SM}$ \\ \hline
$\widetilde{C}_{9}$ & $\hspace{1cm}C_{9}^{SM}$ & $C_{9}^{UED}$ & $\hspace{1cm%
}C_{9}^{Z^{\prime }}$ \\ \hline
$\widetilde{C}_{10}$ & $\hspace{1cm}C_{10}^{SM}$ & $C_{10}^{UED}$ & $\hspace{%
1cm}C_{10}^{Z^{\prime }}$ \\ \hline\hline
\end{tabular}%
\end{table}

\section{Formula for Observables}

In this section we will present the calculations of the physical observables
such as the branching ratios $\mathcal{BR}$, the forward-backward
asymmetries $\mathcal{A}_{FB}$, the single lepton polarization asymmetries $%
P_{L,N,T}$, the double lepton polarization asymmetries $P_{ij}$ $%
(i,j=L,N,T)$, the helicity fractions $f_{L,T}$ of the $\phi $ meson and
the polarized and un-polarized $CP$ asymmetries of the final state lepton in $%
B_{s}\rightarrow \phi \ell ^{+}\ell ^{-}$ decay.

\subsection{The Differential Decay Rate}

In the rest frame of $B_{s}$ meson the differential decay width of $%
B_{s}\rightarrow \phi \ell ^{+}\ell ^{-}$ can be written as
\begin{equation}
\frac{d\Gamma (B_{s}\rightarrow \phi \ell ^{+}\ell ^{-})}{dq^{2}}=\frac{1}{%
\left( 2\pi \right) ^{3}}\frac{1}{32M_{B_{s}}^{3}}%
\int_{-u(q^{2})}^{+u(q^{2})}du\left\vert \mathcal{M}\right\vert ^{2}
\label{62a}
\end{equation}%
where
\begin{eqnarray}
q^{2} &=&(p_{+}+p_{-})^{2} \\
u &=&\left( p-p_{-}\right) ^{2}-\left( p-p_{+}\right) ^{2}.
\end{eqnarray}%
The limits on $q^{2}$ and $u$ are
\begin{eqnarray}
4m_{l}^{2} &\leq &q^{2}\leq (M_{B_{s}}-M_{\phi })^{2}  \label{62d} \\
-u(q^{2}) &\leq &u\leq u(q^{2})  \label{62e}
\end{eqnarray}%
with%
\begin{equation}
u(q^{2})=\sqrt{\lambda \left( 1-\frac{4m_{l}^{2}}{q^{2}}\right) }
\label{62f}
\end{equation}%
and%
\begin{equation}
\lambda \equiv \lambda (M_{B_{s}}^{2},M_{\phi
}^{2},q^{2})=M_{B_{s}}^{4}+M_{\phi }^{4}+q^{4}-2M_{B_{s}}^{2}M_{\phi
}^{2}-2M_{\phi }^{2}q^{2}-2q^{2}M_{B_{s}}^{2}  \label{lambda}
\end{equation}%
In above expressions , $m_{l}$ corresponds to the mass of the lepton which for
our case can be $\mu $ or $\tau $. The total decay rate for the decay $%
B_{s}\rightarrow \phi \ell ^{+}\ell ^{-}$ can take the form\
\begin{equation}
\frac{d\Gamma }{dq^{2}}=\frac{G_{F}^{2}\left\vert V_{tb}V_{ts}^{\ast
}\right\vert ^{2}\alpha ^{2}}{2^{11}\pi ^{5}3M_{B_{s}}^{3}M_{\phi }^{2}q^{2}}%
u(q^{2})\times \mathcal{A}\left( q^{2}\right)  \label{Drate}
\end{equation}%
The function $u(q^{2})$ is defined in Eq. (\ref{62f}) and $\mathcal{A}(q^{2})$
is given by
\begin{align}
\mathcal{A}(q^{2})& =8M_{\phi }^{2}q^{2}\lambda \bigg\{(2m_{l}^{2}+q^{2})%
\left\vert f_{1}(q^{2})\right\vert ^{2}-(4m_{l}^{2}-q^{2})\left\vert
f_{4}(q^{2})\right\vert ^{2}\bigg\}+4M_{\phi }^{2}q^{2}\bigg\{%
(2m_{l}^{2}+q^{2})  \notag \\
& \times \left( 3\left\vert f_{2}(q^{2})\right\vert ^{2}-\lambda \left\vert
f_{3}(q^{2})\right\vert ^{2}\right) -(4m_{l}^{2}-q^{2})\left( 3\left\vert
f_{5}(q^{2})\right\vert ^{2}-\lambda \left\vert f_{6}(q^{2})\right\vert
^{2}\right) \bigg\}  \notag \\
& +\lambda (2m_{l}^{2}+q^{2})\left\vert f_{2}(q^{2})+\left(
M_{B_{s}}^{2}-M_{\phi }^{2}-q^{2}\right) f_{3}(q^{2})\right\vert
^{2}+24m_{l}^{2}M_{\phi }^{2}\lambda \left\vert f_{7}(q^{2})\right\vert ^{2}
\notag \\
& -(4m_{l}^{2}-q^{2})\left\vert f_{5}(q^{2})+\left( M_{B_{s}}^{2}-M_{\phi
}^{2}-q^{2}\right) f_{6}(q^{2})\right\vert ^{2}
\notag \\
& -12m_{l}^{2}q^{2}\left[ \Re (f_{5}f_{7}^{\ast })-\Re (f_{6}f_{7}^{\ast })%
\right] .  \label{63b}
\end{align}

\subsection{Forward-Backward Asymmetries}

The differential forward-backward asymmetry $\mathcal{A}_{FB}$ of final
state lepton for the said decay can be written as
\begin{equation}
{\frac{d\mathcal{A}_{FB}(s)}{dq^{2}}}=\int_{0}^{1}\frac{d^{2}\Gamma }{%
dq^{2}d\cos \theta }d\cos \theta -\int_{-1}^{0}\frac{d^{2}\Gamma }{%
dq^{2}d\cos \theta }d\cos \theta  \label{FBformula}
\end{equation}%
From experimental point of view the normalized forward-backward asymmetry is
more useful, defined as
\begin{equation*}
\mathcal{A}_{FB}=\frac{\int_{0}^{1}\frac{d^{2}\Gamma }{dq^{2}d\cos \theta }%
d\cos \theta -\int_{-1}^{0}\frac{d^{2}\Gamma }{dq^{2}d\cos \theta }d\cos
\theta }{\int_{-1}^{1}\frac{d^{2}\Gamma }{dq^{2}d\cos \theta }d\cos \theta }
\end{equation*}%
The normalized $\mathcal{A}_{FB}$ for $B_{s}\rightarrow \phi \ell ^{+}\ell
^{-}$ can be obtained from Eq. (\ref{62a}), as%
\begin{eqnarray}
\mathcal{A}_{FB} &=&-\frac{1}{d\Gamma /dq^{2}}\frac{G_{F}^{2}\alpha ^{2}}{%
2^{11}\pi ^{5}M_{B_{s}}^{3}}\left\vert V_{tb}V_{ts}^{\ast }\right\vert
^{2}q^{2}u(q^{2})\times \bigg\{4{Re}[f_{2}^{\ast }f_{4}+f_{1}^{\ast }f_{5}]\bigg\}  \label{FBexpression}
\end{eqnarray}%
where $d\Gamma /dq^{2}$ is given in Eq. (\ref{Drate}). Confining ourself
to the SM, the above expression of the FB asymmetry in terms of the Wilson
coefficients becomes%
\begin{eqnarray}
\mathcal{A}_{FB} &=&-\frac{1}{d\Gamma /dq^{2}}\frac{G_{F}^{2}\alpha ^{2}}{%
2^{8}\pi ^{5}M_{B_{s}}^{3}}\left\vert V_{tb}V_{ts}^{\ast }\right\vert
^{2}q^{2}u(q^{2})\times C_{10}  \label{FBFform} \\
&&\bigg\{\Re \left( C_{9}^{eff}\right) V\left( q^{2}\right) A_{1}\left(
q^{2}\right) \frac{m_{b}}{q^{2}}C_{7}^{eff}\left( V\left( q^{2}\right)
T_{2}\left( q^{2}\right) \left( M_{B_{s}}-M_{\phi }\right) +A_{1}\left(
q^{2}\right) T_{1}\left( q^{2}\right) \left( M_{B_{s}}+M_{\phi }\right)
\right) \bigg\}  \notag
\end{eqnarray}
which in agreement with the the one obtained for $B\rightarrow K^{\ast
}l^{+}l^{-}$decay in \cite{10}.

\subsection{Lepton Polarization Asymmetries}

In the rest frame of the lepton $\ell ^{-}$, the unit vectors along
longitudinal, normal and transversal component of the $\ell ^{-}$ can be
defined as \cite{jam,jama,Aliev}:
\begin{subequations}
\begin{eqnarray}
s_{L}^{-\mu } &=&(0,\vec{e}_{L}^{-})=\left( 0,\frac{\vec{p}_{-}}{\left\vert
\vec{p}_{-}\right\vert }\right) ,  \label{p-vectorsa} \\
s_{N}^{-\mu } &=&(0,\vec{e}_{N}^{-})=\left( 0,\frac{\vec{k}\times \vec{p}_{-}%
}{\left\vert \vec{k}\times \vec{p}_{-}\right\vert }\right) ,
\label{p-vectorsb} \\
s_{T}^{-\mu } &=&(0,\vec{e}_{T}^{-})=\left( 0,\vec{e}_{N}\times \vec{e}%
_{L}\right) ,  \label{p-vectorsc}
\end{eqnarray}%
where $\vec{p}_{-}$ and $\vec{k}$ are the three-momenta of the lepton $\ell
^{-}$ and $\phi $ meson respectively in the center mass (c.m.) frame of $%
\ell ^{+}\ell ^{-}$ system. Lorentz transformation is used to boost the
longitudinal component of the lepton polarization to the c.m. frame of the
lepton pair as
\end{subequations}
\begin{equation}
\left( s_{L}^{-\mu }\right) _{CM}=\left( \frac{|\vec{p}_{-}|}{m_{l}},\frac{E%
\vec{p}_{-}}{m_{l}\left\vert \vec{p}_{-}\right\vert }\right)
\label{bossted component}
\end{equation}%
where $E$ and $m_{l}$ are the energy and mass of the lepton. The normal and
transverse components remain unchanged under the Lorentz boost. The
longitudinal ($P_{L}$), normal ($P_{N}$) and transverse ($P_{T}$)
polarizations of lepton can be defined as:
\begin{equation}
P_{i}^{(\mp )}(q^{2})=\frac{\frac{d\Gamma }{dq^{2}}(\vec{\xi}^{\mp }=\vec{e_{i}}%
^{\mp })-\frac{d\Gamma }{dq^{2}}(\vec{\xi}^{\mp }=-\vec{e_{i}}^{\mp })}{\frac{%
d\Gamma }{dq^{2}}(\vec{\xi}^{\mp }=\vec{e_{i}}^{\mp })+\frac{d\Gamma }{dq^{2}}(%
\vec{\xi}^{\mp }=-\vec{e_{i}}^{\mp })}  \label{polarization-defination}
\end{equation}%
where $i=L,\;N,\;T$ and $\vec{\xi}^{\mp }$ is the spin direction along the
leptons $\ell ^{\mp }$. The differential decay rate for polarized lepton $%
\ell ^{\mp }$ in $B_{s}\rightarrow \phi \ell ^{+}\ell ^{-}$ decay along any
spin direction $\vec{\xi}^{\mp }$ is related to the unpolarized decay rate (%
\ref{Drate}) with the following relation
\begin{equation}
\frac{d\Gamma (\vec{\xi}^{\mp })}{dq^{2}}=\frac{1}{2}\left( \frac{d\Gamma }{%
dq^{2}}\right) \left[ 1+(P_{L}^{\mp }\vec{e}_{L}^{\mp }+P_{N}^{\mp }\vec{e}%
_{N}^{\mp }+P_{T}^{\mp }\vec{e}_{T}^{\mp })\cdot \vec{\xi}^{\mp }\right] .
\label{polarized-decay}
\end{equation}%
The expressions of the longitudinal, normal and transverse lepton
polarizations can be written as
\begin{align}
P_{L}(q^{2})\propto & \frac{4\lambda }{3M_{\phi }^{2}}\sqrt{\frac{%
q^{2}-4m_{l}^{2}}{q^{2}}}\times \bigg\{2\Re (f_{2}f_{5}^{\ast })+\lambda \Re
(f_{3}f_{6}^{\ast })+4\sqrt{q^{2}}\Re (f_{1}f_{4}^{\ast })\left( 1+\frac{%
12q^{2}M_{\phi }^{2}}{\lambda }\right)   \notag \\
& +\left( -M_{B_{s}}^{2}+M_{\phi }^{2}+q^{2}\right) \left[ \Re
(f_{3}f_{5}^{\ast })+\Re (f_{2}f_{6}^{\ast })\right]\bigg\}  \label{long-polarization}
\end{align}%
\begin{align}
P_{N}(q^{2})\propto & -\frac{m_{l}\pi }{M_{\phi }^{2}}\sqrt{\frac{\lambda }{%
q^{2}}}\times \bigg\{\lambda q^{2}\Re (f_{3}f_{7}^{\ast })-\lambda
(M_{B_{s}}^{2}-M_{\phi }^{2})\Re (f_{3}f_{6}^{\ast })+\lambda \Re
(f_{3}f_{5}^{\ast })  \notag \\
& +\left( M_{B_{s}}^{2}-M_{\phi }^{2}-q^{2}\right) \left[ q^{2}\Re
(f_{2}f_{7}^{\ast })+(M_{B_{s}}^{2}-M_{\phi }^{2})\Re (f_{2}f_{5}^{\ast
})\right]+8q^{2}M_{\phi }^{2}\Re (f_{1}f_{2}^{\ast })\bigg\}  \label{norm-polarization}
\\
P_{T}\left( q^{2}\right) \propto & i\frac{m_{l}\pi \sqrt{\left( q^{2}-\frac{%
4m_{l}^{2}}{q^{2}}\right) \lambda }}{M_{\phi }^{2}}\bigg\{M_{\phi }\left[
4\Im (f_{2}f_{4}^{\ast })+4\Im (f_{1}f_{5}^{\ast })+3\Im (f_{5}f_{6}^{\ast })%
\right]   \notag \\
& -\lambda \Im (f_{6}f_{7}^{\ast })+\left( -M_{B_{s}}^{2}+M_{\phi
}^{2}+q^{2}\right) \Im (f_{7}f_{5}^{\ast })%
-q^{2}\Im (f_{5}f_{6}^{\ast })\bigg\}  \label{Transverse-polarization}
\end{align}%
where $f_{1},f_{2},\cdots ,f_{7}$ are the auxiliary functions defined above.
Here we have dropped out the constant factors which are however understood.

\subsection{Double Lepton Polarization Asymmetries}

To calculate the double--polarization asymmetries, we consider the
polarizations of both lepton and anti-lepton, simultaneously and introduce
the following spin projection operators for the lepton $\ell ^{-}$ and the
anti-lepton $\ell ^{+}$\cite{DPLBashiry}:%
\begin{eqnarray}
\Lambda _{1} &=&\frac{1}{2}\left( 1+\gamma _{5}\slashed{s}_{i}^{-}\right)
\notag \\
\Lambda _{2} &=&\frac{1}{2}\left( 1+\gamma _{5}\slashed{s}_{i}^{+}\right)
\label{DLPV}
\end{eqnarray}%
where $i=L,T\,\ $and $N$ corresponds to the longitudinal, transverse and
normal lepton polarizations, respectively. In the rest frame of the
lepton-anti-lepton one can define the following set of orthogonal vectors $%
s^{\mu }$:%
\begin{eqnarray}
s_{L}^{-\mu } &=&(0,\vec{e}_{L}^{-})=\left( 0,\frac{\vec{p}_{-}}{\left\vert
\vec{p}_{-}\right\vert }\right) ,  \notag \\
s_{N}^{-\mu } &=&(0,\vec{e}_{N}^{-})=\left( 0,\frac{\vec{k}\times \vec{p}_{-}%
}{\left\vert \vec{k}\times \vec{p}_{-}\right\vert }\right) ,  \notag \\
s_{T}^{-\mu } &=&(0,\vec{e}_{T}^{-})=\left( 0,\vec{e}_{N}\times \vec{e}%
_{L}\right) ,  \label{DLvectors} \\
s_{L}^{+\mu } &=&(0,\vec{e}_{L}^{+})=\left( 0,\frac{\vec{p}_{+}}{\left\vert
\vec{p}_{+}\right\vert }\right) ,  \notag \\
s_{N}^{+\mu } &=&(0,\vec{e}_{N}^{+})=\left( 0,\frac{\vec{k}\times \vec{p}_{+}%
}{\left\vert \vec{k}\times \vec{p}_{+}\right\vert }\right) ,  \notag \\
s_{T}^{+\mu } &=&(0,\vec{e}_{T}^{+})=\left( 0,\vec{e}_{N}\times \vec{e}%
_{L}\right) .  \notag
\end{eqnarray}%
Just like the single lepton polarization, through Lorentz transformations we
can boost the longitudinal component in the CM frame of $\ell ^{-}\ell ^{+}$
as%
\begin{eqnarray}
\left( s_{L}^{-\mu }\right) _{CM} &=&\left( \frac{|\vec{p}_{-}|}{m_{l}},%
\frac{E\vec{p}_{-}}{m_{l}\left\vert \vec{p}_{-}\right\vert }\right)  \notag
\\
\left( s_{L}^{+\mu }\right) _{CM} &=&\left( \frac{|\vec{p}_{+}|}{m_{l}},-%
\frac{E\vec{p}_{+}}{m_{l}\left\vert \vec{p}_{+}\right\vert }\right)
\label{DLboosted}
\end{eqnarray}%
The normal and transverse component remains the same under Lorentz boost. We
now define the double lepton polarization asymmetries as%
\begin{equation}
P_{ij}(q^{2})=\frac{\left( \frac{d\Gamma }{dq^{2}}\left( \vec{s}_{i}^{-},%
\vec{s}_{i}^{+}\right) -\frac{d\Gamma }{dq^{2}}\left( -\vec{s}_{i}^{-},\vec{s%
}_{i}^{+}\right) \right) -\left( \frac{d\Gamma }{dq^{2}}\left( \vec{s}%
_{i}^{-},-\vec{s}_{i}^{+}\right) -\frac{d\Gamma }{dq^{2}}\left( -\vec{s}%
_{i}^{-},-\vec{s}_{i}^{+}\right) \right) }{\left( \frac{d\Gamma }{dq^{2}}%
\left( \vec{s}_{i}^{-},\vec{s}_{i}^{+}\right) -\frac{d\Gamma }{dq^{2}}\left(
-\vec{s}_{i}^{-},\vec{s}_{i}^{+}\right) \right) +\left( \frac{d\Gamma }{%
dq^{2}}\left( \vec{s}_{i}^{-},-\vec{s}_{i}^{+}\right) -\frac{d\Gamma }{dq^{2}%
}\left( -\vec{s}_{i}^{-},-\vec{s}_{i}^{+}\right) \right) }
\label{DLdefinition}
\end{equation}%
where the subscripts $i$ and $j$ corresponds to the lepton and anti-lepton
polarizations, respectively. Using these definitions the various double lepton polarization
asymmetries as a function of $q^{2}$ can be written as

\begin{eqnarray}
P_{LL}(q^{2}) &\propto &\frac{1}{3m_{l}^{2}}\{4\left\vert f_{1}\right\vert
^{2}\left( 8m_{l}^{4}\lambda -q^{2}\mathcal{U}_{2}\right) +4\left\vert
f_{4}\right\vert ^{2}\mathcal{U}_{1}+\frac{1}{M_{\phi }^{2}}[24m_{l}^{4}\lambda
\left( M_{B_{s}}^{2}-M_{\phi }^{2}\right) \left( f_{6}f_{7}^{\ast
}+f_{7}f_{6}^{\ast }\right) \notag \\
&& -12m_{l}^{4}\lambda \left( 2f_{5}f_{7}^{\ast
}+2f_{7}f_{5}^{\ast }\right)+12\left\vert f_{7}\right\vert ^{2}m_{l}^{4}q^{2}\lambda +(\left(
q^{2}-M_{B_{s}}^{2}+M_{\phi }^{2}\right) (\frac{4m_{l}^{4}\lambda }{q^{2}}%
-6m_{l}^{2}\lambda +\mathcal{U}_{2}))\left( f_{1}f_{2}^{\ast }+f_{2}f_{1}^{\ast
}\right)   \notag \\
&&+\left\vert f_{2}\right\vert ^{2}(\frac{4m_{l}^{4}\lambda }{q^{2}}\left(
\lambda +12M_{\phi }^{2}q^{2}\right) -3m_{l}^{2}\left( 2\lambda +8M_{\phi
}^{2}q^{2}\right) +\mathcal{U}_{2})  \label{PLL} \\
&&-\frac{4m_{l}^{4}}{q^{2}}\mathcal{U}_{3}(f_{5}^{\ast }f_{6}+f_{6}^{\ast
}f_{5})+\lambda \left\vert f_{6}\right\vert ^{2}(\frac{4m_{l}^{4}}{q^{2}}%
\mathcal{U}_{4}+2m_{l}^{2}\lambda +\mathcal{U}_{2})  \notag \\
&&+\lambda \left\vert f_{5}\right\vert ^{2}(\frac{4m_{l}^{4}}{q^{2}}(6\left(
q^{2}-M_{B_{s}}^{2}+M_{\phi }^{2}\right) ^{2}-\lambda )-3m_{l}^{2}\left(
6\lambda +8M_{\phi }^{2}q^{2}\right) +q^{2}\mathcal{U}_{2})]\}  \notag
\end{eqnarray}%
\begin{eqnarray}
P_{LT} &\propto &\frac{1}{4M_{\phi }^{2}q^{2}}\pi \sqrt{\lambda }\sqrt{%
q^{2}-4m_{l}^{2}}\{8m_{l}M_{\phi }^{2}q^{2}\left( f_{2}^{\ast
}f_{4}+f_{2}f_{4}^{\ast }\right) +2m_{l}(M_{B_{s}}^{2}q^{2}\left(
f_{5}^{\ast }f_{7}+3f_{5}^{\ast }f_{6}\right)   \notag \\
&&+M_{\phi }^{2}q^{2}\left( 4f_{5}^{\ast }f_{1}-f_{5}^{\ast }f_{7}\right)
-q^{4}\left( f_{5}^{\ast }f_{7}+f_{5}^{\ast }f_{6}\right) -2f_{5}^{\ast
}f_{6}\left( M_{B_{s}}^{2}-M_{\phi }^{2}\right) ^{2}  \label{PLT} \\
&&+2\left( M_{B_{s}}^{2}-M_{\phi }^{2}\right) (\left\vert f_{5}\right\vert
^{2}-f_{5}f_{6}^{\ast }(M_{B_{s}}^{2}-M_{\phi }^{2}))+q^{2}(f_{5}f_{6}^{\ast
}\left( 3M_{B_{s}}^{2}+M_{\phi }^{2}\right) -2\left\vert f_{5}\right\vert
^{2})-f_{5}f_{6}^{\ast }q^{4})  \notag \\
&&+4m_{l}\lambda (q^{2}\left( f_{7}f_{6}^{\ast }+f_{6}f_{7}^{\ast }\right)
+\left\vert f_{6}\right\vert ^{2}\left( M_{B_{s}}^{2}-M_{\phi }^{2}\right)
)\}  \notag
\end{eqnarray}%
\begin{eqnarray}
P_{LN} &\propto &-i\frac{1}{4M_{\phi }^{2}\sqrt{q^{2}}}\pi \sqrt{\lambda }%
\{2m_{l}\left( M_{B}^{2}-M_{\phi }^{2}-q^{2}\right) \left( q^{2}f_{2}^{\ast
}f_{7}+f_{2}^{\ast }f_{6}\left( M_{B}^{2}-M_{\phi }^{2}\right) -f_{2}^{\ast
}f_{5}\right)   \notag \\
&&+2m_{l}\lambda \left( f_{3}^{\ast }f_{5}-f_{3}^{\ast }f_{6}\left(
M_{B}^{2}-M_{\phi }^{2}\right) -q^{2}f_{3}^{\ast }f_{7}\right)\} \label{PLN}
\end{eqnarray}%
\begin{eqnarray}
P_{TL} &\propto &\frac{1}{4M_{\phi }^{2}q^{2}}\pi \sqrt{\lambda }\sqrt{%
q^{2}-4m_{l}^{2}}\{4m_{l}\lambda \left( q^{2}(f_{6}^{\ast }f_{7}+f_{7}^{\ast
}f_{6})+\left\vert f_{7}\right\vert ^{2}(M_{B_{s}}^{2}-M_{\phi }^{2})\right)-8m_{l}M_{\phi }^{2}q^{2}\left( f_{4}^{\ast }f_{2}+f_{2}^{\ast
}f_{4}\right)   \notag \\
&&+2m_{l}(q^{2}\left( M_{\phi }^{2}-M_{B_{s}}^{2}\right) f_{5}^{\ast
}f_{7}+q^{2}\left( M_{\phi }^{2}+3M_{B_{s}}^{2}\right) f_{5}^{\ast }f_{6}+q^{4}\left( f_{5}^{\ast }f_{7}-f_{5}^{\ast }f_{6}\right)
\notag \\
&&-2f_{5}^{\ast
}f_{6}\left( M_{B_{s}}^{2}-M_{\phi }^{2}\right) ^{2}+q^{2}f_{5}^{\ast
}f_{6}\left( 3M_{B_{s}}^{2}+M_{\phi }^{2}\right) +2q^{2}f_{5}^{\ast
}f_{7}(M_{\phi }^{2}-M_{B_{s}}^{2})-2q^{2}\left\vert f_{5}\right\vert ^{2}
\label{PTL} \\
&&+q^{4}(2f_{5}f_{7}^{\ast }-f_{5}f_{6}^{\ast })+2(M_{B_{s}}^{2}-M_{\phi
}^{2})(\left\vert f_{5}\right\vert ^{2}+f_{5}f_{6}^{\ast }(M_{\phi
}^{2}-M_{B_{s}}^{2}))-4M_{\phi }^{2}q^{2}f_{1}f_{5}^{\ast }(2\pi m_{l}\sqrt{%
\lambda }\sqrt{q^{2}-4m_{l}^{2}})\}  \notag
\end{eqnarray}%
\begin{eqnarray}
P_{TN} &\propto &i\{\frac{4}{3}q\sqrt{q^{2}-4m_{l}^{2}}\left( 4\lambda
-3M_{B_{s}}^{4}+6M_{B}^{2}(M_{\phi }^{2}+q^{2})-3(M_{\phi
}^{2}-q^{2})^{2}\right) \notag \\
&&+\sqrt{q^{2}\lambda }u\left( f_{1}f_{4}^{\ast }+f_{1}^{\ast }f_{4}\right)\} \label{PTN}
\end{eqnarray}%
\begin{eqnarray}
P_{TT} &\propto &\frac{1}{3M_{\phi }^{2}q^{2}}\{48\left\vert
f_{2}\right\vert ^{2}m_{l}^{2}M_{\phi }^{2}q^{2}-4m_{l}^{2}\left(
5\left\vert f_{5}\right\vert ^{2}+(M_{B_{s}}^{2}-M_{\phi
}^{2})(f_{3}f_{2}^{\ast }-5f_{5}^{\ast }f_{6})\right)   \notag \\
&&+(2\left\vert f_{5}\right\vert ^{2}+12f_{5}f_{7}^{\ast
}m_{l}^{2}+(M_{B_{s}}^{2}-M_{\phi }^{2})(f_{3}f_{2}^{\ast }-f_{6}f_{5}^{\ast
}))  \label{PTT} \\
&&+4m_{l}^{2}(3+f_{3}f_{2}^{\ast }+8M_{\phi }^{2}(\left\vert
f_{1}\right\vert ^{2}+\left\vert f_{4}\right\vert ^{2})-2f_{6}f_{5}^{\ast
})-2\left( 3(M_{B_{s}}^{2}-M_{\phi }^{2})f_{6}f_{7}^{\ast
}+3(M_{B_{s}}^{2}+M_{\phi }^{2})\left\vert f_{6}\right\vert ^{2}\right)
\notag \\
&&+6q^{2}\left( f_{7}f_{5}^{\ast }-2f_{7}f_{6}^{\ast }(M_{B_{s}}^{2}-M_{\phi
}^{2})\right) +2q^{4}(f_{6}f_{5}^{\ast }-f_{3}f_{2}^{\ast
}+6m_{l}^{2}(\left\vert f_{6}\right\vert ^{2}-\left\vert f_{7}\right\vert
^{2})  \notag \\
&&+4M_{\phi }^{2}(\left\vert f_{1}\right\vert ^{2}-\left\vert
f_{4}\right\vert ^{2}))+2\lambda (q^{2}-2m_{l}^{2})\left( \left\vert
f_{2}\right\vert ^{2}+f_{2}f_{3}^{\ast }(q^{2}-M_{B_{s}}^{2}+M_{\phi
}^{2})\right) +2\lambda ^{2}\left\vert f_{6}\right\vert
^{2}(q^{2}-10m_{l}^{2})\}  \notag
\end{eqnarray}%
\begin{eqnarray}
P_{NN} &\propto &-\frac{1}{3M_{\phi }^{2}q^{2}}\{48\left\vert
f_{2}\right\vert ^{2}m_{l}^{2}M_{\phi
}^{2}q^{2}-4m_{l}^{2}(M_{B_{s}}^{2}-M_{\phi }^{2})(\left\vert
f_{5}\right\vert ^{2}+(f_{3}f_{2}^{\ast }-f_{6}f_{5}^{\ast }))  \notag \\
&&+(4m_{l}^{2}+M_{B}^{2})f_{6}f_{5}^{\ast }+12m_{l}^{2}f_{5}f_{7}^{\ast
}-\left\vert f_{5}\right\vert ^{2}+12m_{l}^{2}M_{B_{s}}^{2}(\left\vert
f_{6}\right\vert ^{2}-f_{6}f_{7}^{\ast })  \label{PNN} \\
&&+4m_{l}^{2}M_{\phi }^{2}\left( 4\left\vert f_{1}\right\vert
^{2}-4\left\vert f_{4}\right\vert ^{2}+3(f_{6}f_{7}^{\ast }+\left\vert
f_{6}\right\vert ^{2})\right) +f_{3}f_{2}^{\ast
}(2m_{l}^{2}-M_{B_{s}}^{2}+M_{\phi }^{2})  \notag \\
&&+6m_{l}^{2}q^{2}(f_{7}f_{5}^{\ast }-2f_{7}f_{6}^{\ast
}(M_{B_{s}}^{2}-M_{\phi }^{2}))+q^{2}(f_{3}f_{2}^{\ast }-f_{6}f_{5}^{\ast
}-6m_{l}^{2}(\left\vert f_{7}\right\vert ^{2}+\left\vert f_{6}\right\vert
^{2})  \notag \\
&&+4(\left\vert f_{4}\right\vert ^{2}-\left\vert f_{1}\right\vert
^{2}))+\lambda (q^{2}+2m_{l}^{2})\left( \left\vert f_{2}\right\vert
^{2}+f_{2}f_{3}^{\ast }(q^{2}-M_{B_{s}}^{2}+M_{\phi }^{2})+2\lambda
^{2}\left\vert f_{6}\right\vert ^{2}\right) \}  \notag
\end{eqnarray}%
with
\begin{eqnarray*}
\mathcal{U}_{1} &=&m_{l}^{2}q^{2}\left( 6M_{B_{s}}^{2}\left( M_{\phi
}^{2}+q^{2}\right) -3\left( M_{\phi }^{2}-q^{2}\right)
^{2}-3M_{B_{s}}^{4}-5\lambda \right) +q^{2}\mathcal{U_{2}} \\
\mathcal{U}_{2} &=&\lambda q^{2}-\sqrt{q^{2}\left( q^{2}-4m_{l}^{2}\right) }u%
\sqrt{\lambda } \\
\mathcal{U}_{3} &=&6\left( M_{B_{s}}^{6}-M_{\phi }^{6}\right) +9M_{\phi
}^{4}q^{2}-3q^{6}-3M_{B_{s}}^{4}\left( 6M_{\phi }^{2}+5q^{2}\right)
+M_{B_{s}}^{2}\left( 18M_{\phi }^{4}+6q^{2}M_{\phi }^{2}+12q^{4}-\lambda
\right)  \\
&&+\lambda \left( M_{B_{s}}^{2}+M_{\phi }^{2}\right) +q^{2}\left(
M_{B_{s}}^{2}-M_{\phi }^{2}-q^{2}\right) \mathcal{U}_{2} \\
\mathcal{U}_{4} &=&6\left( M_{B_{s}}^{2}-M_{\phi }^{2}\right) ^{2}-6q^{2}\left(
M_{B_{s}}^{2}+M_{\phi }^{2}\right) +3q^{4}-\lambda
\end{eqnarray*}%
and $u$ and $\lambda $ are defined in Eqs. (\ref{62f}) and (\ref{lambda}),
respectively.

Just to add few words about the lepton polarization asymmetry: we have seen
that the expressions of various double lepton polarization asymmetries are
function of the $q^{2}$ and of the parameters of NP models. From
experimental point of view, it will be more interesting if we can eliminate
the dependency on one parameter and this we can easily do by performing
integration on $q^{2}$. This will give us the average lepton polarization
asymmetry and it is defined as:
\begin{equation}
\left\langle P_{ij}\right\rangle =\frac{\int\limits_{all\text{ }q^{2}}P_{ij}%
\frac{dB}{dq^{2}}dq^{2}}{\int\limits_{all\text{ }q^{2}}\frac{dB}{dq^{2}}dq^{2}}\label{AvPij}
\end{equation}

\subsection{Helicity Fractions of $\protect\phi$ in $B_{s}\rightarrow
\protect\phi\ell^{+}\ell^{-}$}

We now discuss helicity fractions of $\phi$ in $B_{s}\rightarrow
\phi\ell^{+}\ell^{-}$ which are interesting variable and are as such
independent of the uncertainties arising due to form factors and other input
parameters. The final state meson helicity fractions were already discussed
in literature for $B\rightarrow K^{\ast }\left( K_{1}\right)
\ell^{+}\ell^{-} $ decays \cite{Colangelo, paracha}.

The explicit expression of the decay rate for $B_{s}^{-}\rightarrow \phi
\ell ^{+}\ell ^{-}$ decay can be written in terms of longitudinal $\Gamma
_{L}$ and transverse components $\Gamma _{T}$ as
\begin{equation}
\frac{d\Gamma (q^{2})}{dq^{2}}=\frac{d\Gamma _{L}(q^{2})}{dq^{2}}+\frac{%
d\Gamma _{T}(q^{2})}{dq^{2}}
\end{equation}%
where%
\begin{equation*}
\frac{d\Gamma _{T}(q^{2})}{dq^{2}}=\frac{d\Gamma _{+}(q^{2})}{dq^{2}}+\frac{%
d\Gamma _{-}(q^{2})}{dq^{2}}
\end{equation*}%
and
\begin{eqnarray}
\frac{d\Gamma _{L}(q^{2})}{dq^{2}} &=&\frac{G_{F}^{2}\left\vert
V_{tb}V_{ts}^{\ast }\right\vert ^{2}\alpha ^{2}}{2^{11}\pi ^{5}}\frac{%
u(q^{2})}{M_{B_{s}}^{3}}\times \frac{1}{3}\mathcal{A}_{L}  \label{65d} \\
\frac{d\Gamma _{\pm }(q^{2})}{dq^{2}} &=&\frac{G_{F}^{2}\left\vert
V_{tb}V_{ts}^{\ast }\right\vert ^{2}\alpha ^{2}}{2^{11}\pi ^{5}}\frac{%
u(q^{2})}{M_{B_{s}}^{3}}\times \frac{4}{3}\mathcal{A}_{\pm }.  \label{65h}
\end{eqnarray}

The different functions appearing in Eqs. (\ref{65d}) and (\ref{65h}) can be
expressed in terms of auxiliary functions (cf. Eqs. (\ref{621a}-\ref{621g}))
as
\begin{align}
\mathcal{A}_{L}& =\frac{1}{q^{2}M_{\phi }^{2}}\bigg[24\left\vert
f_{7}(q^{2})\right\vert ^{2}m^{2}M_{\phi }^{2}\lambda
+(2m^{2}+q^{2})\left\vert (M_{B_{s}}^{2}-M_{\phi
}^{2}-q^{2})f_{2}(q^{2})+\lambda f_{3}(q^{2})\right\vert ^{2}  \notag \\
& +(q^{2}-4m^{2})\left\vert (M_{B_{s}}^{2}-M_{\phi
}^{2}-q^{2})f_{5}(q^{2})+\lambda f_{6}(q^{2})\right\vert ^{2}\bigg] \label{65j}
\end{align}%
\begin{equation}
A_{\pm }=(q^{2}-4m^{2})\left\vert f_{5}(q^{2})\mp \sqrt{\lambda }%
f_{4}(q^{2})\right\vert ^{2}+\left( q^{2}+2m^{2}\right) \left\vert
f_{2}(q^{2})\pm \sqrt{\lambda }f_{1}(q^{2})\right\vert ^{2}  \label{65k}
\end{equation}

Finally the longitudinal and transverse helicity amplitude becomes
\begin{eqnarray}
f_{L}(q^{2}) &=&\frac{d\Gamma _{L}(q^{2})/dq^{2}}{d\Gamma (q^{2})/dq^{2}}
\notag \\
f_{\pm }(q^{2}) &=&\frac{d\Gamma _{\pm }(q^{2})/dq^{2}}{d\Gamma
(q^{2})/dq^{2}}  \notag \\
f_{T}(q^{2}) &=&f_{+}(q^{2})+f_{-}(q^{2})  \label{helicity-fractions}
\end{eqnarray}%
so that \ the sum of the longitudinal and transverse helicity amplitudes is
equal to one i.e. $f_{L}(q^{2})+f_{T}(q^{2})=1$ for each value of $q^{2}$%
\cite{22}.

\section{Numerical Analysis}

In this section we will examine the above derived physical observables and
analyze the effects of different new physics scenarios on them. Form factors
which are the non perturbative quantities and for them we rely on the Light
Cone Sum Rule (LCSR) approach for the numerical calculations. The numerical values of the LCSR form factors along with
the different fitting parameters \cite{45} are summarized in Table I. In addition to the parameters corresponding to
different NP models there are some standard inputs which are collected in
Table III.

\begin{table}[ht]
\caption{Default values of input parameters used in the calculations.}
\label{input}\centering
\begin{tabular}{c}
\hline\hline
$m_{B_s}=5.366$ GeV, $m_{b}=4.28$ GeV, $m_{s}=0.13$ GeV, \\
$m_{\mu}=0.105$ GeV, $m_{\tau}=1.77$ GeV, $f_{B}=0.25$ GeV, \\
$|V_{tb}V_{ts}^{\ast}|=45\times 10^{-3}$, $\alpha^{-1}=137$, $%
G_{F}=1.17\times 10^{-5}$ GeV$^{-2}$, \\
$\tau_{B}=1.54\times 10^{-12}$ sec, $m_{\phi}=1.020$ GeV. \\ \hline\hline
\end{tabular}%
\end{table}

The strength of \ the other NP
parameters that corresponds to the UED and $Z^{\prime }$ model are varied
such that they lie inside the bounds given by different flavor decays observed so for. It is emphasis here that
in all the figures the band corresponds to the uncertainties in different input
parameters where form factors are the main contributors (c.f. Table I ) and we defined $q^{2}=s$. The
NP curves are plotted by varying the values of
NP parameters in the range summarized in the Table IV.

Semileptonic $B_{s}$ decay are ideal probes to study the physics in and
beyond the Standard Model. In this context, there are large number of observables which are
accessible in these decays. However, the branching ratio, in general for
semi-leptonic decays like, is prune to
many sources of uncertainties. The major source of uncertainty originate
from the $B\rightarrow \phi $ transition form factors that can bring about $20-30\%$ uncertainty to
the differential branching ratio. This goes to show that differential
branching ratio may not be a suitable observable to look for the NP effects
unless these effects are very drastic. In the absence of the precise form factors,
it is still possible to constraint new physics with the help of observables that exhibits reduced
sensitivity to the form factors. In this
regard the most important observables are the zero position of the
forward-backward asymmetry, different lepton polarization asymmetries and
the helicity fractions of the final state meson. This will become clear from Figs. 2-14, where we will
see that the gray band corresponding to the uncertainties in different input parameters totally shrinks.

The SM predicts the zero crossing of $\mathcal{A}_{FB}(q^{2})$ at a well determined
position which is free from the hadronic uncertainties at the leading order
(LO) in strong coupling $\alpha _{s}$ \cite{9,10,11}. For this reason, the zero position of $%
A_{FB}$, is an important observable in
the search of new physics. In order to make this point clear, the zero
position $\left( q_{0}^{2}\right) $ is just the root of the Eq. (\ref%
{FBFform}), which can be written as%
\begin{equation}
q_{0}^{2}=-\frac{C_{7}^{eff}}{\Re \left( C_{9}^{eff}\left( q_{0}^{2}\right)
\right) }m_{b}\left[ \frac{T_{2}\left( q_{0}^{2}\right) }{A_{1}\left(
q_{0}^{2}\right) }\left( M_{B_{s}}-M_{\phi }\right) +\frac{T_{1}\left(
q_{0}^{2}\right) }{V\left( q_{0}^{2}\right) }\left( M_{B_{s}}+M_{\phi
}\right) \right] .  \label{zposition}
\end{equation}%
It is really spectacular that for the $B\rightarrow Vl^{+}l^{-}$ decays, we
can find that with the use of effective theories like Soft Collinear Effective Theory (SCET) both
ratios of the form factors appearing in Eq. (\ref{zposition}) have no
hadronic uncertainty, i.e., all dependence on the intrinsically
non-perturbative quantities cancels. Therefore, one can simply write%
\begin{eqnarray*}
\frac{T_{2}\left( q^{2}\right) }{A_{1}\left( q^{2}\right) } &=&\frac{%
M_{B_{s}}}{M_{B_{s}}-M_{\phi }} \\
\frac{T_{1}\left( q^{2}\right) }{V\left( q^{2}\right) } &=&\frac{M_{B_{s}}}{%
M_{B_{s}}+M_{\phi }}
\end{eqnarray*}
and using these relations, the short distance expression for the zero
position $\mathcal{A}_{FB}$ is given by \cite{10}%
\begin{equation}
q_{0}^{2}=\frac{2m_{b}M_{B_{s}}}{\Re \lbrack C_{9}^{eff}(q_{0}^{2})]}%
C_{7}^{eff}  \label{sdrelation}
\end{equation}%
Recently LHCb
has published its results on $\mathcal{A}_{FB}(\bar{B}\rightarrow \bar{K}^{\ast }\mu
^{+}\mu ^{-})$ which shows, with small error bars, that the zero position of $\mathcal{A}_{FB}(\bar{B%
}\rightarrow \bar{K}^{\ast }\mu ^{+}\mu ^{-})$ is close to the SM's zero
position. Like $\bar{B}\rightarrow \bar{K}^{\ast }\mu ^{+}\mu ^{-}$ decay,
the semileptonic decay $B_{s}\rightarrow \phi \mu ^{+}\mu ^{-}$ also occurs
through the quark level transition $b\rightarrow s\mu ^{+}\mu ^{-}$.
Therefore, the future measurements of the $\mathcal{A}_{FB}(B_{s}\rightarrow \phi \mu
^{+}\mu ^{-})$ will shed more light on the NP in the flavor sector.

The other "optimized" observables are the various polarization asymmetries
attached to the final state leptons and meson where the uncertainties are also mild.
Regarding this the longitudinal and normal lepton
polarization asymmetries are the good tool to probe the NP. On the other hand, the
transverse lepton polarization asymmetry (c.f. Eq. (\ref{Transverse-polarization}))
is proportional to the imaginary part of the auxiliary functions and hence will be negligible
in the models where we have the real couplings. In
addition to the single lepton polarization, we will also discuss the
dependence of double lepton polarization asymmetries on $q^{2}$ and will
also give the numerical values of their averages that can be obtained after
integration on $q^2$.

Another interesting observable in this list is the study of the spin effects of final
state meson which for our case is the $\phi$ meson. A detailed discussion
about the NP effects on the longitudinal and transverse helicity fractions
has been done in the forthcoming numerical analysis which will provide help to dig out the
potential of various NP scenarios.

Similarly, the polarized and unpolarized $CP$
violating asymmetries are useful tool to find the distinguishing feature from the SM as well as will help us
to segregate the two NP models. It is worth mentioning that the FCNC transitions
are proportional to the CKM matrix elements, $V_{tb}V^{\ast}_{ts}$, $V_{cb}V^{\ast}_{cs}$
and $V_{ub}V^{\ast}_{us}$, where the later two are highly suppressed compared to the
$V_{tb}V^{\ast}_{ts}$. This will eventually suppress the value of $CP$ violation
asymmetries in the SM and also in the UED model. Because of the extra phase in the
$Z^{\prime}$ model we are expecting a prominent deviation. Therefore, the study of
the $CP$ violation asymmetries will provide a key evidence of the NP coming through the
extra $Z^{\prime}$ boson. This will be discussed as a separate study which will be presented in
\cite{13IJ}.

The only free parameter in the UED model is
the inverse of the compactification radius i.e. $1/R$. Taking into account
the leading order (LO) contributions due to the exchange of KK modes as well
as already available next-to-next-to-leading order (NNLO) corrections to $%
B\rightarrow X_{s}\gamma $ the Haisch et al. \cite{Haisch} have obtained the
lower bound on inverse of compactification radius to be $600$ GeV. Using the
electroweak precision measurements and also some cosmological constraints,
the lower limit on the inverse of the the compactification radius is
obtained to be in or above the 500 GeV range \cite{Gogoladze, Feng}. It is well known that
by increasing $1/R$ the values of different physical observables becomes closer to the SM values. Therefore, in our
numerical analysis we take the value of $1/R$ to be $500$ GeV just to see the
maximum possible definition from the SM value.

On the other hand the effects of the family non-universal $Z^{\prime}$ boson
on $b\rightarrow s$
transition have attracted much more attention and been widely studied where
it is argued that the behaviour of a family non-universal $Z^{\prime }$
boson is helpful to resolve many puzzles in $B$ meson decays, such as $\pi K$
puzzle and anomalous $\bar{B}_{s}-B_{s}$ mixing \cite{Zp1, Zp2, Zp3, Zp4,
Zp5}. In literature the differential decay width and forward backward
asymmetry of $B_{s}\rightarrow \phi \mu ^{+}\mu ^{-}$ decay have been
studied in the $Z^{\prime }$ model using three different scenarios which
corresponds to different values of the the left and right handed couplings
of $Z^{\prime }$ with leptons, i.e. $S_{LL}$ and $D_{LL}$, as well as the
right handed coupling with quarks, i.e., $B_{sb}$ and these are collected in Table IV
\cite{Zp5}. In the present study we will use these limits to see their impact on
the branching ratio and to various asymmetries mentioned above.
\begin{table}[ht]
\caption{The numerical values of the $Z^{\prime}$ parameter.}
\label{input}\centering
\begin{tabular}{ccccc}
\hline\hline
& $\left\vert B_{sb}\right\vert \times 10^{-3}$ & $\phi _{sb}[^{o}]$ & $%
S_{LL}\times 10^{-2}$ & $D_{LL}\times 10^{-2}$ \\ \hline
$\mathcal{S}_{1}$ & $1.09\pm 0.22$ & $-72\pm 7$ & $-2.8\pm3.9$ & $-6.7\pm 2.6$ \\ \hline
$\mathcal{S}_{2}$ & $2.20\pm 0.15$ & $-82\pm 4$ & $-1.2\pm 1.4$ & $-2.5\pm 0.9$ \\ \hline\hline
\end{tabular}%
\end{table}

The numerical results of the branching ratios, forward-backward asymmetry,
different polarization asymmetries of the final state leptons,
helicity fractions of the final state $\phi $ meson as a function of $q^{2}$
in $B_{s}\rightarrow \phi l^{+}l^{-}$ decays are presented in Figs. 1-14.
Fig. 1(a,b) describes the differential branching ratio of $B_{s}\rightarrow
\phi \mu ^{+}\mu ^{-}(\tau ^{+}\tau ^{-})$ decay, where one can see that
for the choice of the parameters made in accordance with the current data on
various flavor physics decay modes lies close to the SM predictions. This
can also be summarized in Table VII.  Just to mention, when we have muon's as the final state leptons (c.f. Fig.1a)
the bands for two $Z^{\prime}$ scenarios over lap with each other. We can also see that the value of the
branching ratio lies well with in the range of the experimental limits with
the choice of different values of NP parameters and one can notice that the NP
contributions are over shadowed by the uncertainties involved in different
input parameters. Therefore, to look for NP we have to calculate the observables where
hadronic uncertainties almost have no effect and which are almost
independent of the choice of form factors. Among them the most pertinent are
the zero position of the forward-backward asymmetry, lepton polarization
asymmetries, the helicity fractions of the final state meson and $CP$ asymmetries, which being
almost free from the hadronic uncertainties and serve as handy tools to extract NP signature.

\begin{table}[tbh]
\caption{Branching ratio of $B_{s}\rightarrow \protect\phi l^{+}l^{-}$ in SM
and different NP scenarios. The central values of the form factors and other input
parameters are used.}
\label{input}\centering
\begin{tabular}{cccc}
\hline\hline
Model & $B_{s}\rightarrow \phi \mu ^{+}\mu ^{-}$ & $B_{s}\rightarrow \phi
\tau ^{+}\tau ^{-}$ &  \\ \hline
SM & $1.58\times 10^{-6}$ & $2.37\times 10^{-7}$ &  \\ \hline
UED & $0.91\times 10^{-6}$ & $0.94\times 10^{-6}$ &  \\ \hline
$Z^{\prime }-$ & $1.86\times 10^{-6}$ & $3.07\times 10^{-7}$ & \\ \hline\hline
\end{tabular}%
\end{table}

\begin{figure}[tbp]
\caption{The differential width for the $B_{s} \to \protect\phi l^+l^-$ ($l=%
\protect\mu, \protect\tau$) decays as functions of $q^2=s$. The gray, green and red
bands corresponds to the Standard Model, $Z^{\prime}$ scenarios $S_{1}$ and $S_{2}$ respectively.
The dashed blue line corresponds to the UED model.}
\label{decay rate}
\begin{center}
\begin{tabular}{ccc}
\vspace{-2cm} \includegraphics[scale=0.5]{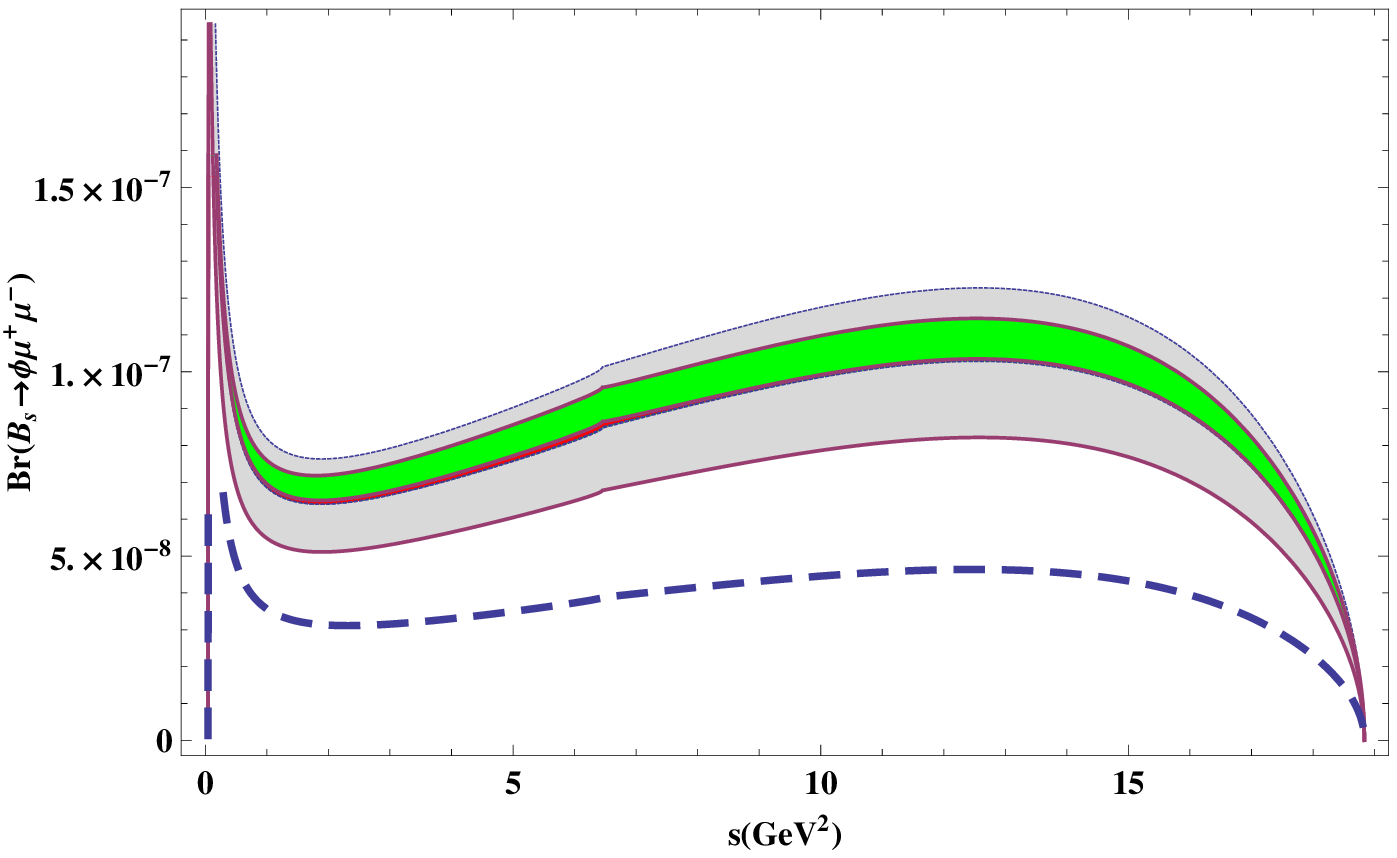} %
\includegraphics[scale=0.5]{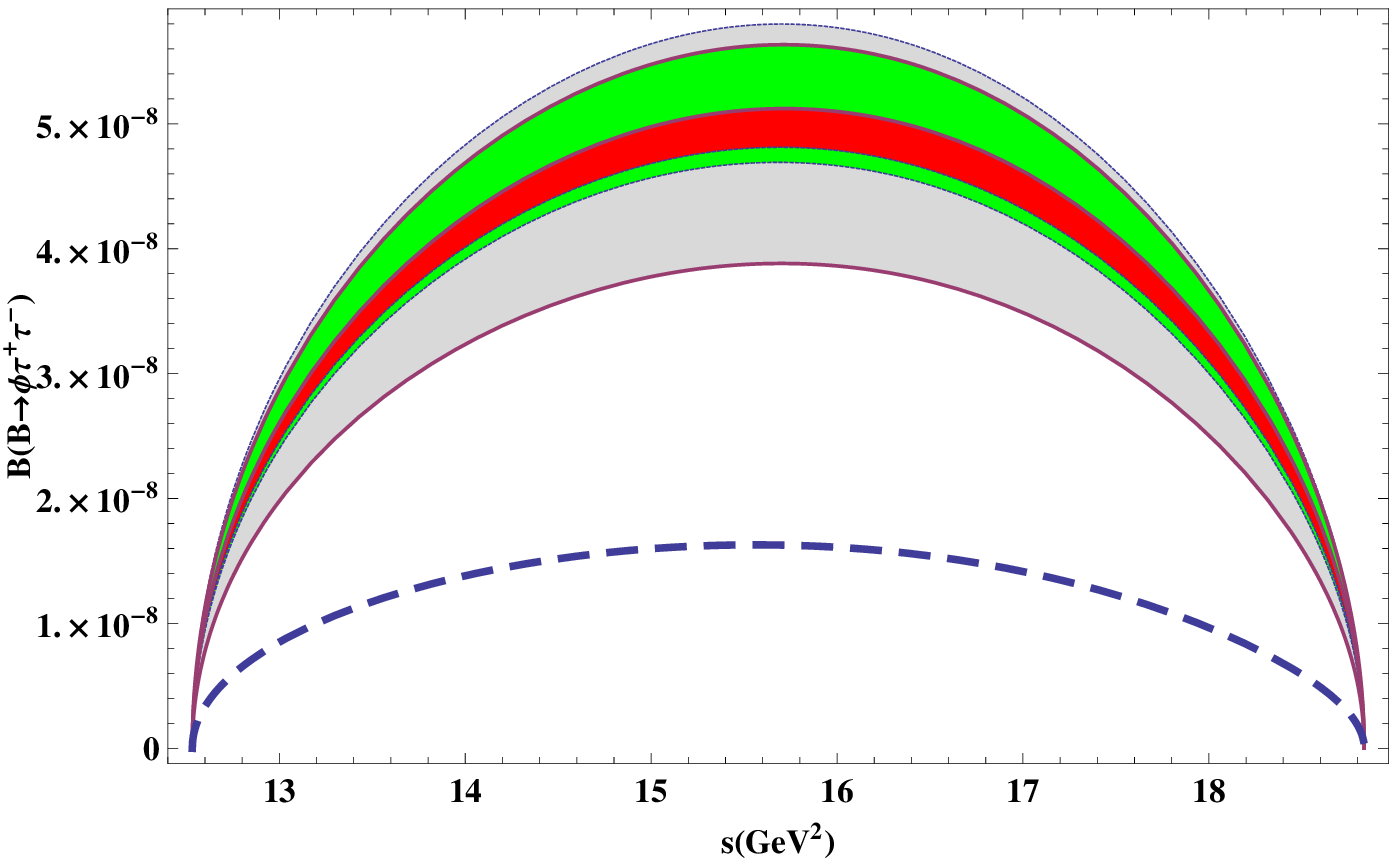} \put (-320,180){(a)} \put
(-100,180){(b)} &  &
\end{tabular}%
\end{center}
\end{figure}

\begin{figure}[tbp]
\begin{tabular}{cc}
\epsfig{file=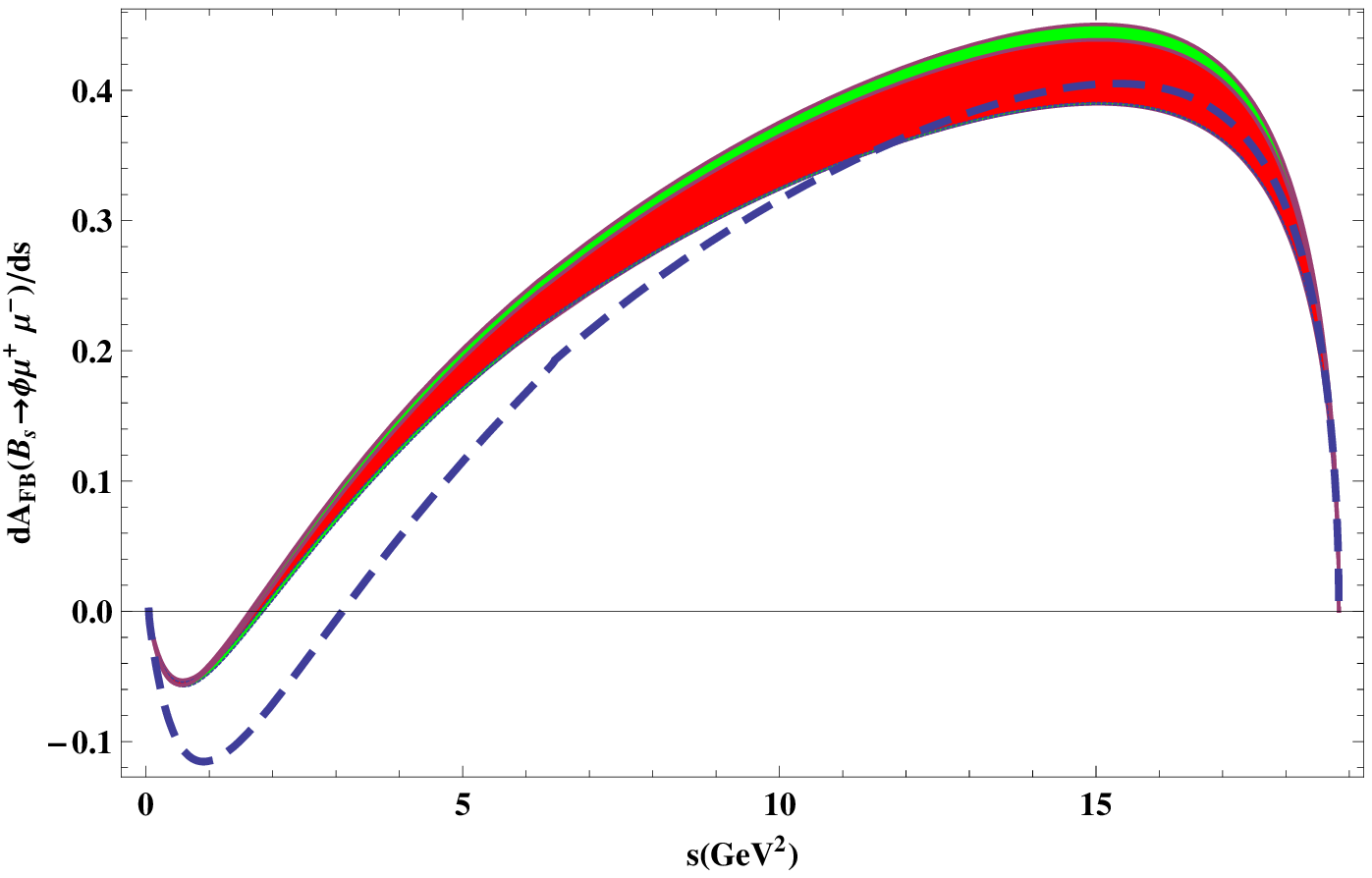,width=0.4\linewidth,clip=a} \put (-100,190){(a)} & %
\epsfig{file=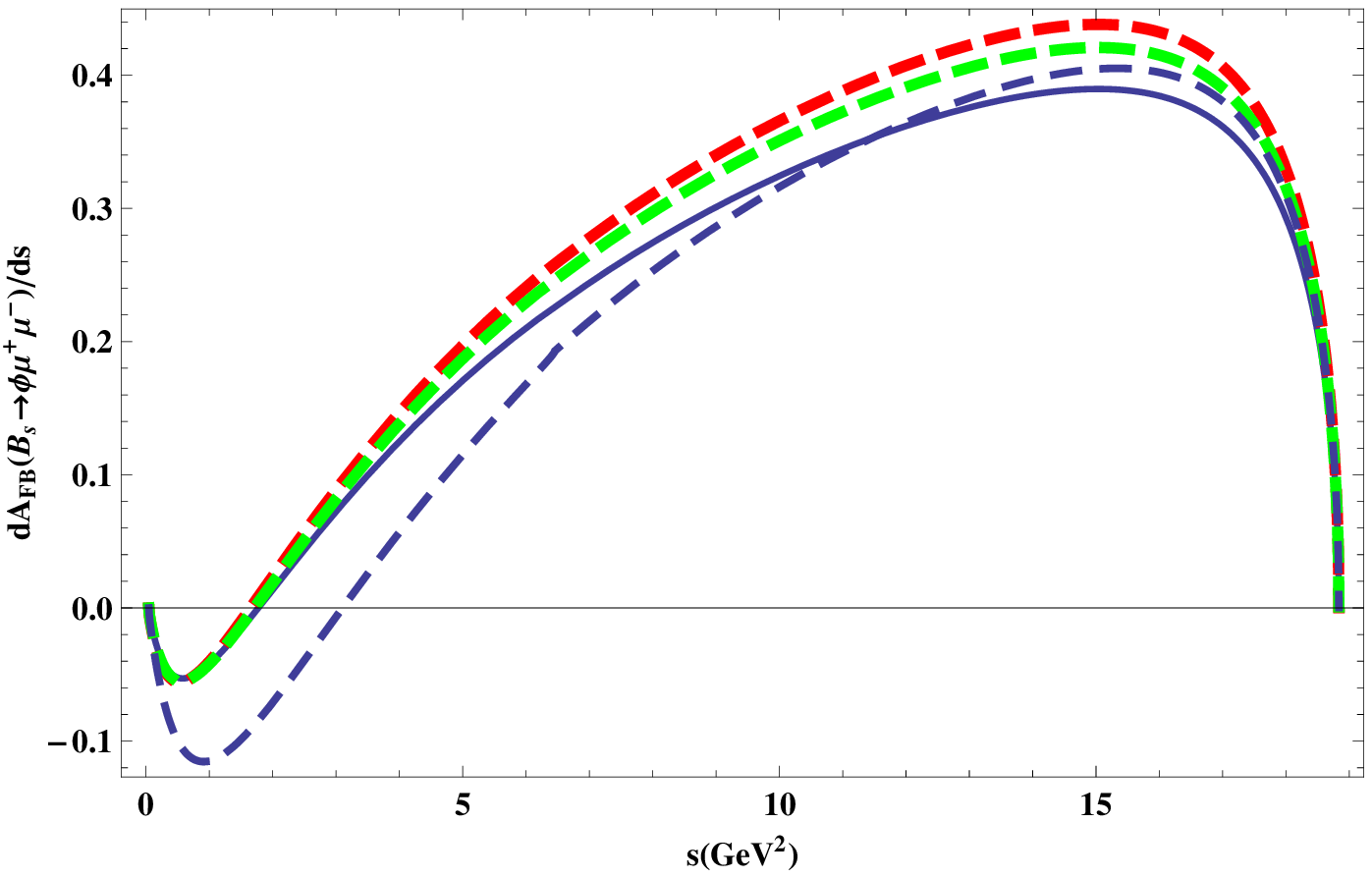,width=0.4\linewidth,clip=b} \put (-100,190){(b)} \\
\epsfig{file=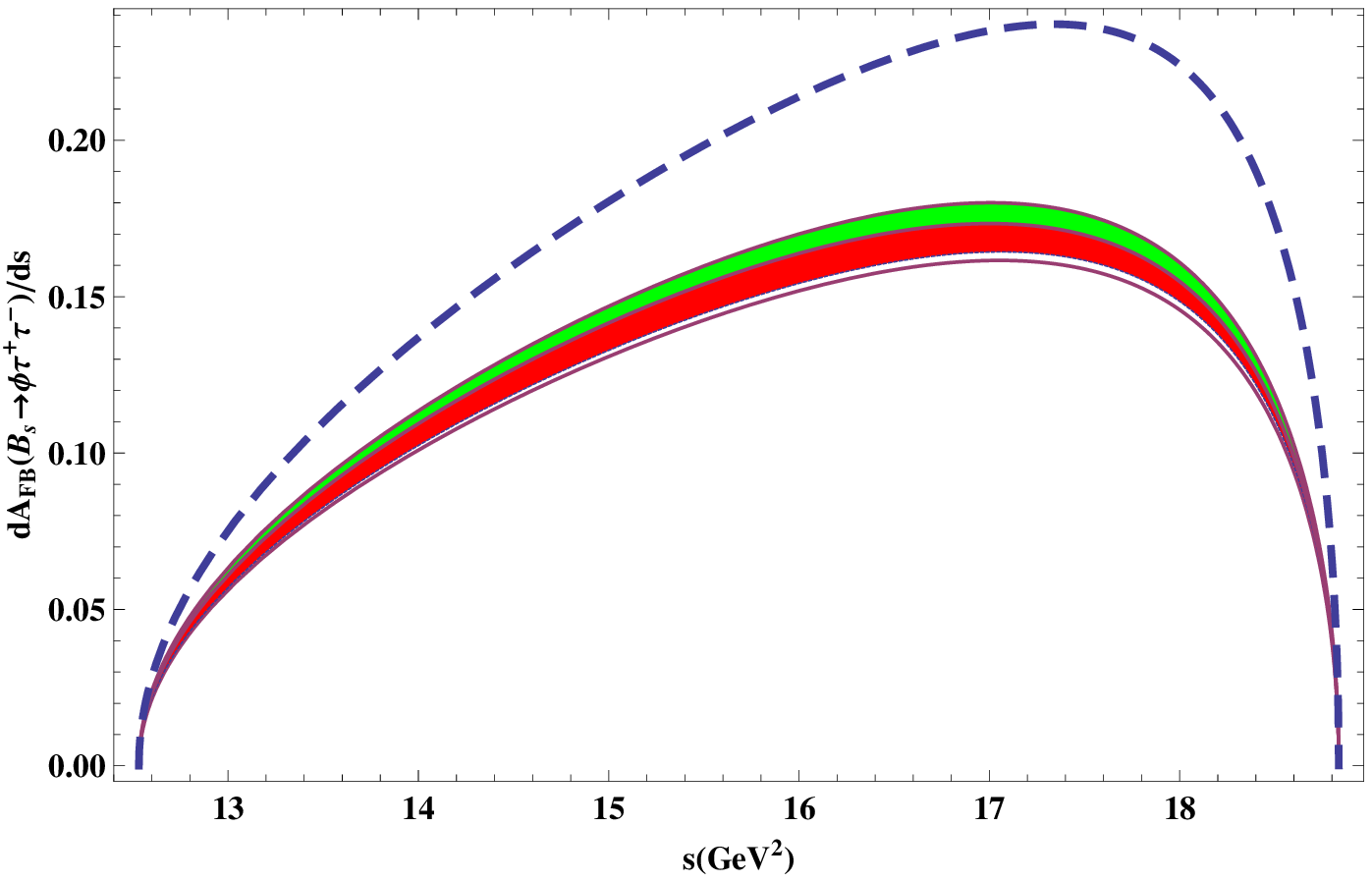,width=0.4\linewidth,clip=c} \put (-100,190){(c)} & %
\epsfig{file=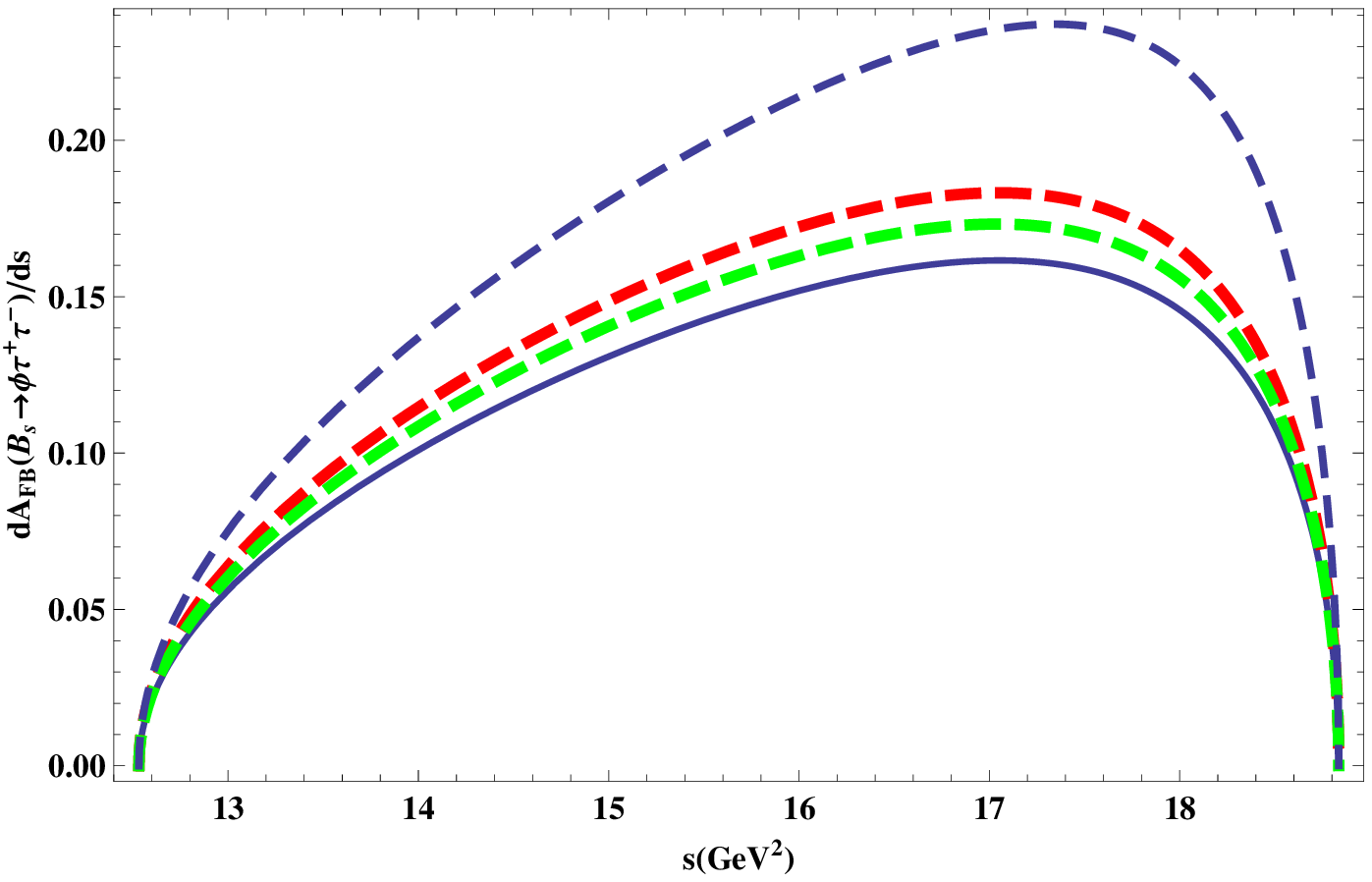,width=0.4\linewidth,clip=d} \put (-100,190){(d)}%
\end{tabular}%
\caption{The differential forward-backward asymmetry for the $B_{s} \to \protect\phi l^+l^-$ ($l=%
\protect\mu, \protect\tau$) decays as functions of $q^2$. The gray, green and red
bands corresponds to the Standard Model, $Z^{\prime}$ scenarios $S_{1}$ and $S_{2}$ respectively.
The dashed blue line corresponds to the UED model. In Fig. 2(b,d) the solid, dashed-blue, dashed-green and dashed-red
lines corresponds to SM, UED model, and $Z^{\prime}$ models scenario-I and II, respectively. Here the central values of the form factors
and other input parameters are used.} \label{FBasymmetry}
\end{figure}

As we have already mentioned that at the leading order in the strong coupling constant $%
\alpha _{s}$ in the SM the destructive interference between the photon
penguin ($C_{7}^{eff}$) and the $Z$ penguin ($C_{9}^{eff}$) make the FBA
equal to zero at a particular position which is independent of the form
factors as depicted in Eq. (\ref{sdrelation}). For the decay $%
B_{s}\rightarrow \phi \mu ^{+}\mu ^{-}$, the value of the the zero crossing
is approximately $(q^{2}\simeq 1.6\text{GeV}^{2})$. The deviation of the
zero crossing from the SM value gives us some clues for the NP. Fig. 2b
shows the effect of various NP scenarios on the zero position of the
forward-backward asymmetry for $B_{s}\rightarrow \phi \mu ^{+}\mu ^{-}$
decay. Playing on the pitch of $B\rightarrow K^{\ast }l^{+}l^{-}$ where the
experimental results of LHCb lies close to the SM value, we can see that
only the small deviation from the SM zero position of $\mathcal{A}_{FB}$ comes in case
of the $Z^{\prime}$ model. In case of the
UED model the value of the Wilson coefficient $C_{7}$ is significantly reduced whereas
$C_{9}$ almost remains unaltered for $1/R=500$GeV. By looking at the Eq. (\ref{sdrelation})
we can see that the zero position is directly proportional to $C_{7}$ therefore, we expect
large deviation in the UED model and this is obvious from Fig. 2b.
We expect that in future when more data will come from the LHCb
the measurement of the forward-backward asymmetry in $B_{s}\rightarrow \phi
\mu ^{+}\mu ^{-}$ will help us in observing new physics and will also give
us an opportunity to distinguish between various NP scenarios.

It has been pointed out by Beneke et al. in ref. \cite{11} that the Next-to-Leading Order (NLO)
corrections to the lepton invariant mass spectrum in $B\to K^{\ast}l^{+}l^{-}$ is small but there is
a large correction to the predicted location of the zero position of the forward-backward asymmetry which is
estimated to be $30\%$. Such calculation is still been waited for the $B_{s}\to \phi l^{+}l^{-}$ decays before
one can say anything about the NP by measuring forward-backward asymmetry in these decays.

Fig. 3(a,b,c,d) shows the dependence of longitudinal lepton
polarization asymmetry for the $B_{s}\rightarrow \phi l^{+}l^{-}$ decay on
the square of momentum transfer for different NP models. In case of the UED model,
the value of the longitudinal lepton polarizations lies close to the SM value where
as significant deviation is obtained in case of the $%
Z^{\prime }$ model. This can also be seen quantitatively from Table VI, where $11\%$ deviation
is observed in case of the $Z^{\prime}$ model for the central values of its parameters.

\begin{figure}[tbp]
\begin{tabular}{cc}
\epsfig{file=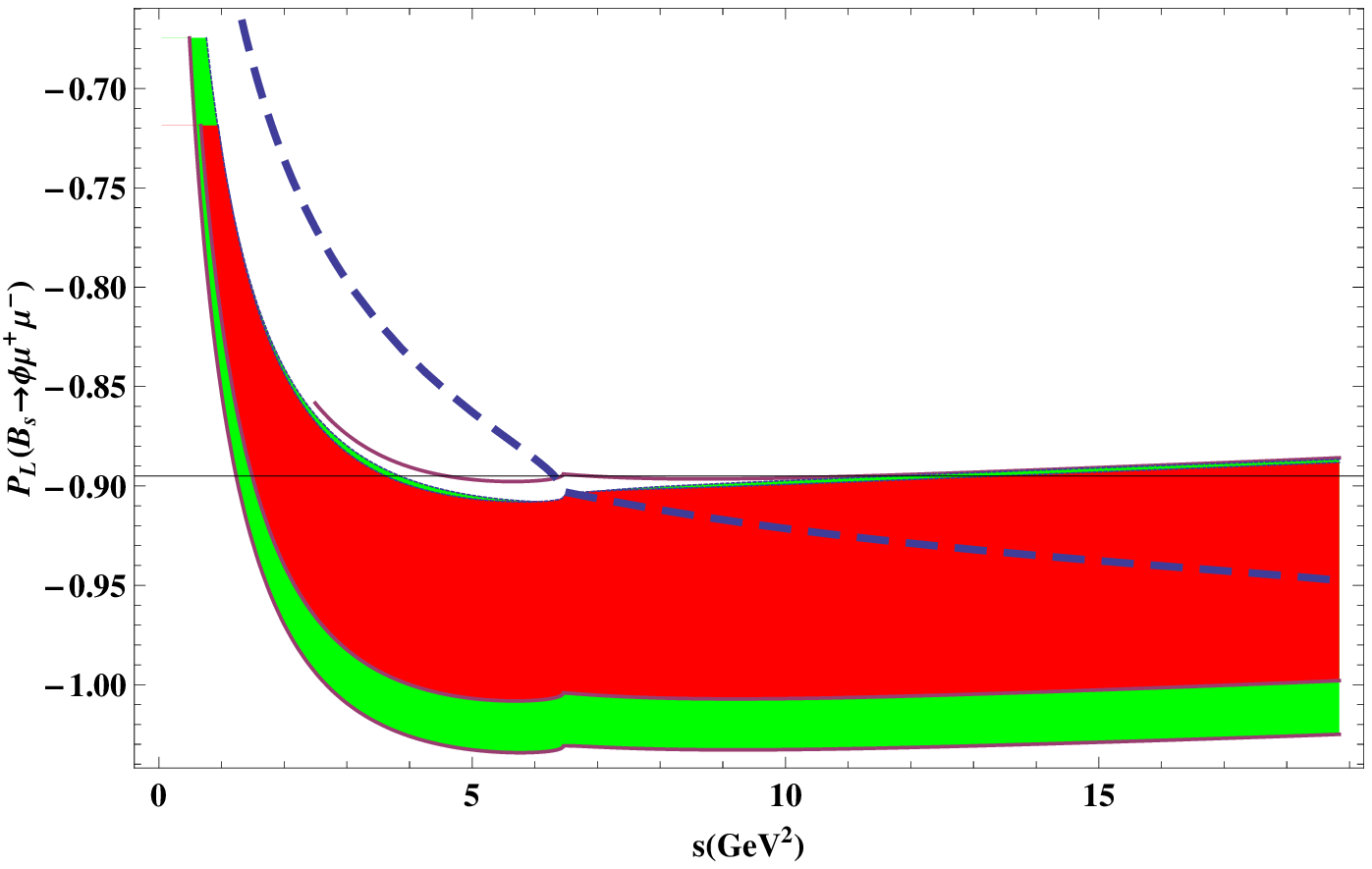,width=0.4\linewidth,clip=a} \put (-100,190){(a)} & %
\epsfig{file=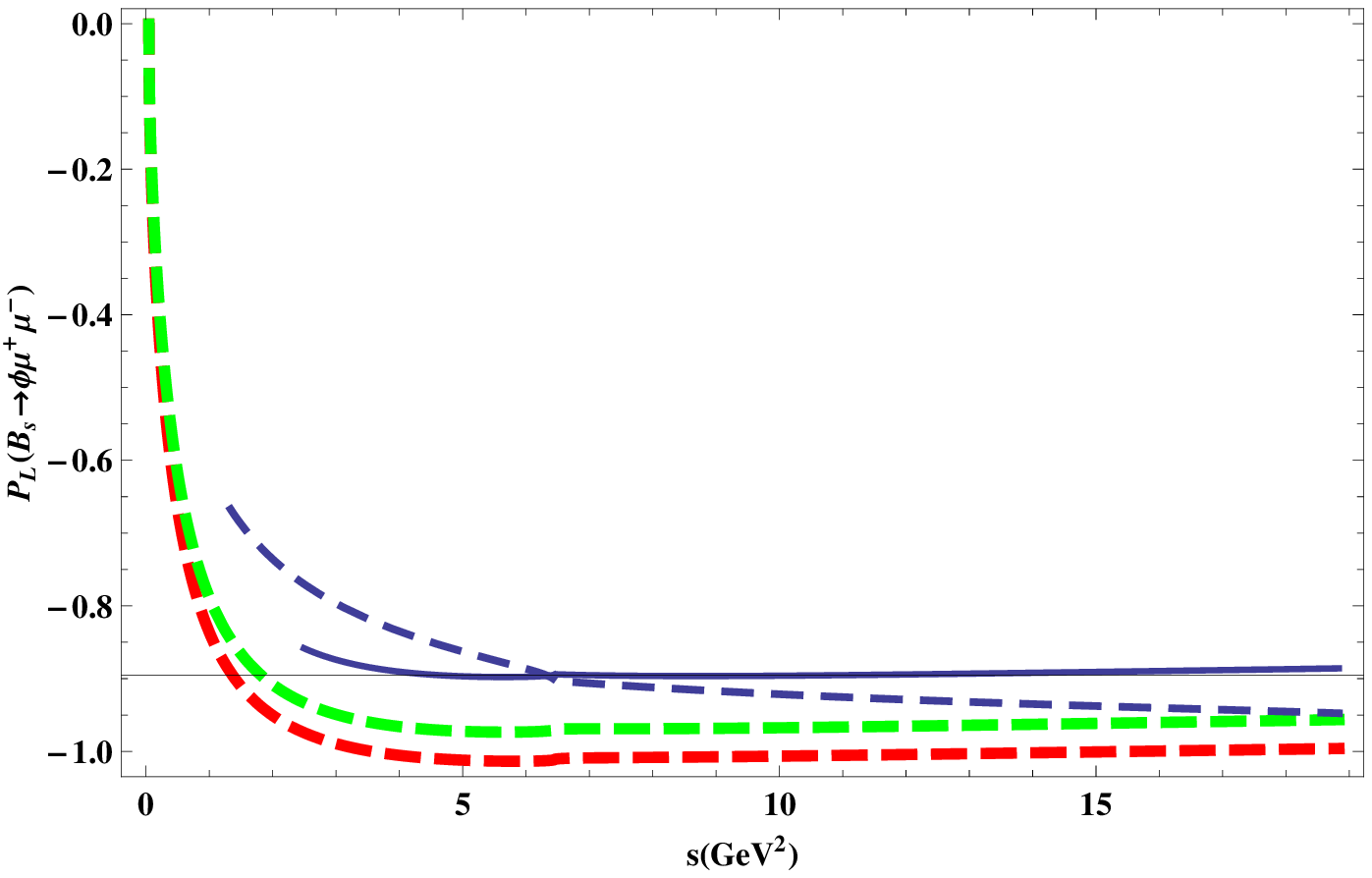,width=0.4\linewidth,clip=b} \put (-100,190){(b)} \\
\epsfig{file=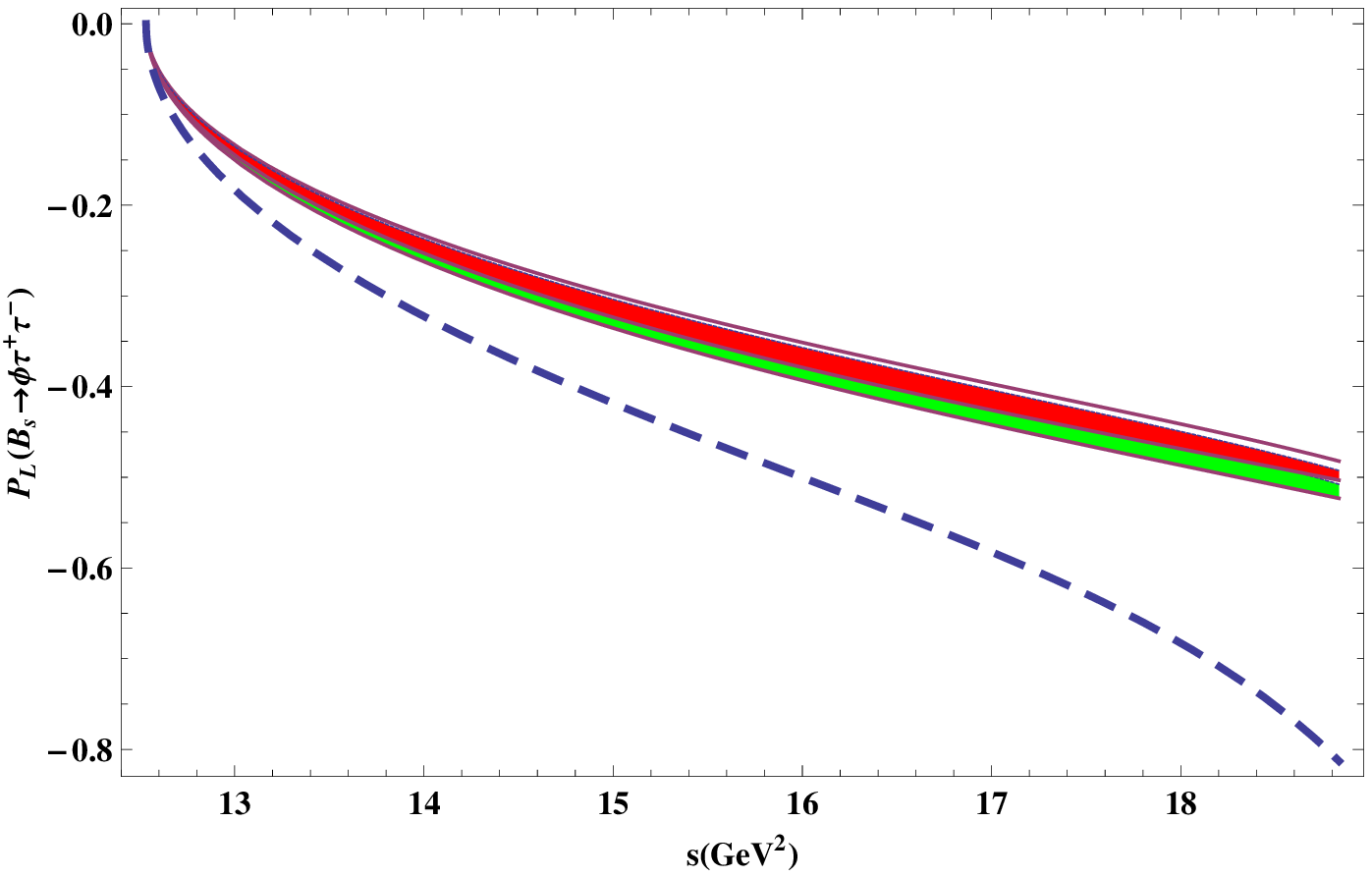,width=0.4\linewidth,clip=c} \put (-100,190){(c)} & %
\epsfig{file=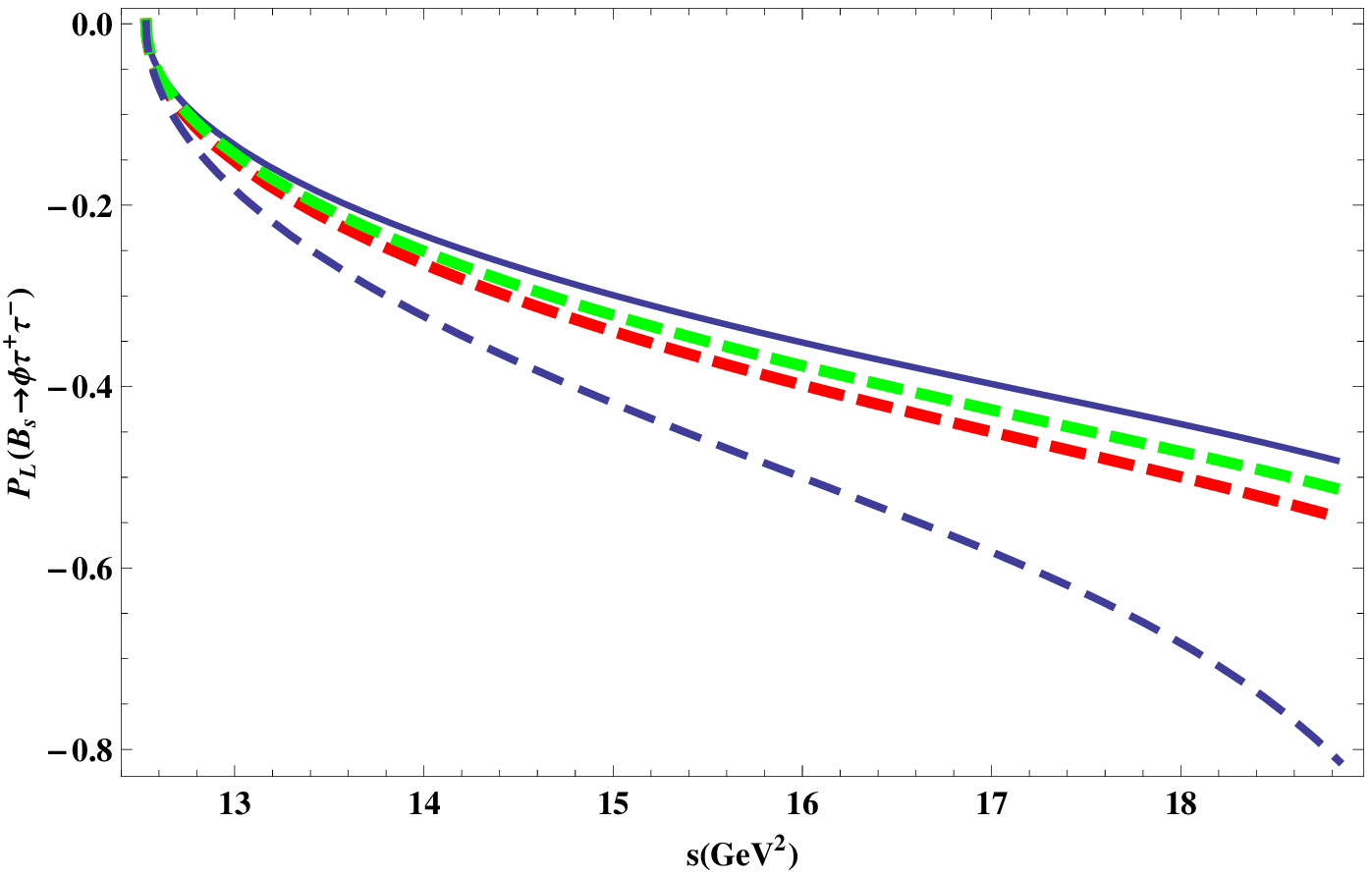,width=0.4\linewidth,clip=d} \put (-100,190){(d)}%
\end{tabular}%
\caption{The longitudinal lepton polarization asymmetry for the
$B_{s} \to \phi l^+l^-$ ($l=\protect\mu, \protect\tau$) decays as functions of $%
q^2$. The legends are same as in Fig.2. }
\label{PL}
\end{figure}

Fig. 4(a,b,c,d) displays the behavior of normal lepton polarization asymmetry
for the $B_{s}\rightarrow \phi l^{+}l^{-}$ with square of momentum transfer
in SM and in NP models. From Eq. (\ref{norm-polarization}) one can see that
it is proportional to the mass of leptons. Therefore, when we have the muons as
a final state leptons we can see that its SM value and the deviation from this value
through NP are only significant at low $q^2$ region and these effects are almost vanishes when we increase the values
of $q^2$. is small. Similarly, we have drawn normal lepton polarization asymmetry when tau's are the final state leptons in
Fig. 6(c,d). We can see that in SM the value of normal lepton polarization
asymmetry is positive almost throughout the kinematical region. It can be easily seen that the value in the $Z^{\prime }$ model is also quite
different from that of the SM value. The most interesting effects comes in
the UED model where the value of this asymmetry is negative in almost all
the available $q^{2}$ range. Hence it will be a clear signal of new physics
and by measuring its sign we can distinguish between the under consideration NP models.

\begin{figure}[tbp]
\begin{tabular}{cc}
\epsfig{file=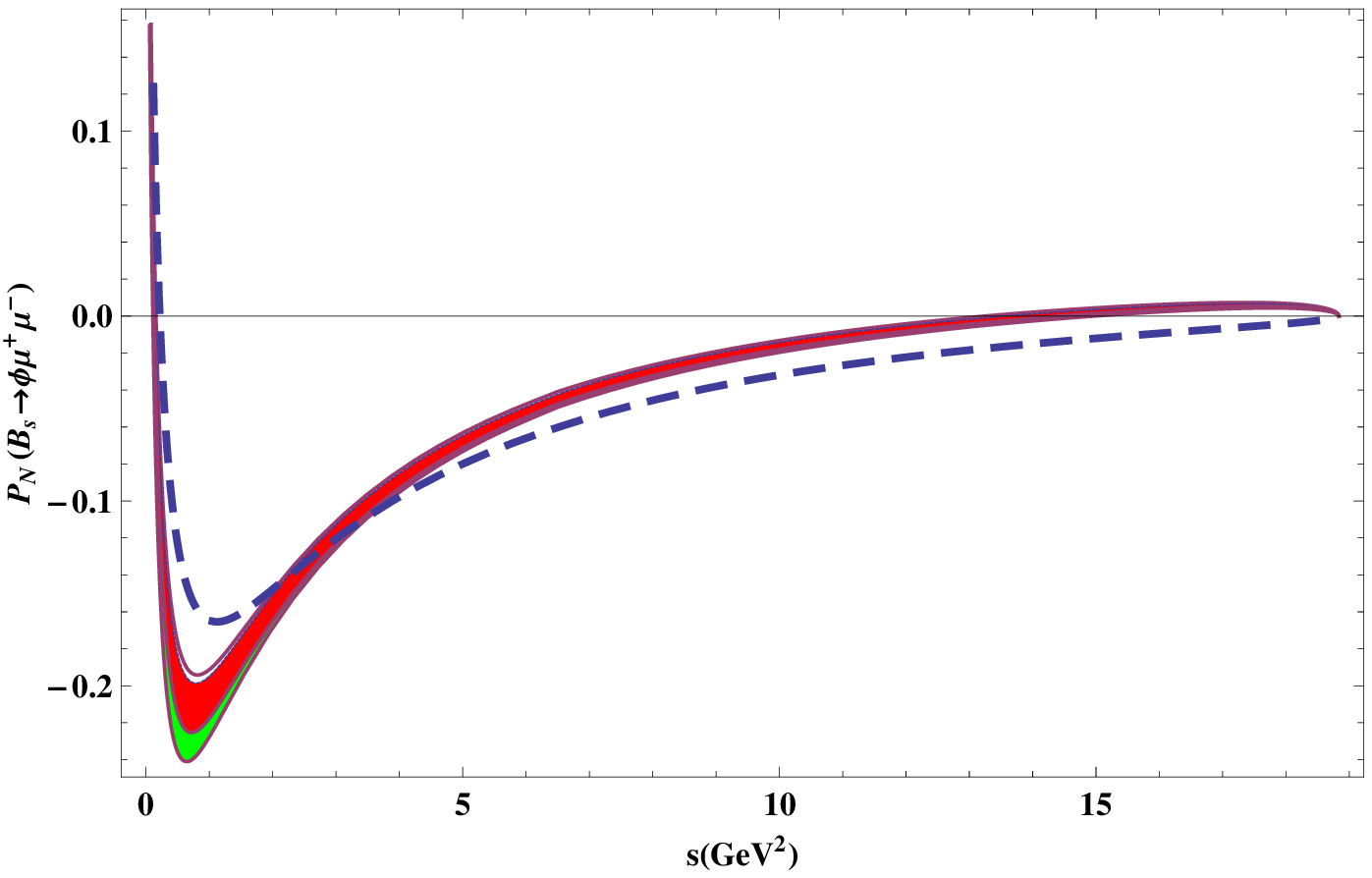,width=0.4\linewidth,clip=a} \put (-100,190){(a)} & %
\epsfig{file=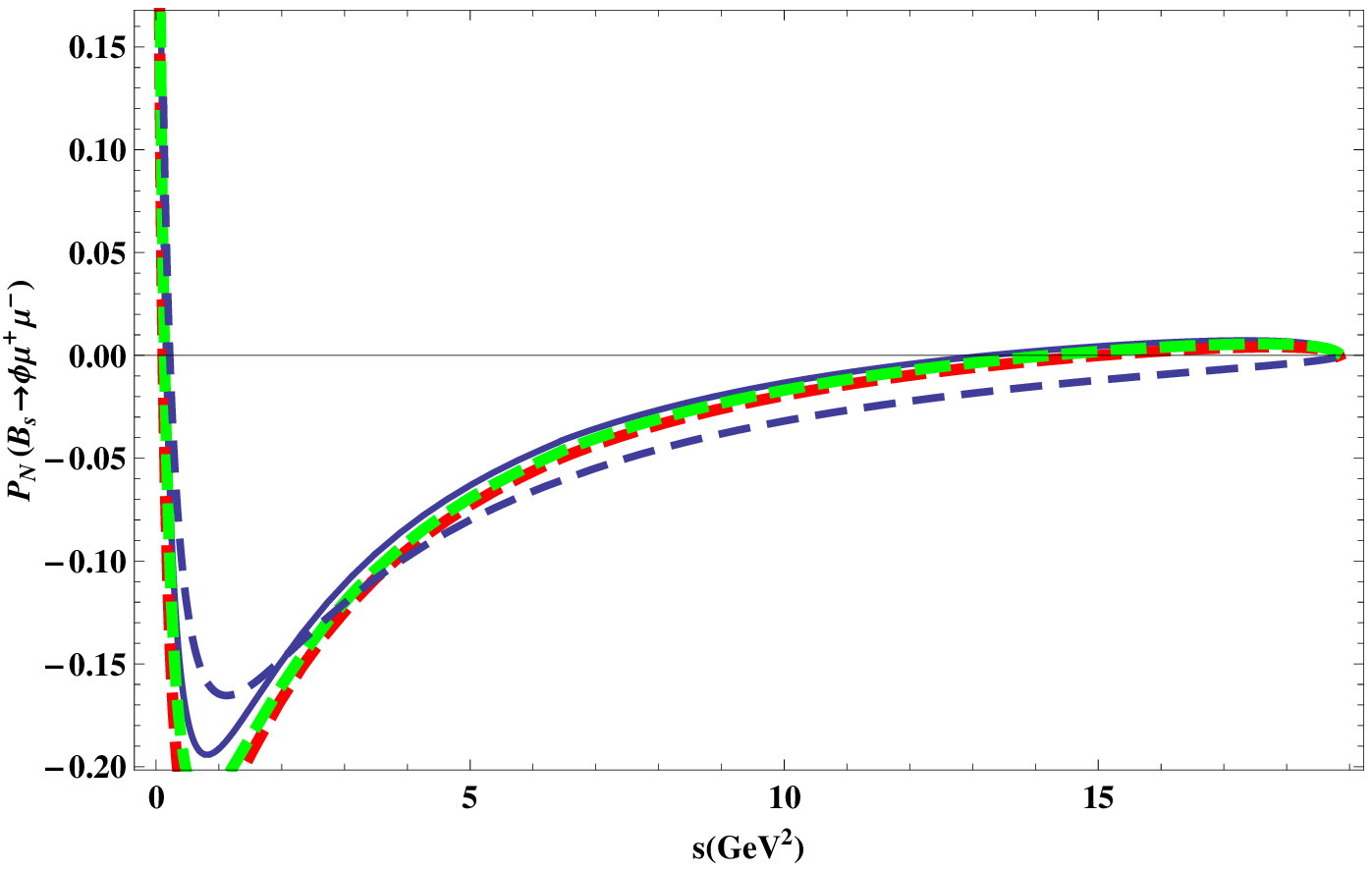,width=0.4\linewidth,clip=b} \put (-100,190){(b)} \\
\epsfig{file=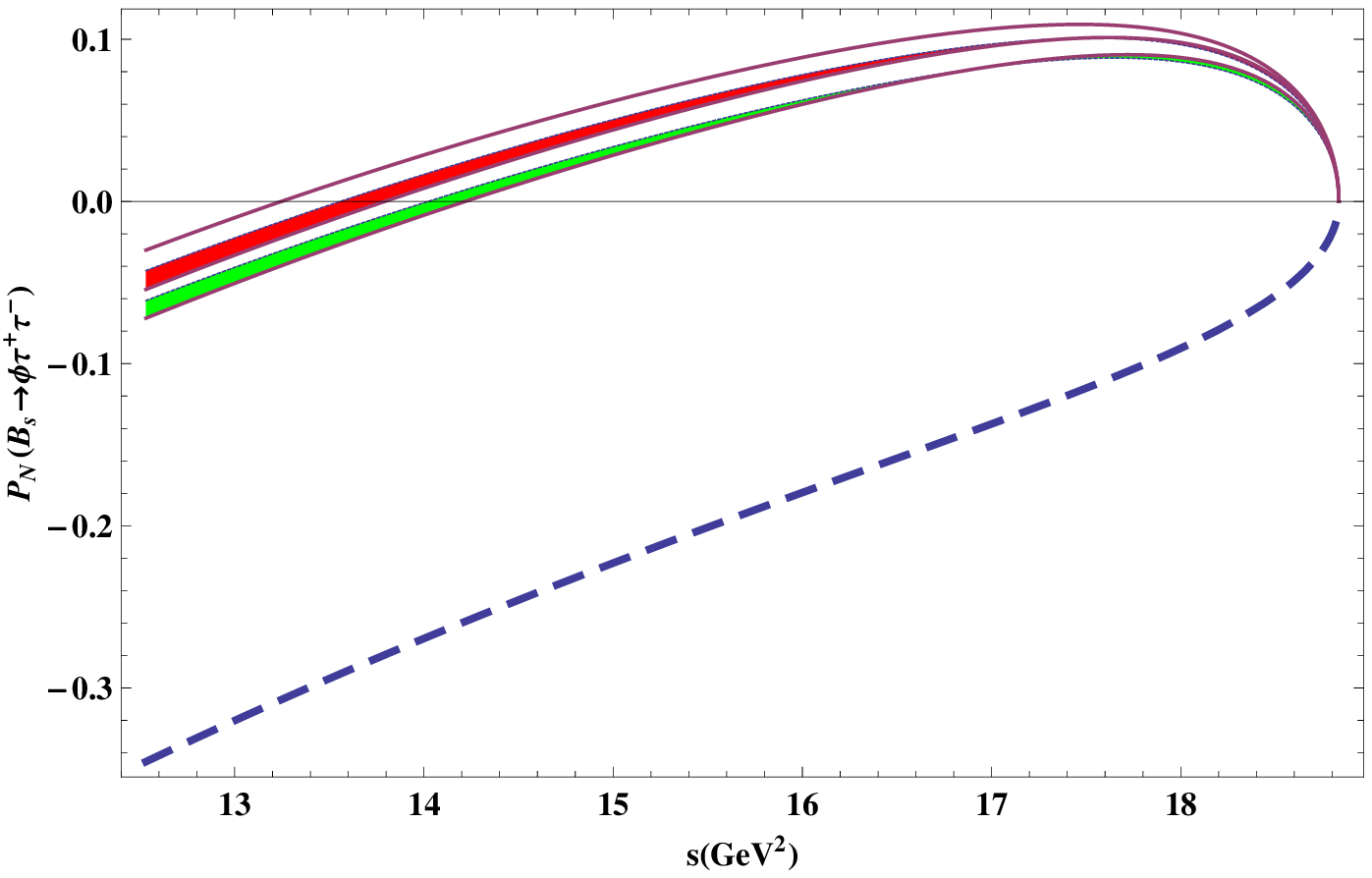,width=0.4\linewidth,clip=c} \put (-100,190){(c)} & %
\epsfig{file=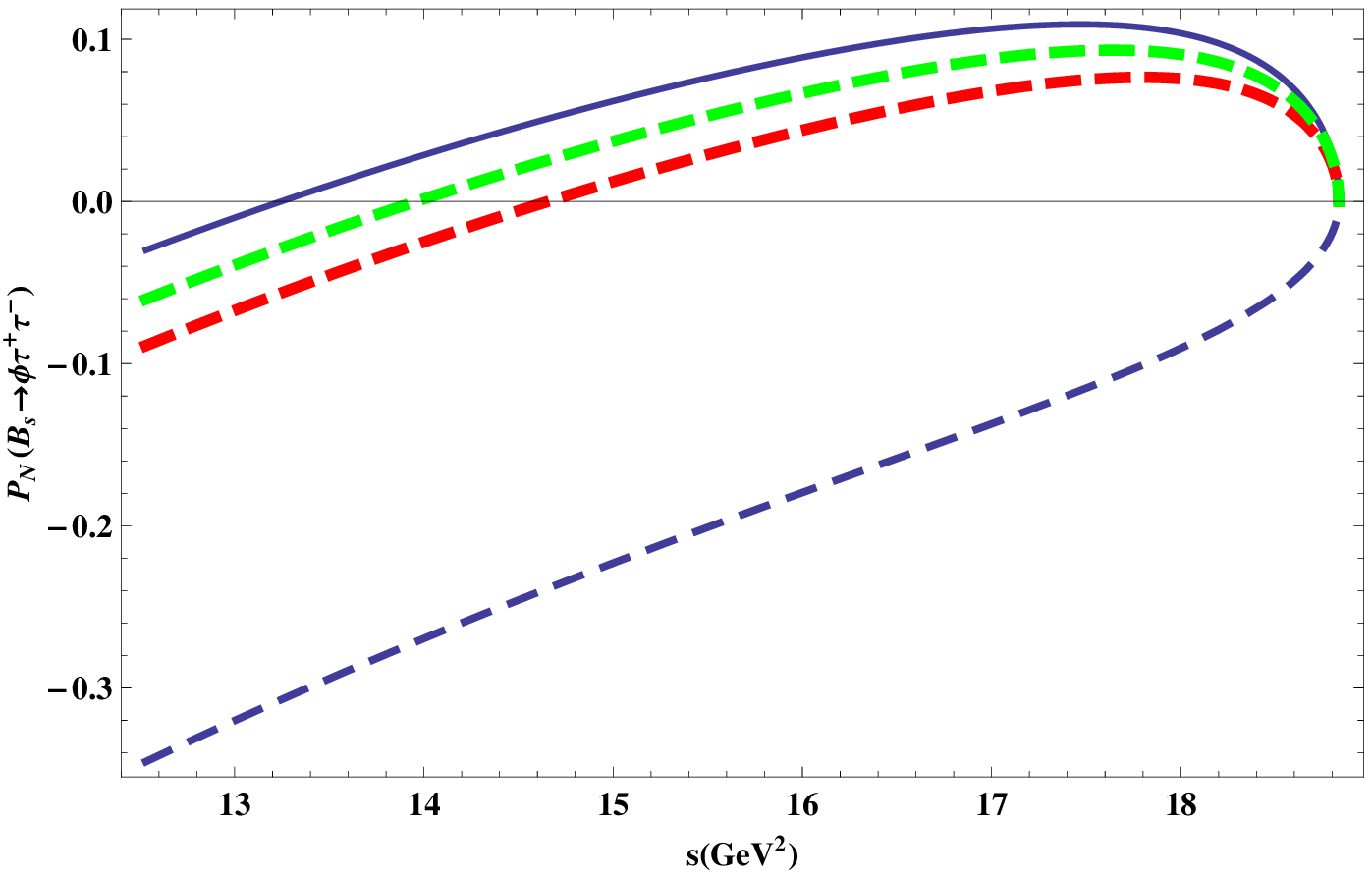,width=0.4\linewidth,clip=d} \put (-100,190){(d)}%
\end{tabular}%
\caption{The Normal lepton polarization asymmetry $B_{s} \to \protect\phi %
l^+l^-$ ($l=\protect\mu, \protect\tau$) decays as functions of $q^2$. The
legends are same as in Fig.2.}
\label{PN}
\end{figure}

Just like the normal lepton polarization asymmetry the transverse lepton
polarization asymmetry is also proportional to the lepton mass. In addition
to this it is also proportional to the imaginary part of the combination of
different auxiliary functions and so of the Wilson coefficients. The Wilson coefficients
remains real in the SM and UED model but not in the $Z^{\prime }$ model. However,
in this model imaginary part is also too small. Hence the value of transverse
lepton polarization asymmetry remains very small to be measured and Fig.
5(a,b,c,d) portrays this fact.

\begin{figure}[tbp]
\begin{tabular}{cc}
\epsfig{file=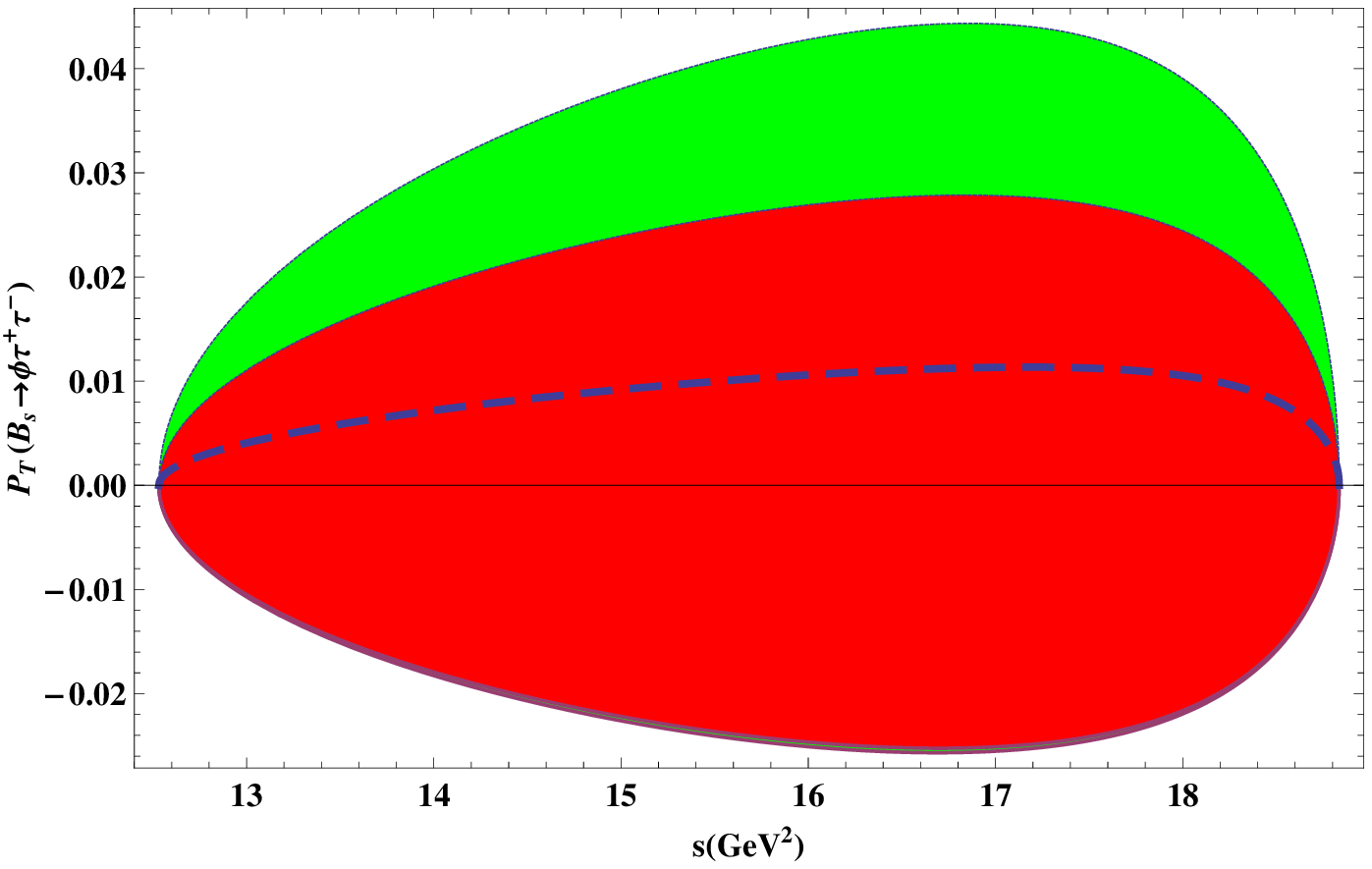,width=0.4\linewidth,clip=a} \put (-100,190){(a)} & %
\epsfig{file=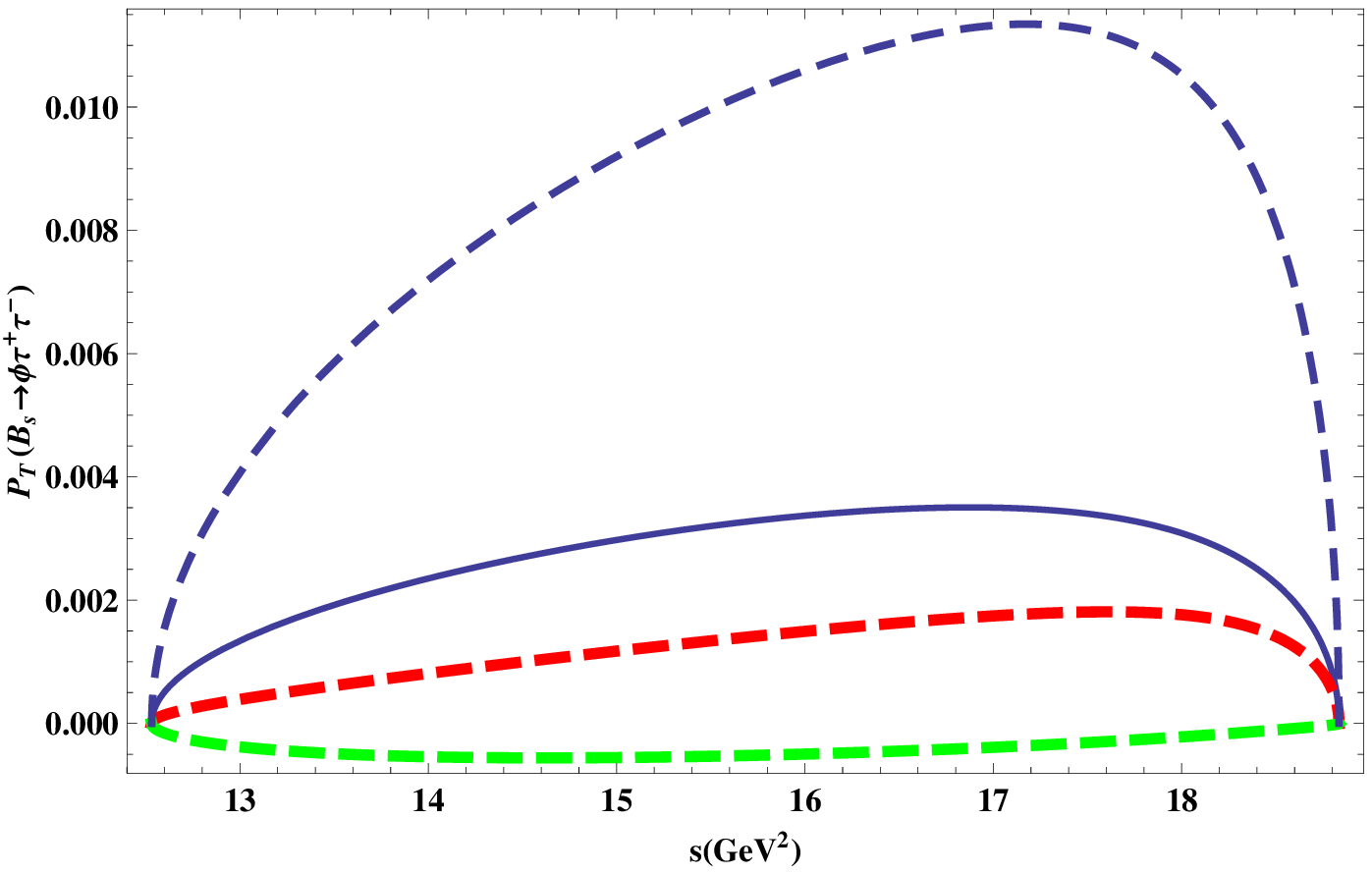,width=0.4\linewidth,clip=b} \put (-100,190){(b)} %
\end{tabular}%
\caption{The transverse lepton polarization asymmetry for the $%
B_{s}\rightarrow \protect\phi l^{+}l^{-}$ ($l=\protect\mu ,\protect\tau $)
decays as functions of $q^{2}$. he legends are same as in Fig.2.}
\end{figure}

\begin{table}[ht]
\caption{Average values of various single lepton polarizations for the central values of the
input parameters. The values in the bracket are for the $\tau^{+}\tau^{-}$ channel.}
\label{input}\centering
\begin{tabular}{ccccc}
\hline\hline
Model& $\left\langle P_{L}\right\rangle$ &$%
\left\langle P_{N}\right\rangle$& $\left\langle P_{T}\right\rangle$ &  \\ \hline
SM & $-0.859(-0.322)$ & $-0.0334(0.068)$ & $0.0003(0.0028)$ & \\ \hline
UED & $-0.857(-0.462)$ & $-0.0451(-0.197)$ & $0.0007(0.0091)$ & \\ \hline
$Z^{\prime}(S_{1})$ & $-0.972^{+0.108}_{-0.024}(-0.365^{+0.031}_{+0.075})$ & $-0.0445^{+0.007}_{-0.001}(0.023^{+0.019}_{+0.016})$ & $-0.0005^{+0.005}_{-0.004}(0.0012^{+0.035}_{-0.023})$ & \\ \hline
$Z^{\prime}(S_{2})$ & $-0.931^{0.066}_{-0.037}(-0.346^{+0.018}_{-0.001})$ & $-0.040^{+0.001}_{-0.002}(0.046^{+0.012}_{+0.009})$ & $-0.0006^{+0.004}_{-0.0036}(-0.0004^{+0.0023}_{-0.021})$ & \\ \hline\hline
\end{tabular}%
\end{table}

The dependence of various double lepton polarization asymmetries on $q^{2}$
for the aforementioned decay in SM and various NP scenarios is given in Fig.
(6-12). In Fig. 6(a,b,c,d) we have plotted $P_{LL}$ as a function of $q^{2}$. It is
clear from\ Eq. (\ref{PLL}) that double longitudinal lepton polarization
asymmetry is proportional to inverse of the mass of lepton, therefore it is
expected to have a large value when final state leptons are the muons
compared to the case when we have tauons and it is also clear from Fig.
6(a,b,c,d). We can also see that the dependency of $P_{LL}$ on NP parameters is
small for the $\mu $-channel. However for the $\tau $-channel the maximum
shift comes in the $Z^{\prime}$ model where $\left\langle
P_{LL}\right\rangle $ deviate almost an order of magnitude from the SM value (c.f. Table VII).
Its measurement will help us in identifying the NP effects arising due to the
extra gauge boson in the $Z^{\prime}$ model.

\begin{table}[tbh]
\caption{{}The values of double lepton polarizations for $\protect\mu (%
\protect\tau )$ in $B_{s}\rightarrow \protect\phi \ell ^{+}\ell ^{-}$ decays.
}%
\begin{tabular}{|c|c|c|c|c|c|c|c|}
\hline
& SM & UED & $Z^{\prime }\left( S_{1}\text{-Scenario}\right) $ & $Z^{\prime
}\left( S_{2}\text{-Scenario}\right) $ \\ \hline
$\left\langle P_{LL}\right\rangle $ & $-0.956\left( -0.046\right) $ & $%
-0.939\left( -0.017\right) $ & $-0.961^{+0.003}_{-0.001}\left( -0.30^{+0.02}_{-0.025}\right) $ & $-0.959^{+0.001}_{-0.006}\left(
-0.313^{+0.011}_{-0.058}\right) $ \\ \hline
$\left\langle P_{NN}\right\rangle $ & $0.014\left( 0.153\right) $ & $%
-0.066\left( 0.192\right) $ & $-0.241^{+0.002}_{-0.014}\left( -0.166^{+0.021}_{-0.029}\right) $ & $-0.265^{+0.005}_{-0.005}\left(
-0.181^{+0.014}_{-0.023}\right) $ \\ \hline
$\left\langle P_{TT}\right\rangle $ & $0.025\left( -0.207\right) $ & $%
-0.037\left( -0.237\right) $ & $-0.218^{+0.005}_{-0.013}\left( 0.102^{-0.021}_{+0.026}\right) $ & $-0.238^{+0.005}_{-0.042}\left(
0.114^{-0.013}_{+0.046}\right) $ \\ \hline
$\left\langle P_{LN}\right\rangle $ & $0.0027\left( 0.057\right) $ & $%
0.0023\left( 0.055\right) $ & $0.0028^{+0.021}_{-0.0138}\left( 0.017^{+0.104}_{-0.067}\right) $ & $0.0014^{+0.014}_{-0.002}\left(
0.011^{+0.068}_{-0.060}\right) $ \\ \hline
$\left\langle P_{LT}\right\rangle $ & $-0.069\left( -0.104\right) $ & $%
-0.070\left( -0.104\right) $ & $-0.022^{-0.004}_{+0.003}\left( 0.030^{-0.037}_{+0.015}\right) $ & $-0.019^{-0.003}_{+0.002}\left(
0.031^{-0.014}_{+0.014}\right) $ \\ \hline
$\left\langle P_{TL}\right\rangle $ & $0.085\left( 0.356\right) $ & $%
0.084\left( 0.356\right) $ & $0.050^{-0.001}_{-0.003}\left( 0.180^{-0.001}_{-0.016}\right) $ & $0.045^{+0.0001}_{-0.0001}\left(
0.167^{+0.0001}_{-0.011}\right) $ \\ \hline
$\left\langle P_{TN}\right\rangle $ & $0.0016\left( -0.0018\right) $ & $%
-0.007\left( -0.0015\right) $ & $0.039^{+0.023}_{-0.011}\left( 0.0058^{-0.0057}_{+0.004}\right) $ & $0.037^{+0.007}_{-0.006}\left(
0.0055^{-0.0054}_{+0.004}\right) $ \\ \hline
\end{tabular}%
\label{DLP}
\end{table}

\begin{figure}[tbp]
\begin{tabular}{cc}
\epsfig{file=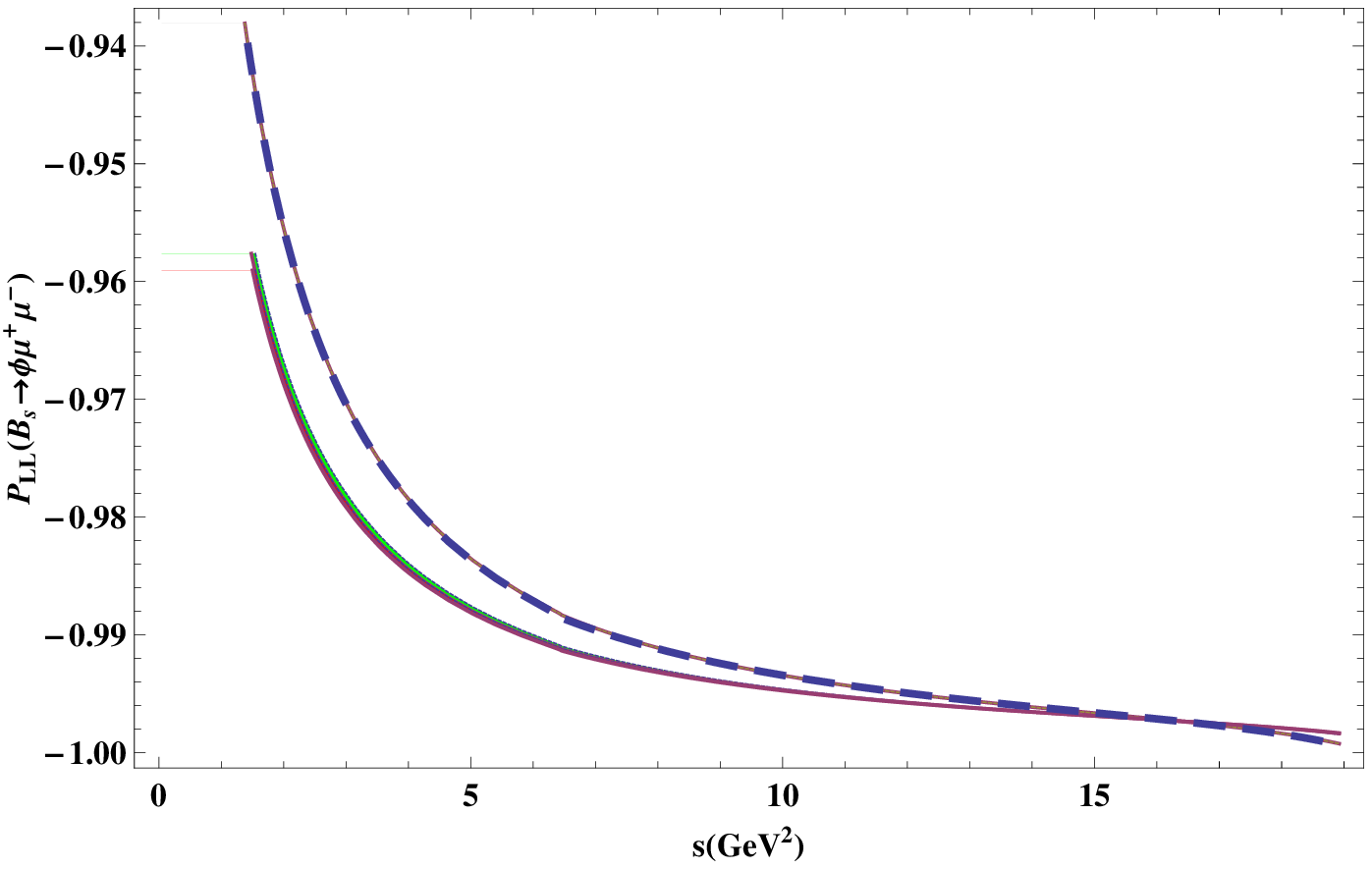,width=0.4\linewidth,clip=a} \put (-100,190){(a)} & %
\epsfig{file=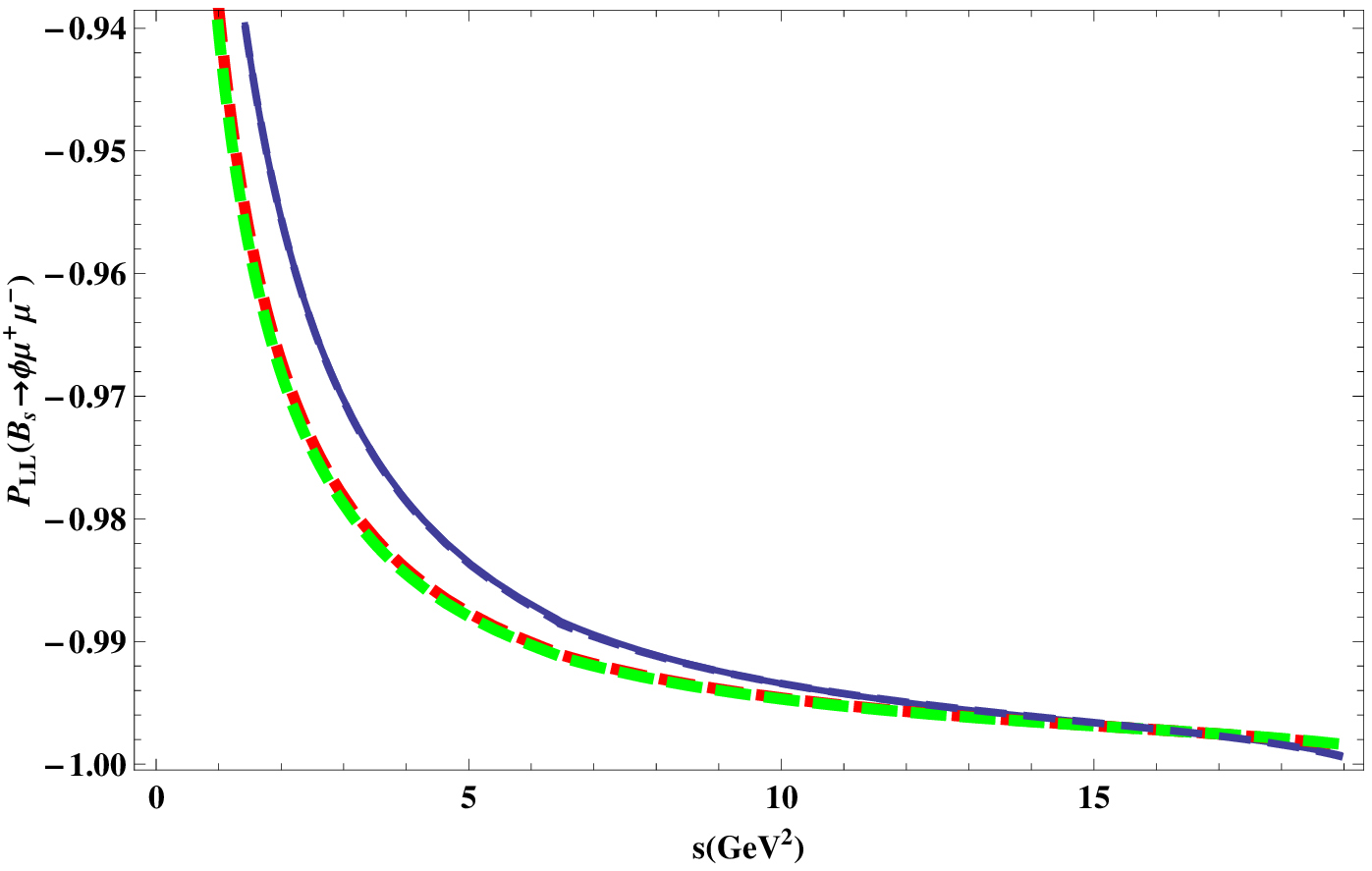,width=0.4\linewidth,clip=b} \put (-100,190){(b)} \\
\epsfig{file=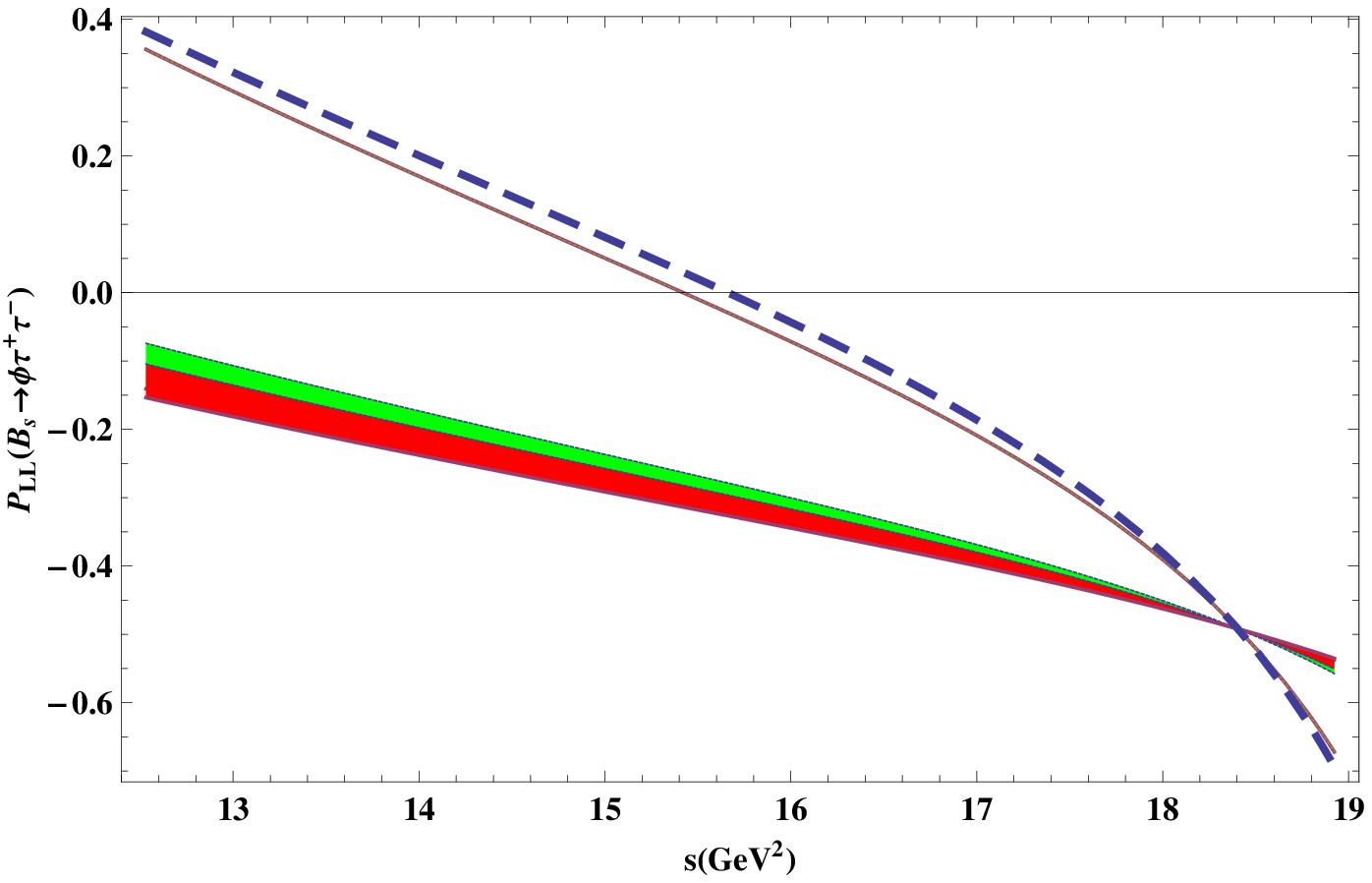,width=0.4\linewidth,clip=c} \put (-100,190){(c)} & %
\epsfig{file=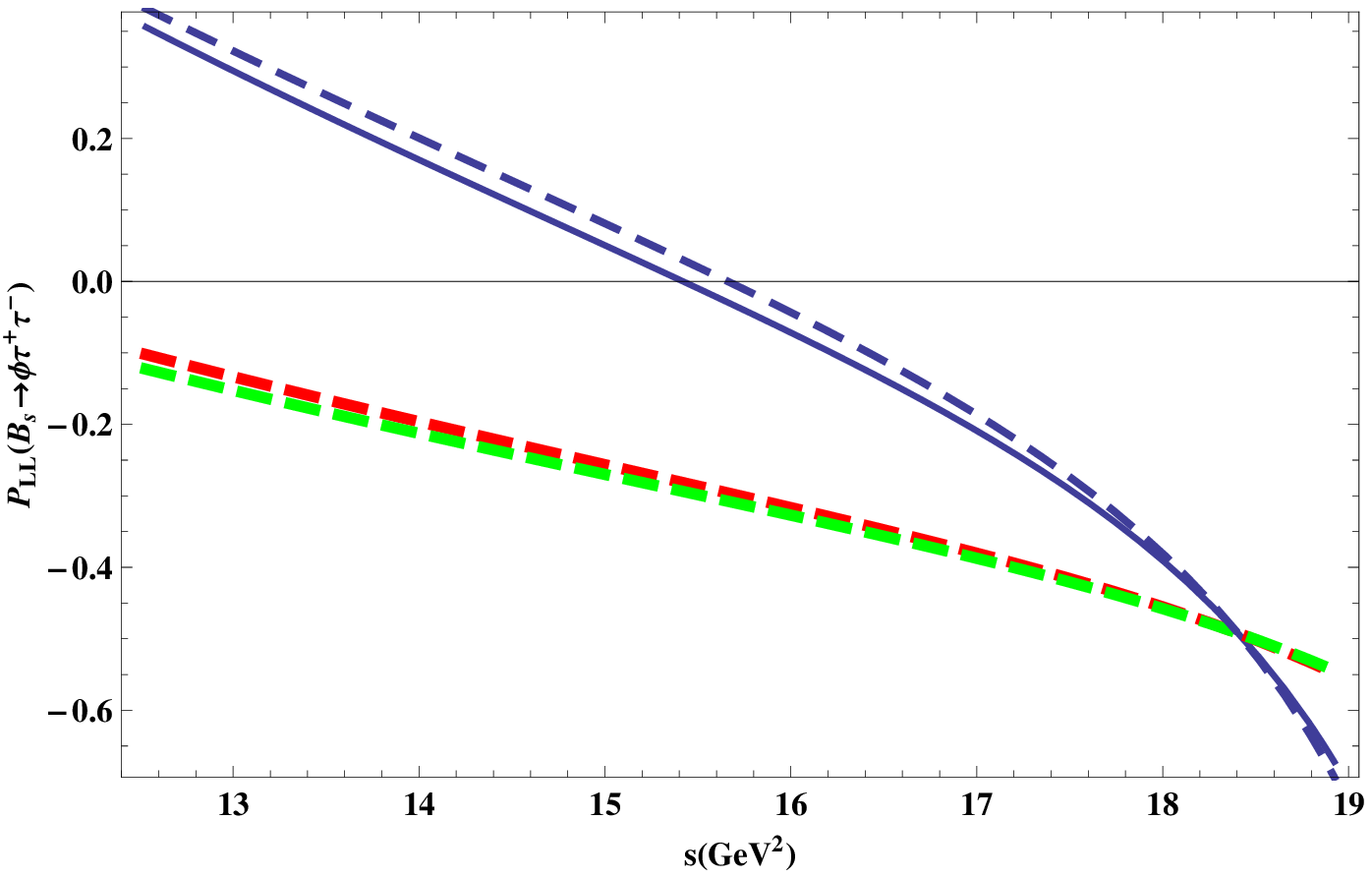,width=0.4\linewidth,clip=d} \put (-100,190){(d)}%
\end{tabular}%
\caption{$P_{LL}$ for the $B_{s}\rightarrow \protect\phi l^{+}l^{-}$ ($l=%
\protect\mu ,\protect\tau $) decays as functions of $q^{2}$. The legends are
same as in Fig.2.}
\end{figure}

By looking at Eq. (\ref{PLN}), we can see that $P_{LN}$ is proportional to the
imaginary part of the different auxiliary functions and so of the Wilson
coefficients, therefore, its non zero value is expected only in the
$Z^{\prime }$ model. However the imaginary part in this
cases is small, therefore its values is expected to be small and Fig.
7(a,b) displays this fact which quantitatively can also be seen in Table VII.

\begin{figure}[tbp]
\begin{tabular}{cc}
\epsfig{file=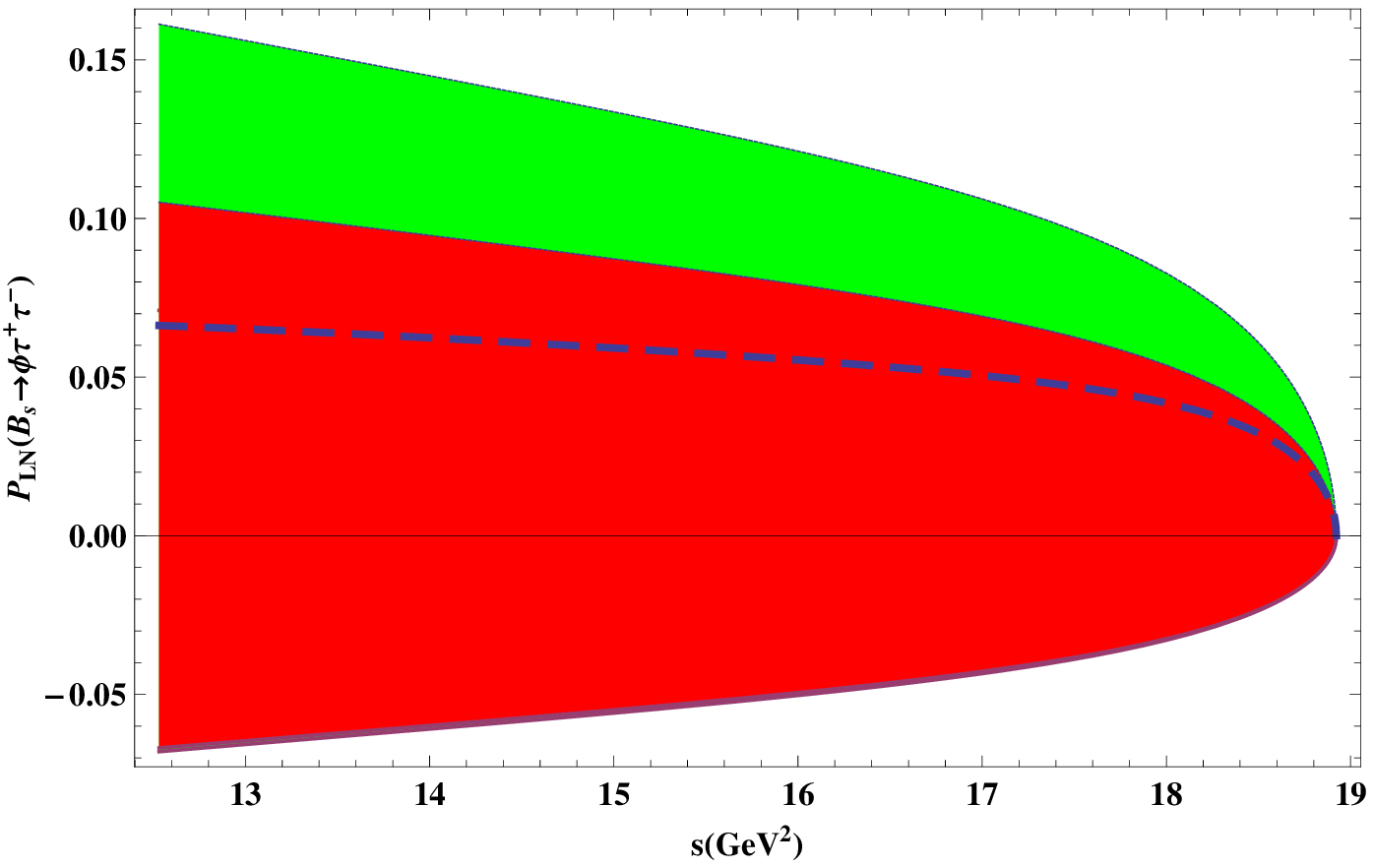,width=0.4\linewidth,clip=a} \put (-100,190){(a)} & %
\epsfig{file=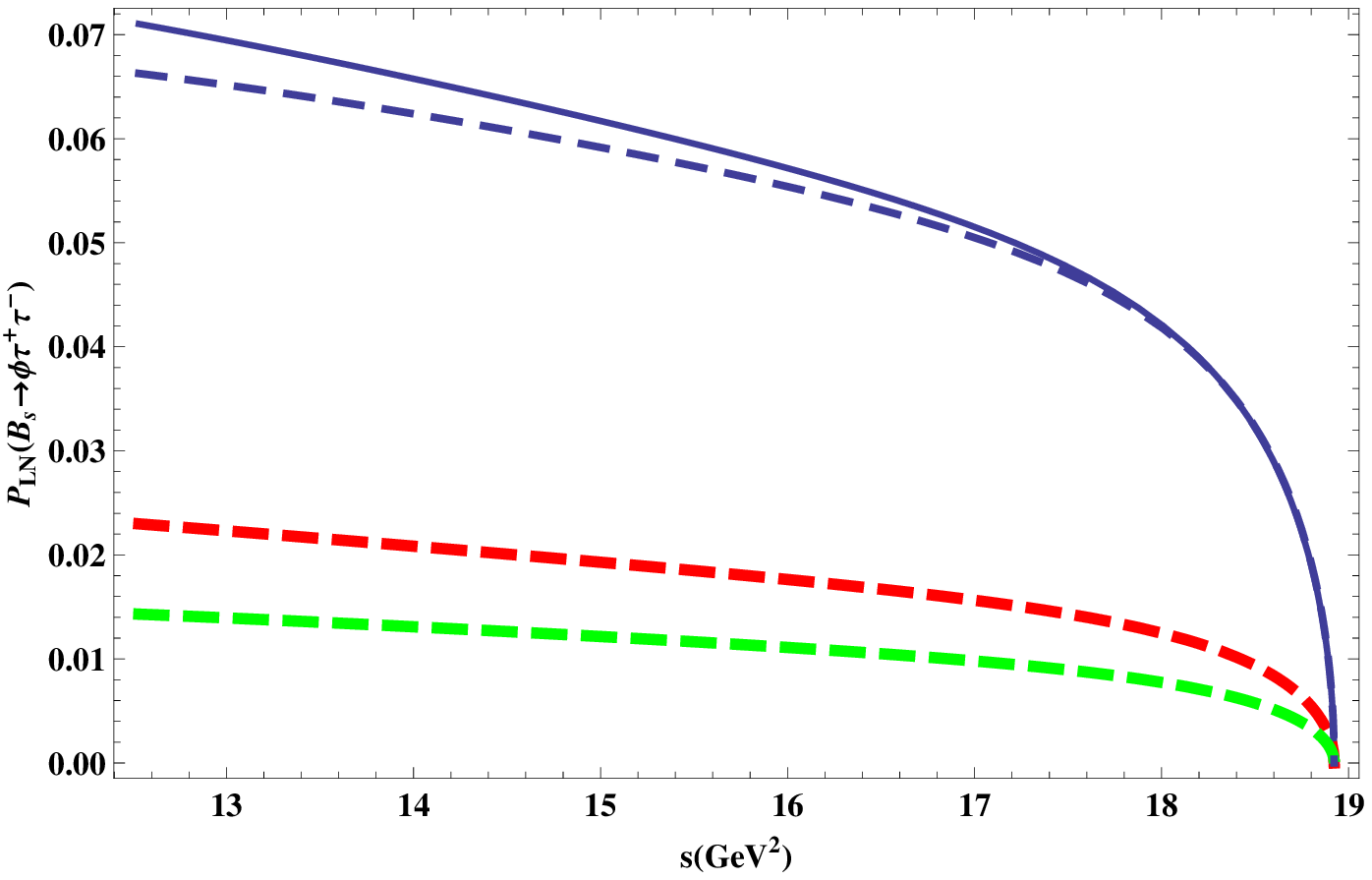,width=0.4\linewidth,clip=b} \put (-100,190){(b)}%
\end{tabular}%
\caption{$P_{LN}$ for the $B_{s}\rightarrow \protect\phi l^{+}l^{-}$ ($l=%
\protect\mu ,\protect\tau $) decays as functions of $q^{2}$. The legends are
same as in Fig.2. }
\label{PLN}
\end{figure}

Fig. 8(a,b,c,d) depicts the behavior of $P_{LT}$ with $q^{2}$ where we can see
that the new physics effects are quite promising both for the $\mu$ and $\tau $%
-channels in almost whole range of available $q^{2}$. Quantitatively we can
see from Table VII that the value of average $\left\langle P_{LT}\right\rangle
$ in UED model is closed to the SM value where as in the $Z^{\prime}$ model
the average value of $P_{LT}$ is significantly suppressed in magnitude from its SM value
as well as it flips its sign for the
$\tau$ channel.

\begin{figure}[tbp]
\begin{tabular}{cc}
\epsfig{file=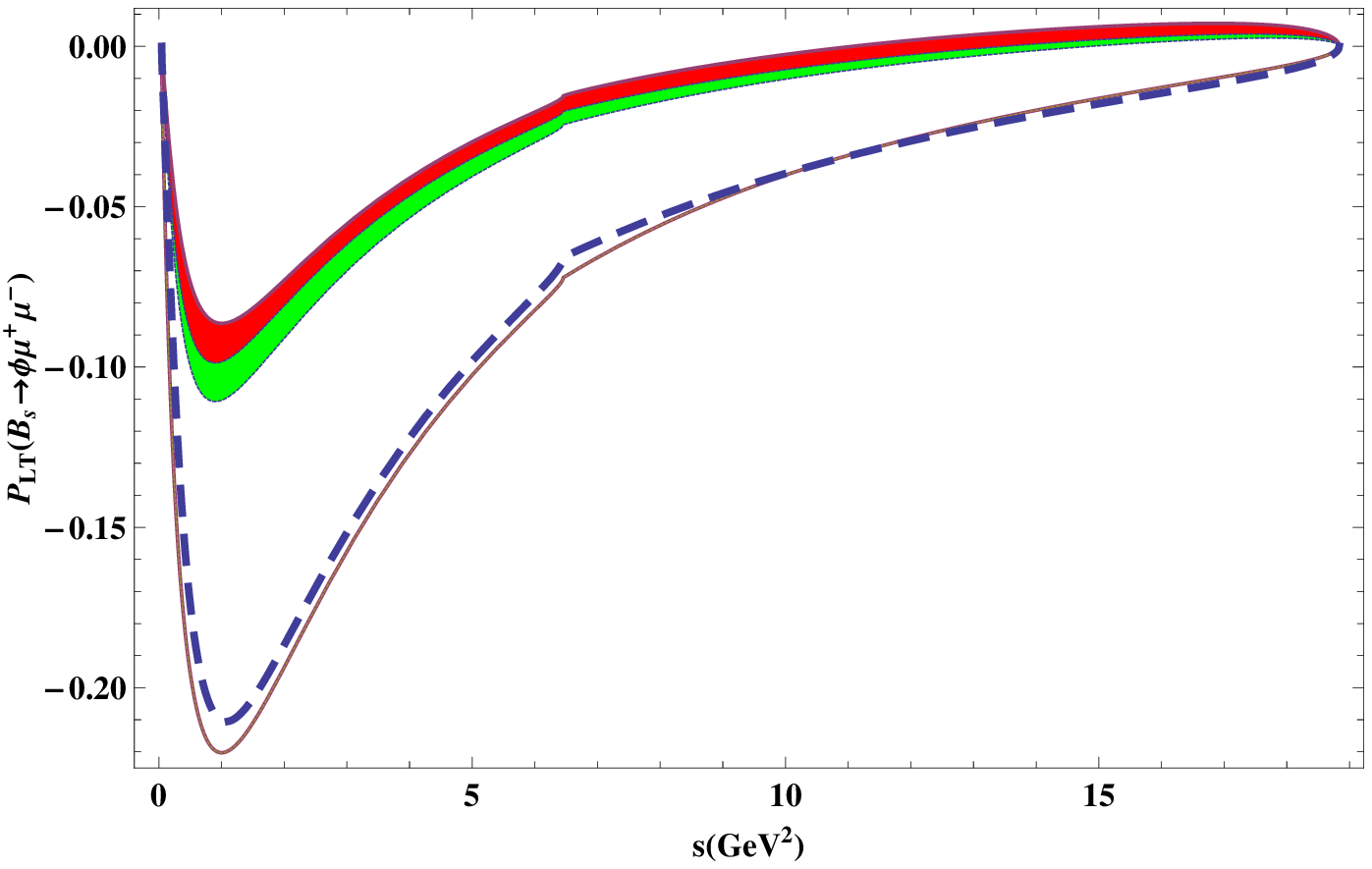,width=0.4\linewidth,clip=a} \put (-100,190){(a)} & %
\epsfig{file=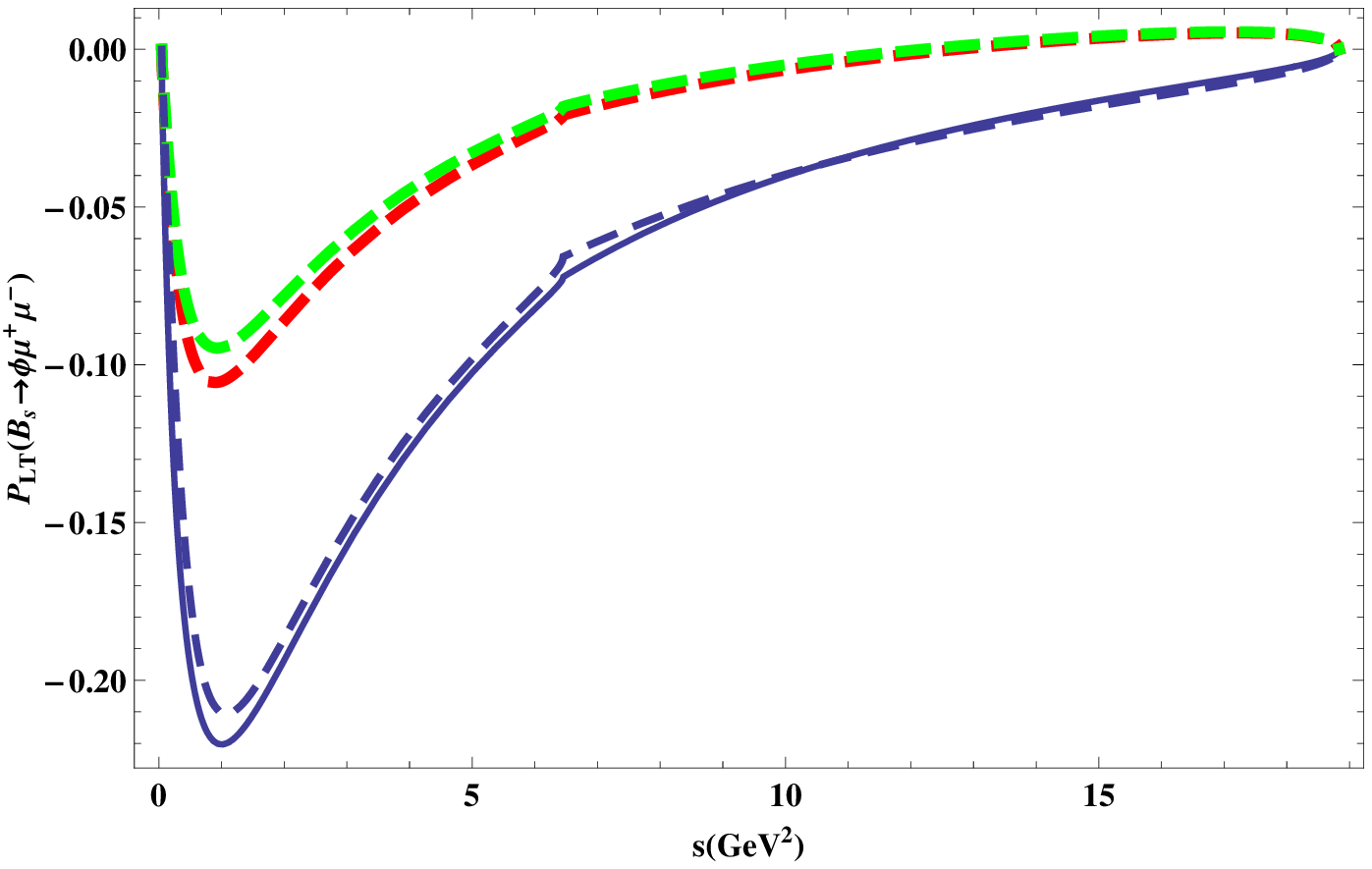,width=0.4\linewidth,clip=b} \put (-100,190){(b)} \\
\epsfig{file=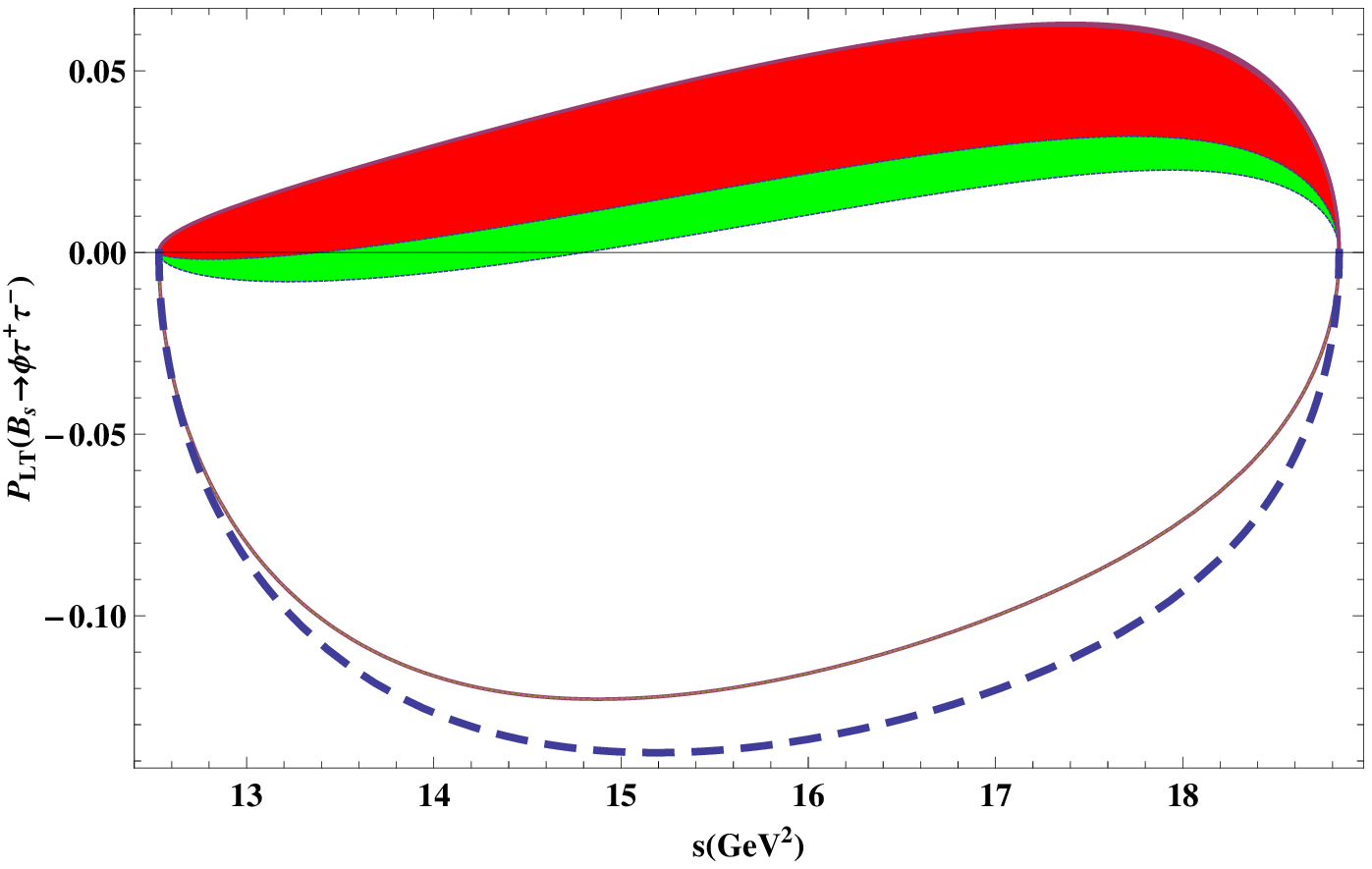,width=0.4\linewidth,clip=c} \put (-100,190){(c)} & %
\epsfig{file=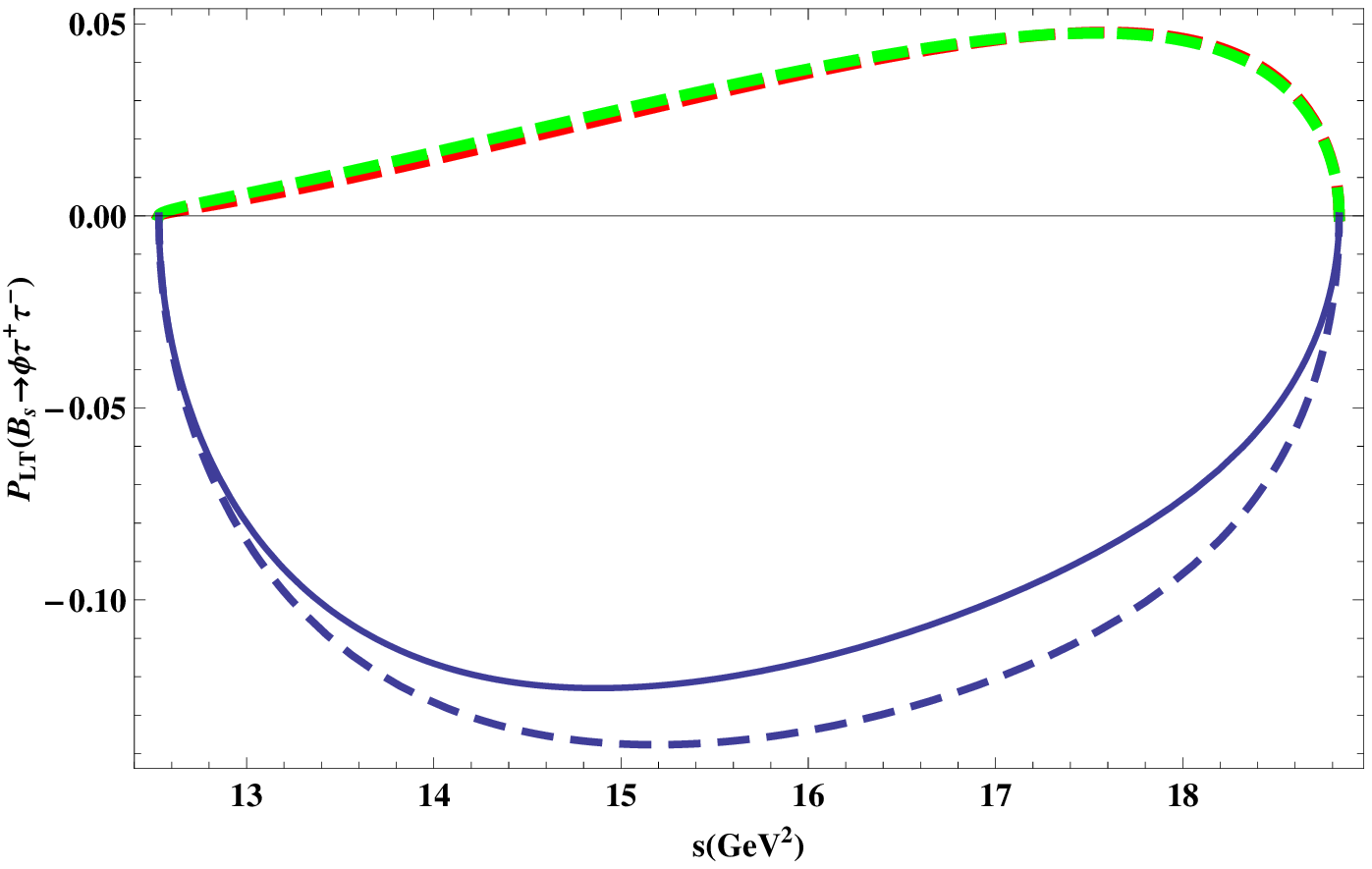,width=0.4\linewidth,clip=d} \put (-100,190){(d)}%
\end{tabular}%
\caption{$P_{LT}$ for the $B_{s} \to \protect\phi l^+l^-$ ($l=\protect\mu,
\protect\tau$) decays as functions of $q^2$. The legends are same as in
Fig.2. }
\label{PLT}
\end{figure}

In Fig. 9(a,b,c,d) we displayed the effects of various NP models on the $P_{NN}$. We can
see that the value of $P_{NN}$ shows strong dependence on the parameters of NP which
is quite prominent in both $\mu$ and $\tau$-channels. From Table VII,
it is clear that in the $Z^{\prime}$ model the value of $P_{NN}$ is significantly different
not only from the SM value but also from the UED model. Therefore the experimental observation of
this observable will help us to segregate the $Z^{\prime}$ model both from the SM as well as from UED
model.

\begin{figure}[tbp]
\begin{tabular}{cc}
\epsfig{file=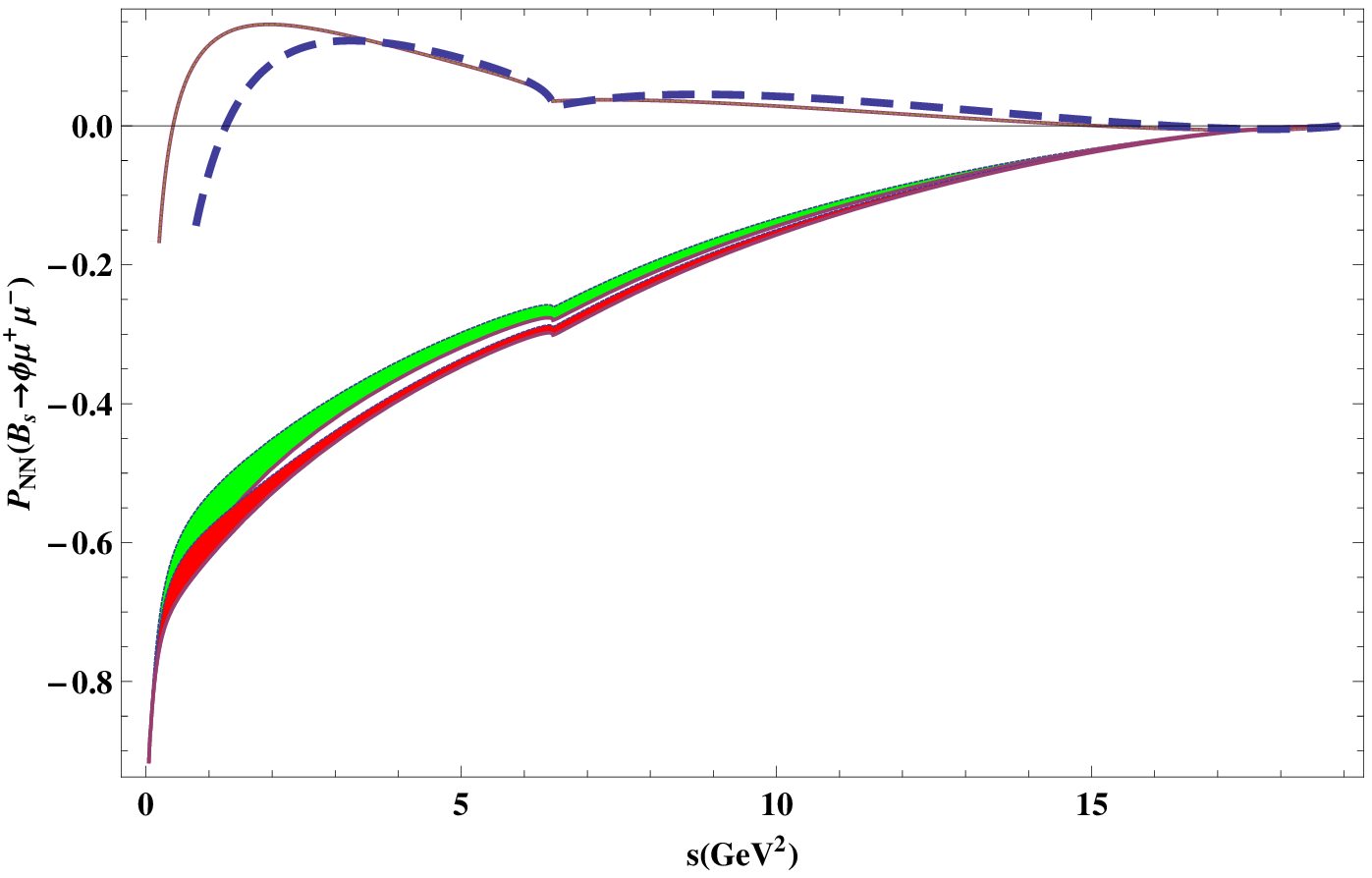,width=0.4\linewidth,clip=a} \put (-100,190){(a)} & %
\epsfig{file=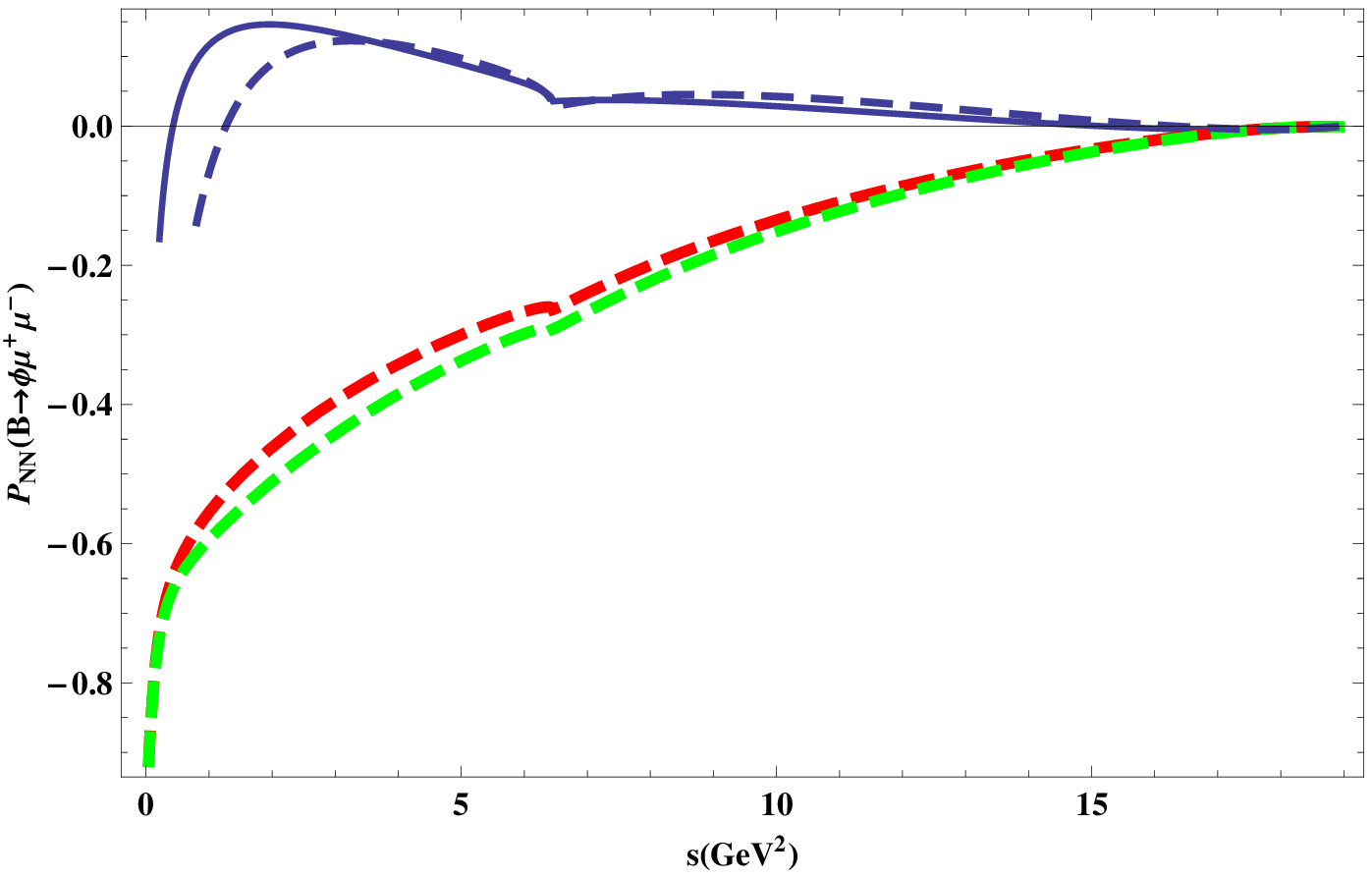,width=0.4\linewidth,clip=b} \put (-100,190){(b)} \\
\epsfig{file=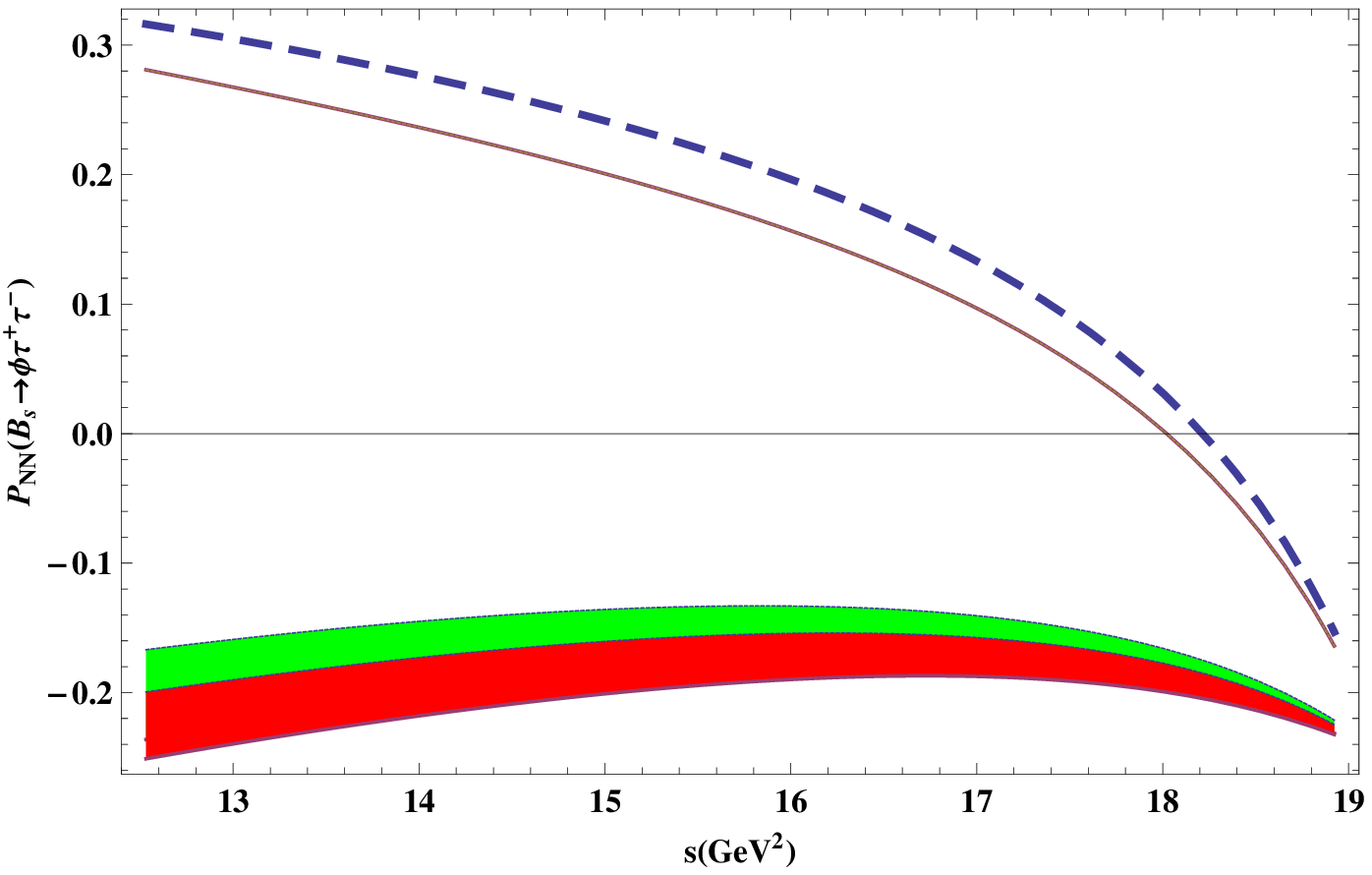,width=0.4\linewidth,clip=c} \put (-100,190){(c)} & %
\epsfig{file=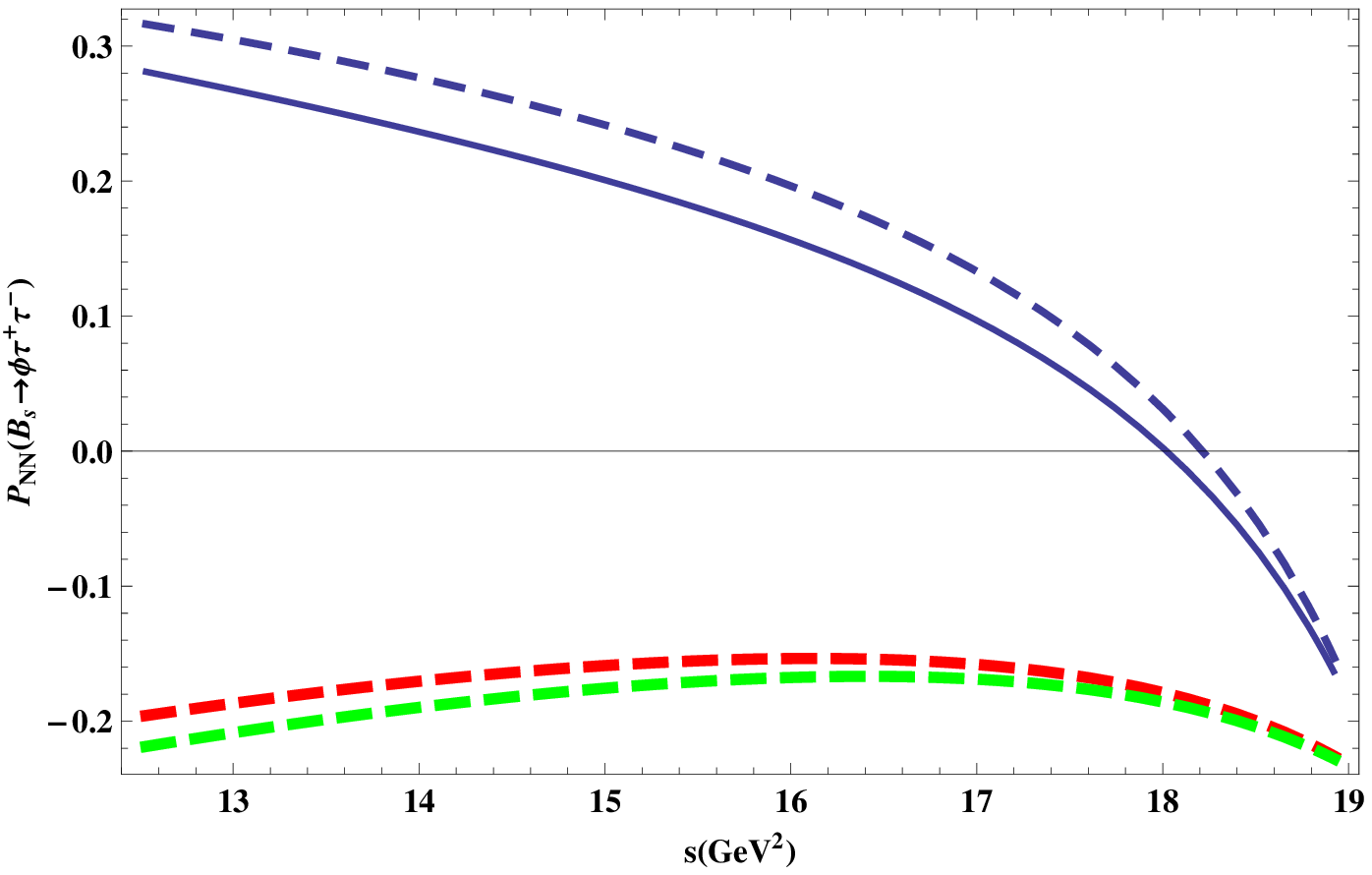,width=0.4\linewidth,clip=d} \put (-100,190){(d)}%
\end{tabular}%
\caption{$P_{NN}$ for the $B_{s} \to \protect\phi l^+l^-$ ($l=\protect\mu,
\protect\tau$) decays as functions of $q^2$. The legends are same as in
Fig.2. }
\label{PNN}
\end{figure}

The situation for the $P_{TL}$ is not so interesting for $\mu$- channel because its average
value is small in this case. However when we have $\tau$'s as final state leptons, the effects of
extra gauge boson in $Z^{\prime}$ model reduce the average value of $P_{TL}$ by $50\%$. This can be
seen quantitatively from Table VII and it is also depicted in Fig. 10(a,b,c,d).

\begin{figure}[tbp]
\begin{tabular}{cc}
\epsfig{file=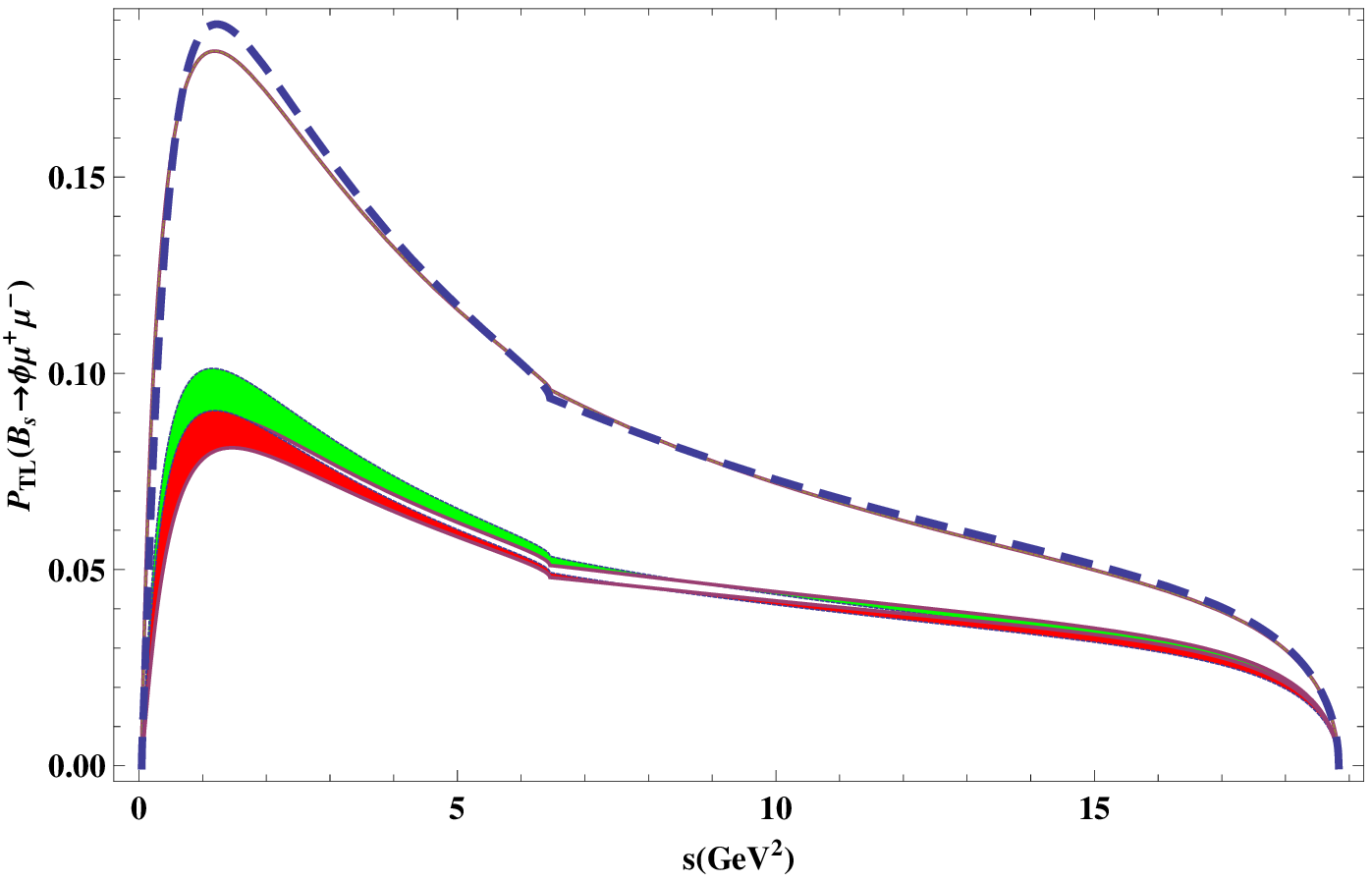,width=0.4\linewidth,clip=a} \put (-100,190){(a)} & %
\epsfig{file=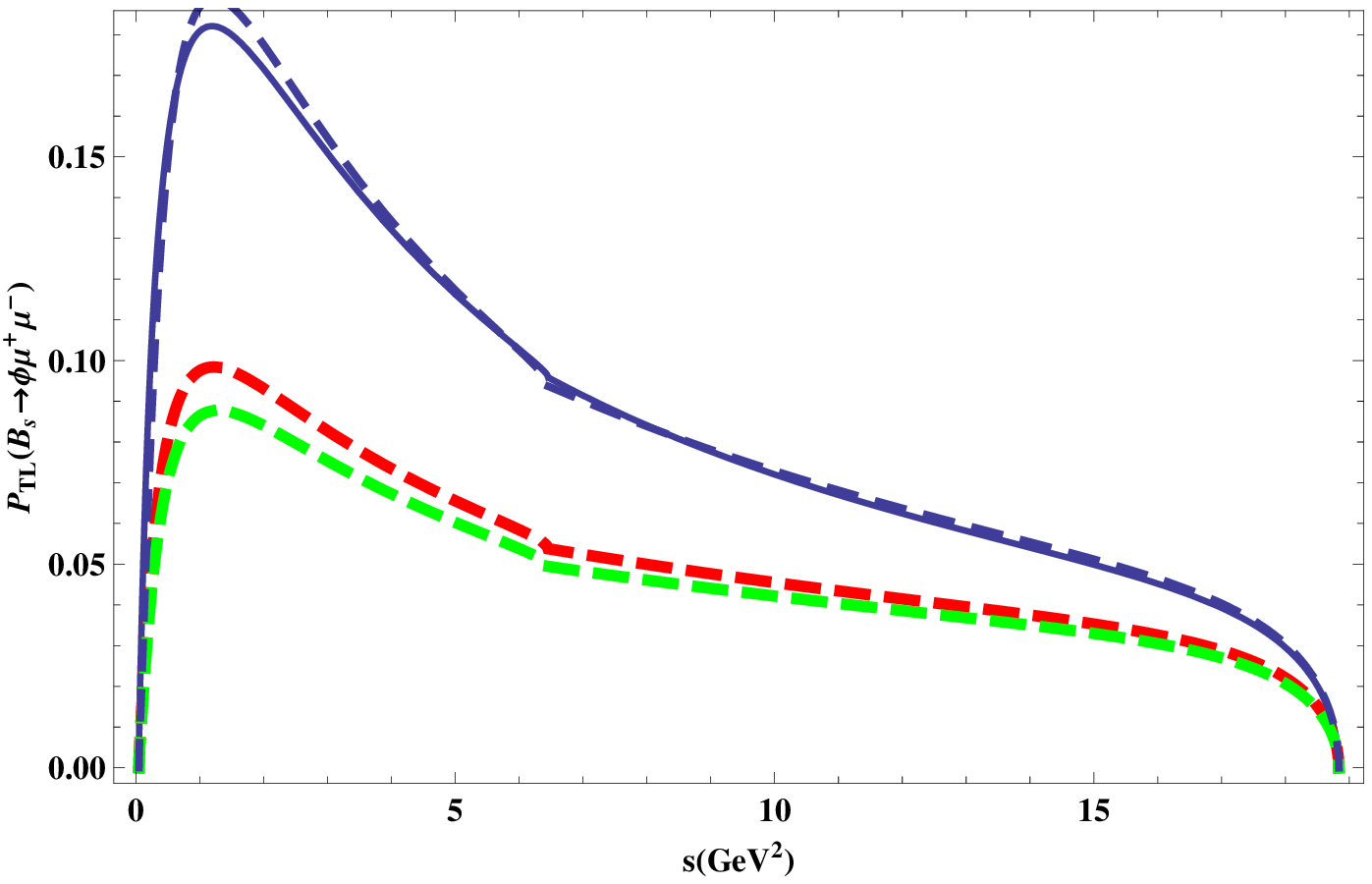,width=0.4\linewidth,clip=b} \put (-100,190){(b)} \\
\epsfig{file=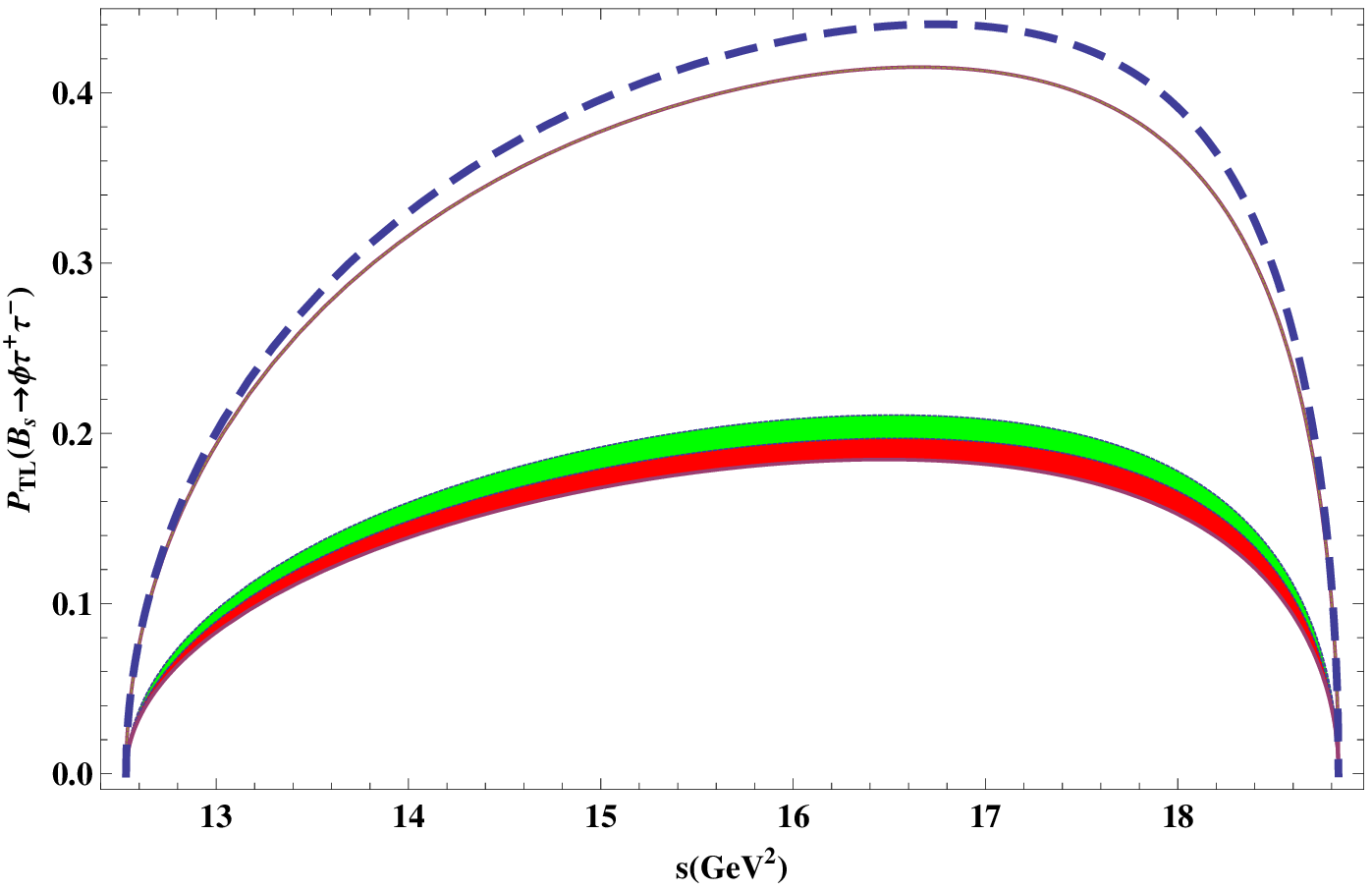,width=0.4\linewidth,clip=c} \put (-100,190){(c)} & %
\epsfig{file=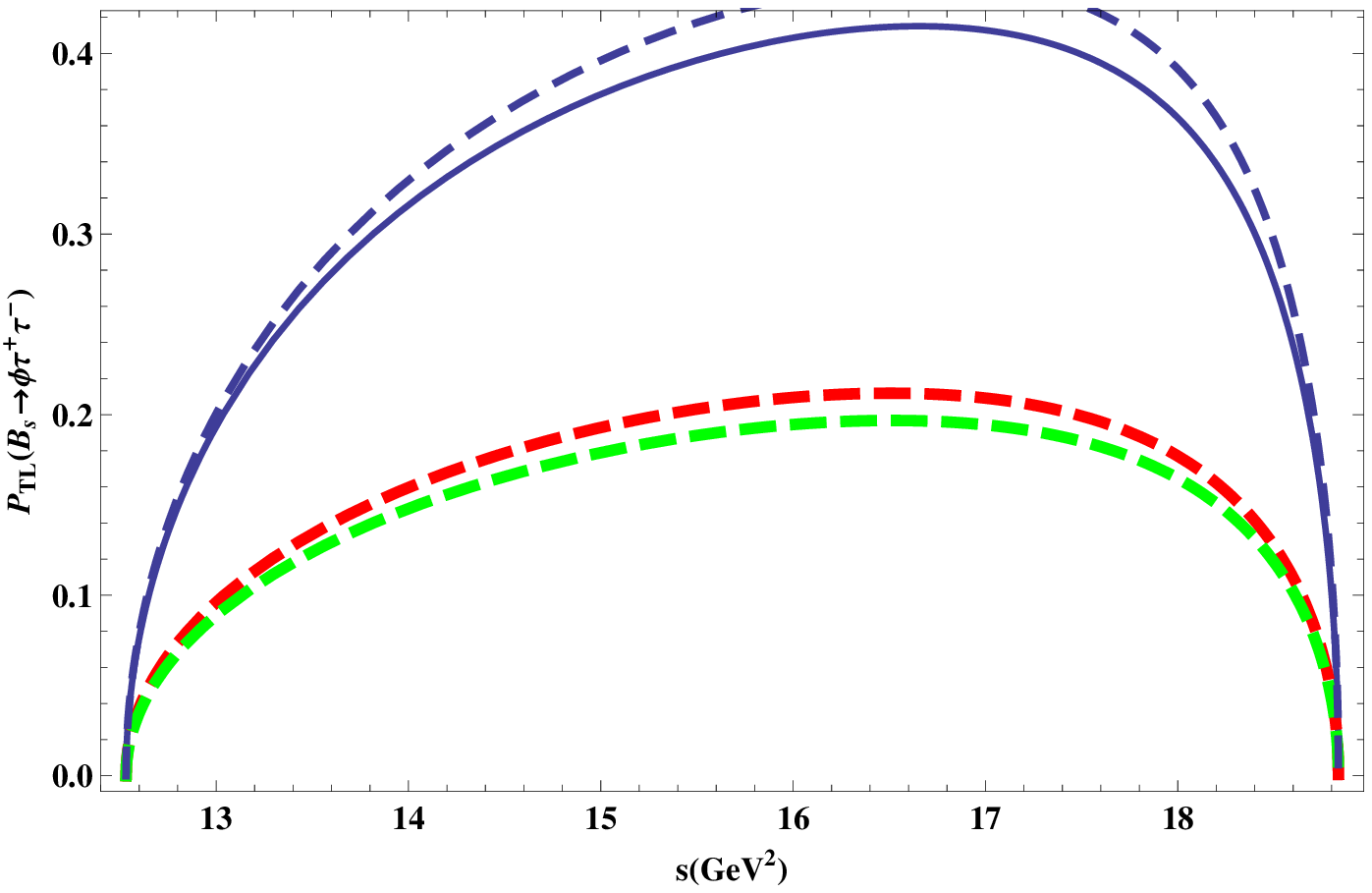,width=0.4\linewidth,clip=d} \put (-100,190){(d)}%
\end{tabular}%
\caption{$P_{TL}$ for the $B_{s} \to \protect\phi l^+l^-$ ($l=\protect\mu,
\protect\tau$) decays as functions of $q^2$. The legends are same as in
Fig.2. }
\label{PTL}
\end{figure}

In Eq. (\ref{PTN}) we can see that the value of $P_{TN}$ comes from the imaginary part
of various Wilson Coefficients, therefore, as expected its value is too small to measure. This is obvious from
Fig. 10(a,b,c,d) and also from Table VII.

\begin{figure}[tbp]
\begin{tabular}{cc}
\epsfig{file=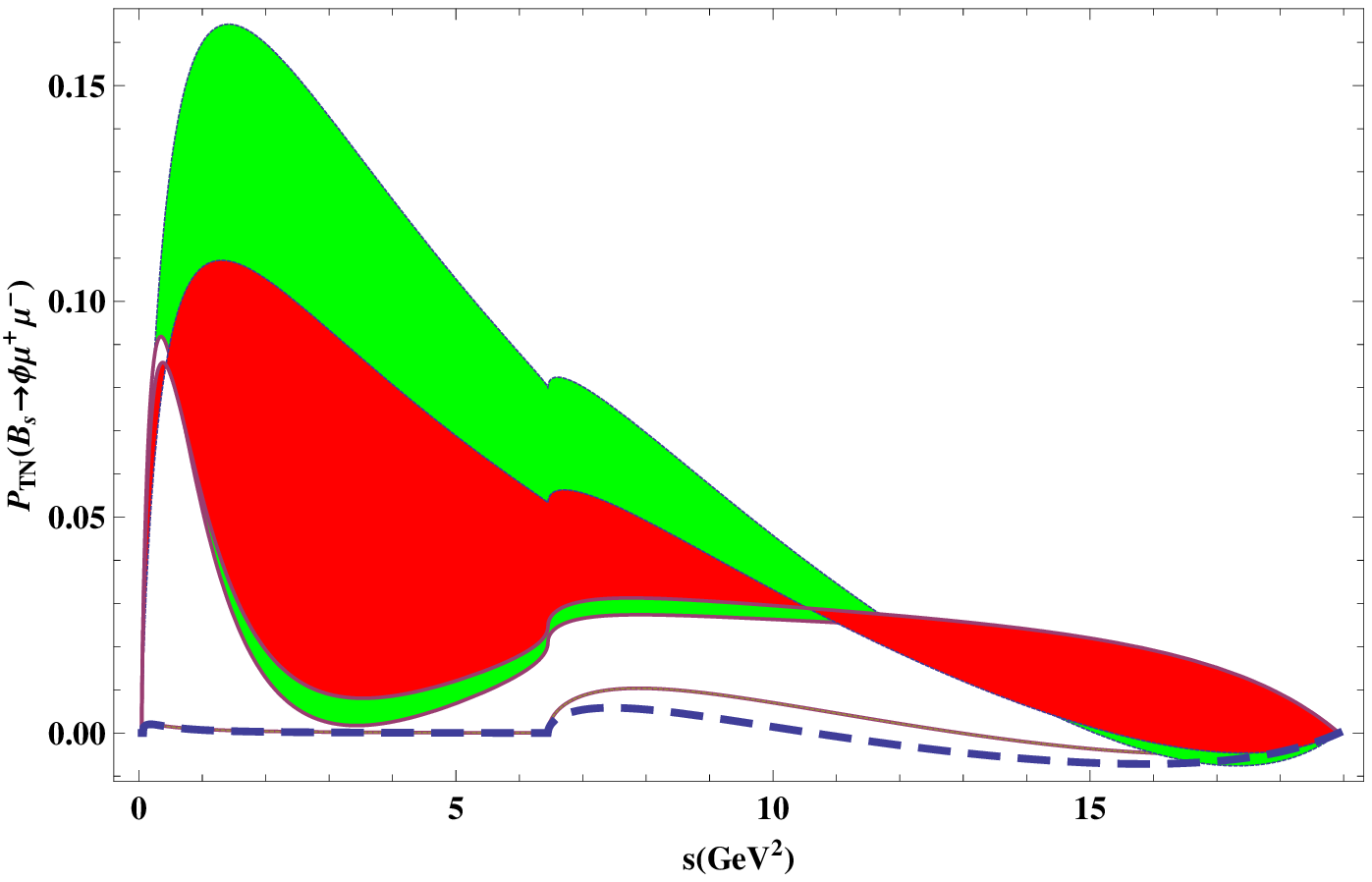,width=0.4\linewidth,clip=a} \put (-100,190){(a)} & %
\epsfig{file=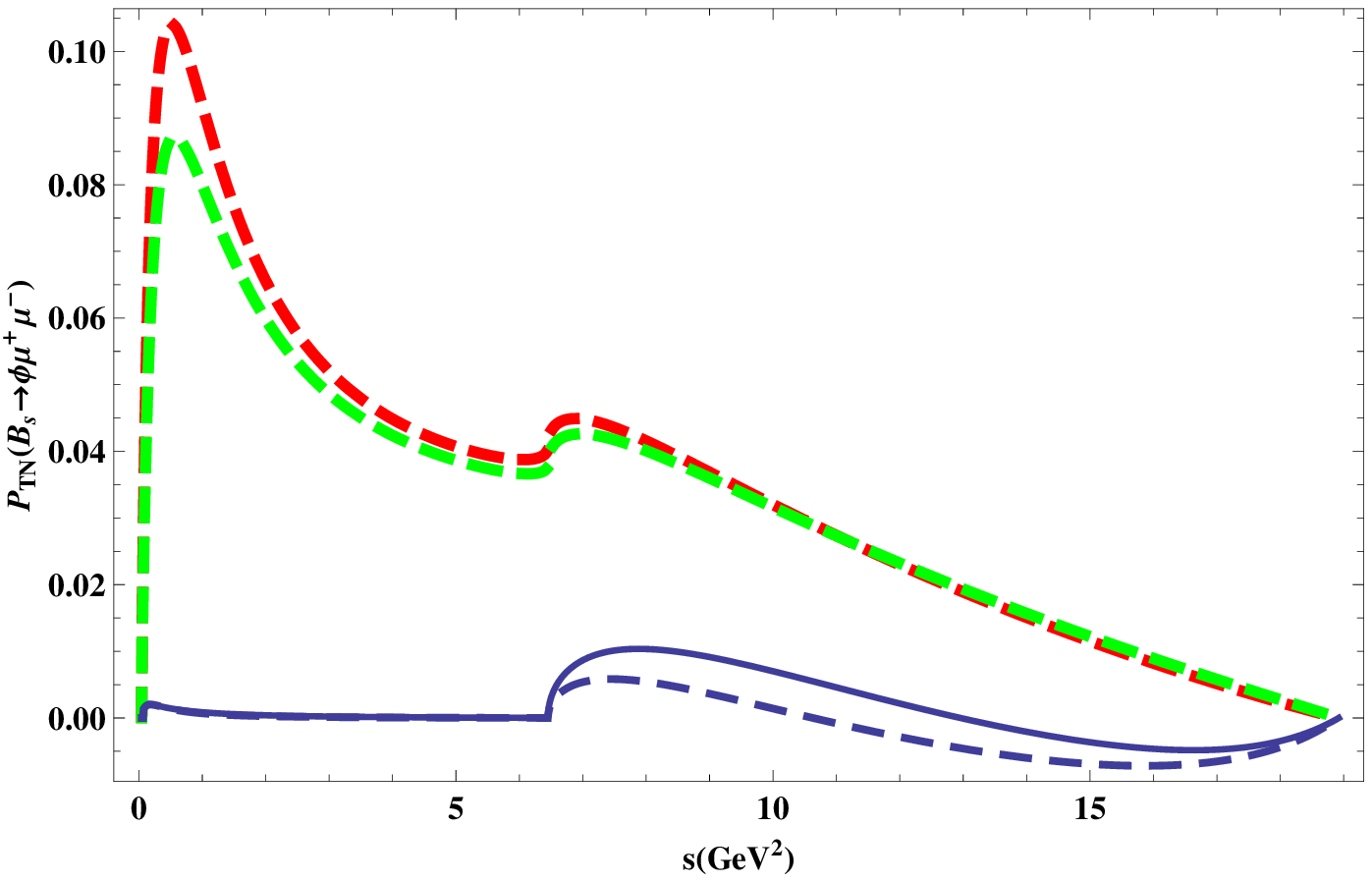,width=0.4\linewidth,clip=b} \put (-100,190){(b)} \\
\epsfig{file=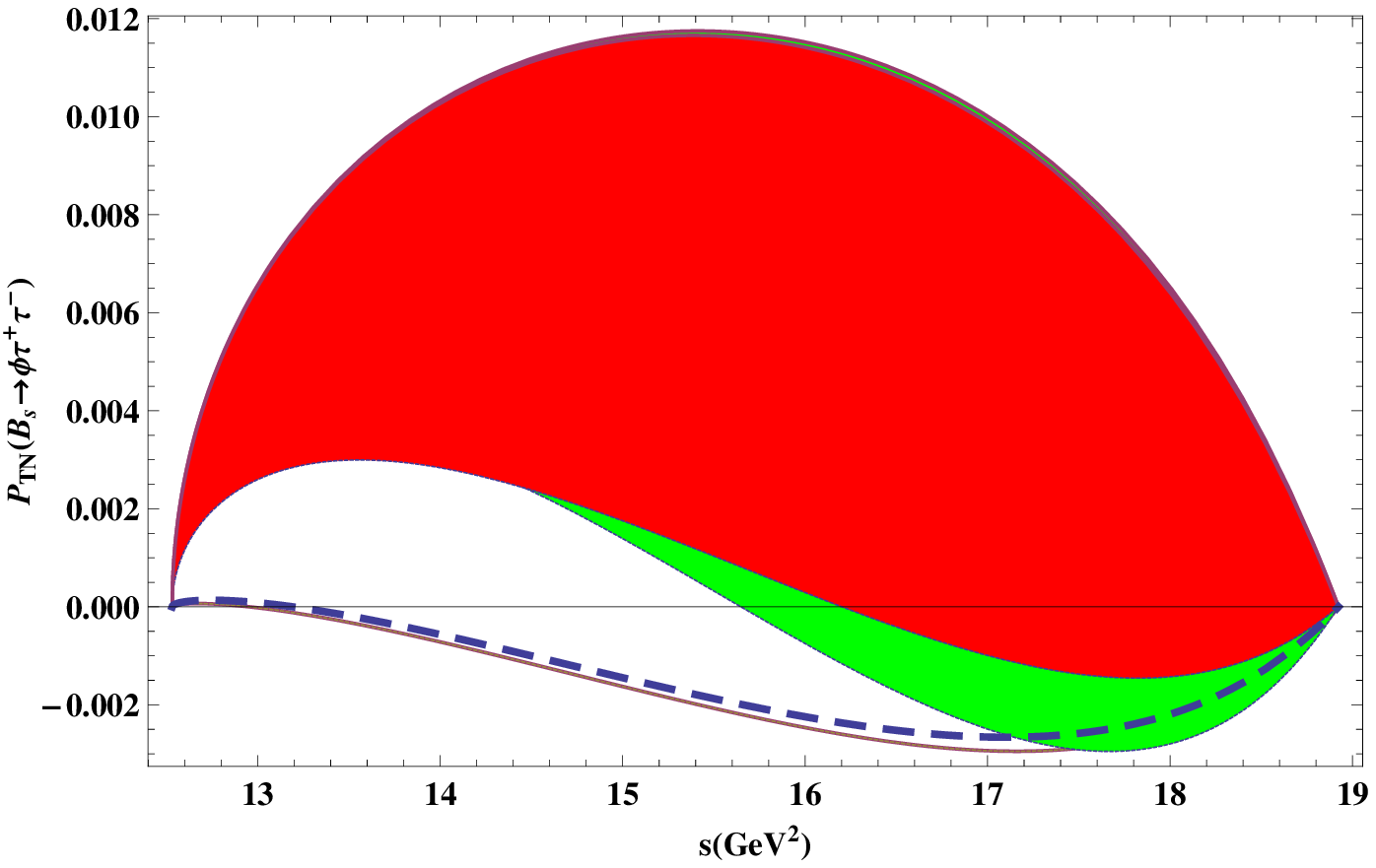,width=0.4\linewidth,clip=c} \put (-100,190){(c)} & %
\epsfig{file=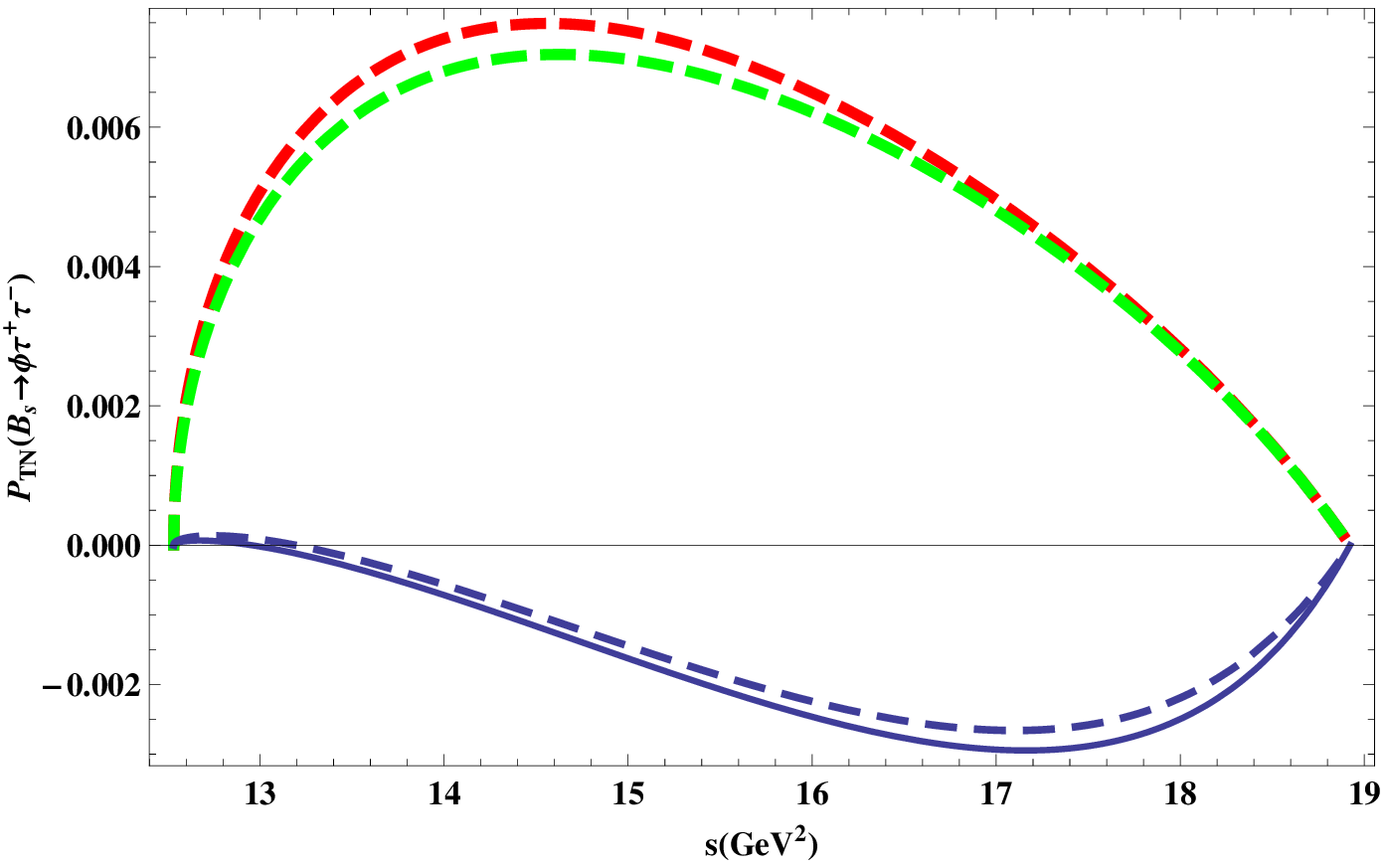,width=0.4\linewidth,clip=d} \put (-100,190){(d)}%
\end{tabular}%
\caption{$P_{TN}$ for the $B_{s} \to \protect\phi l^+l^-$ ($l=\protect\mu,
\protect\tau$) decays as functions of $q^2$. The legends are same as in
Fig.2. }
\label{PTN}
\end{figure}

The case in which both the leptons are transversely polarized that is $P_{TT}$
becomes important for the $\tau$ channel. Here we can see that its behavior with $q^2$
is very different in the $Z^{\prime}$ model where it has positive values compared to its values in
the SM and in UED model where the value of $P_{TT}$ is negative. This fact
is depicted in Fig. 12(a,b,c,d) and numerically given in Table VII.

\begin{figure}[tbp]
\begin{tabular}{cc}
\epsfig{file=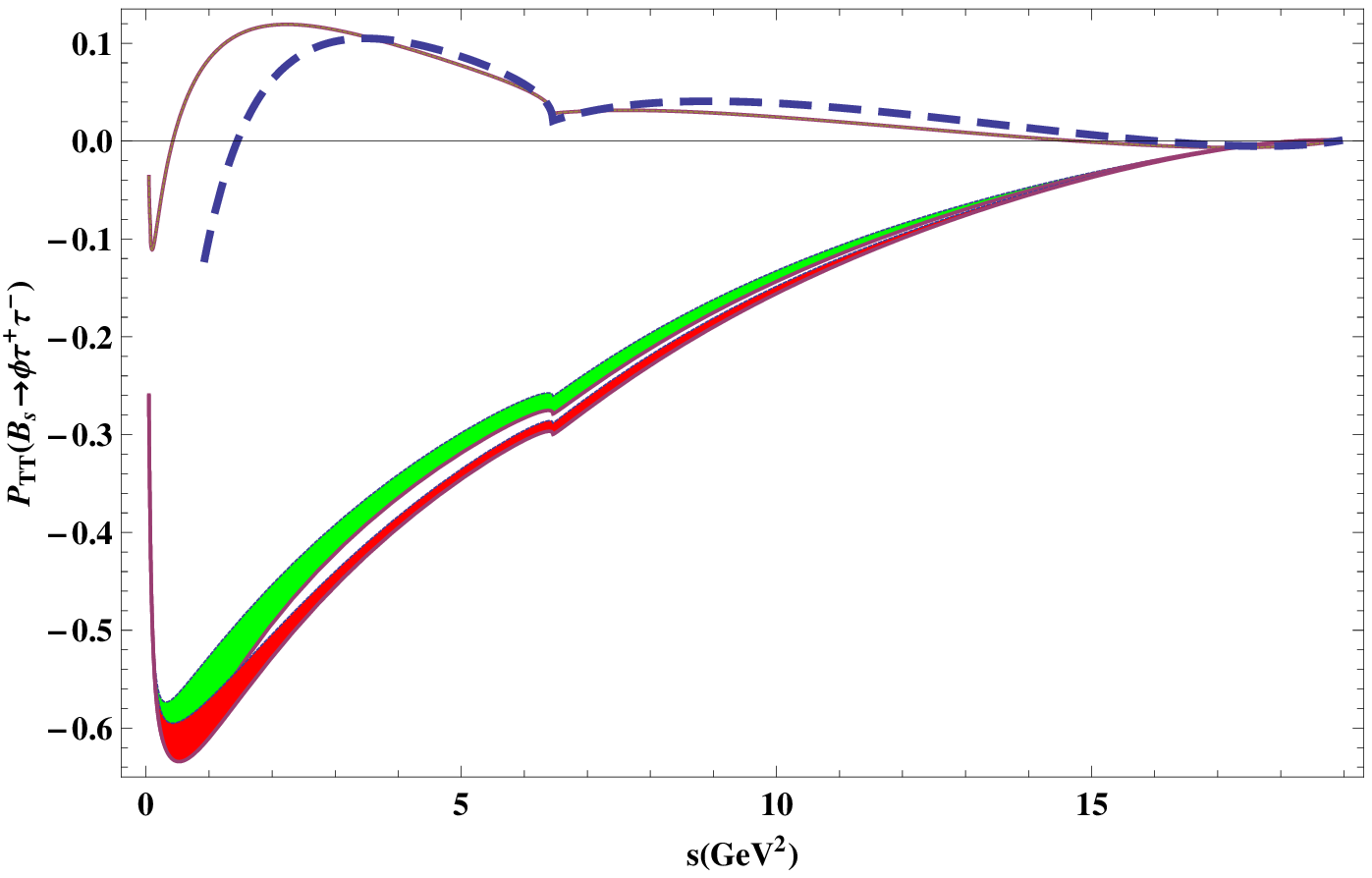,width=0.4\linewidth,clip=a} \put (-100,190){(a)} & %
\epsfig{file=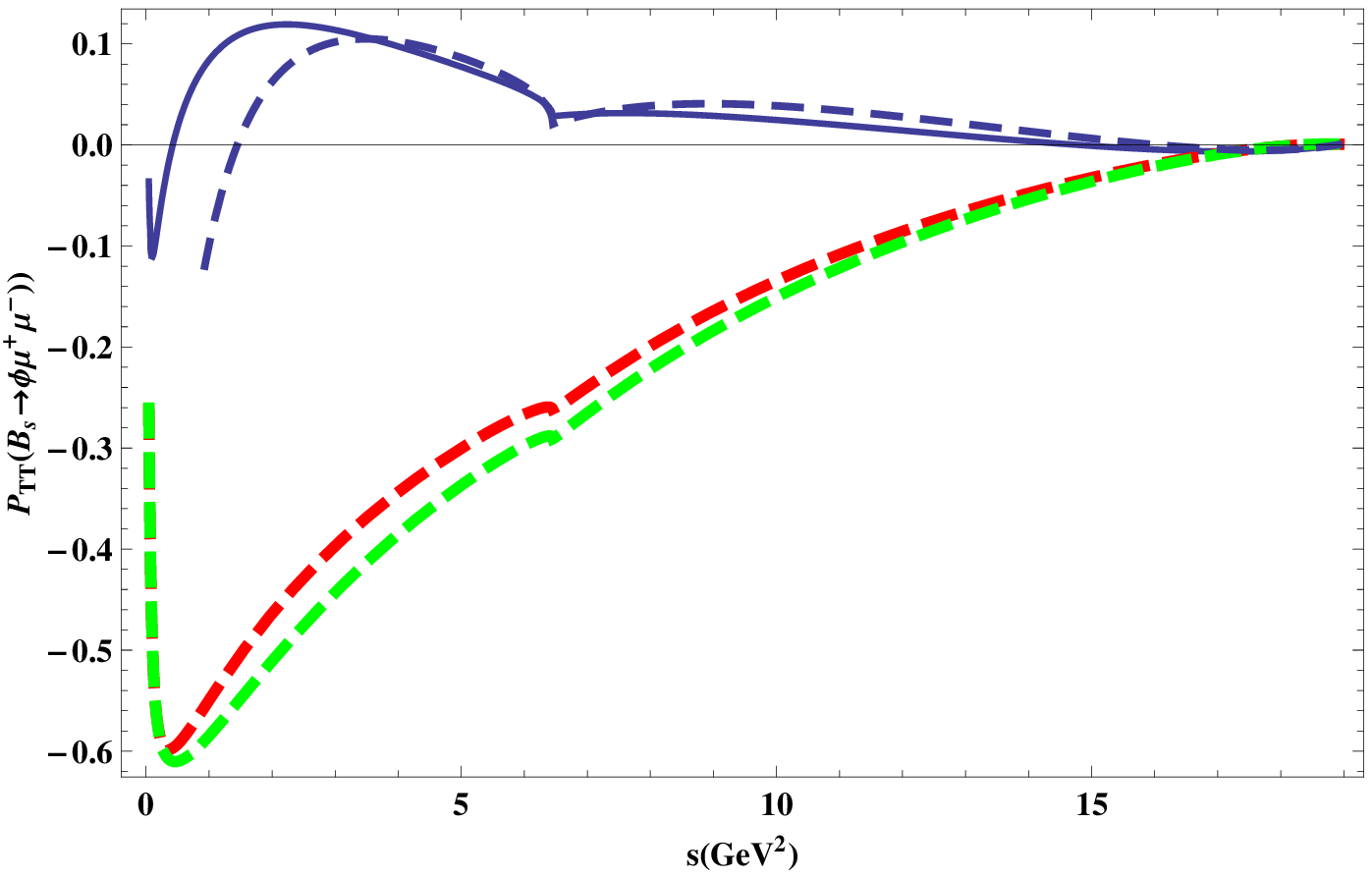,width=0.4\linewidth,clip=b} \put (-100,190){(b)} \\
\epsfig{file=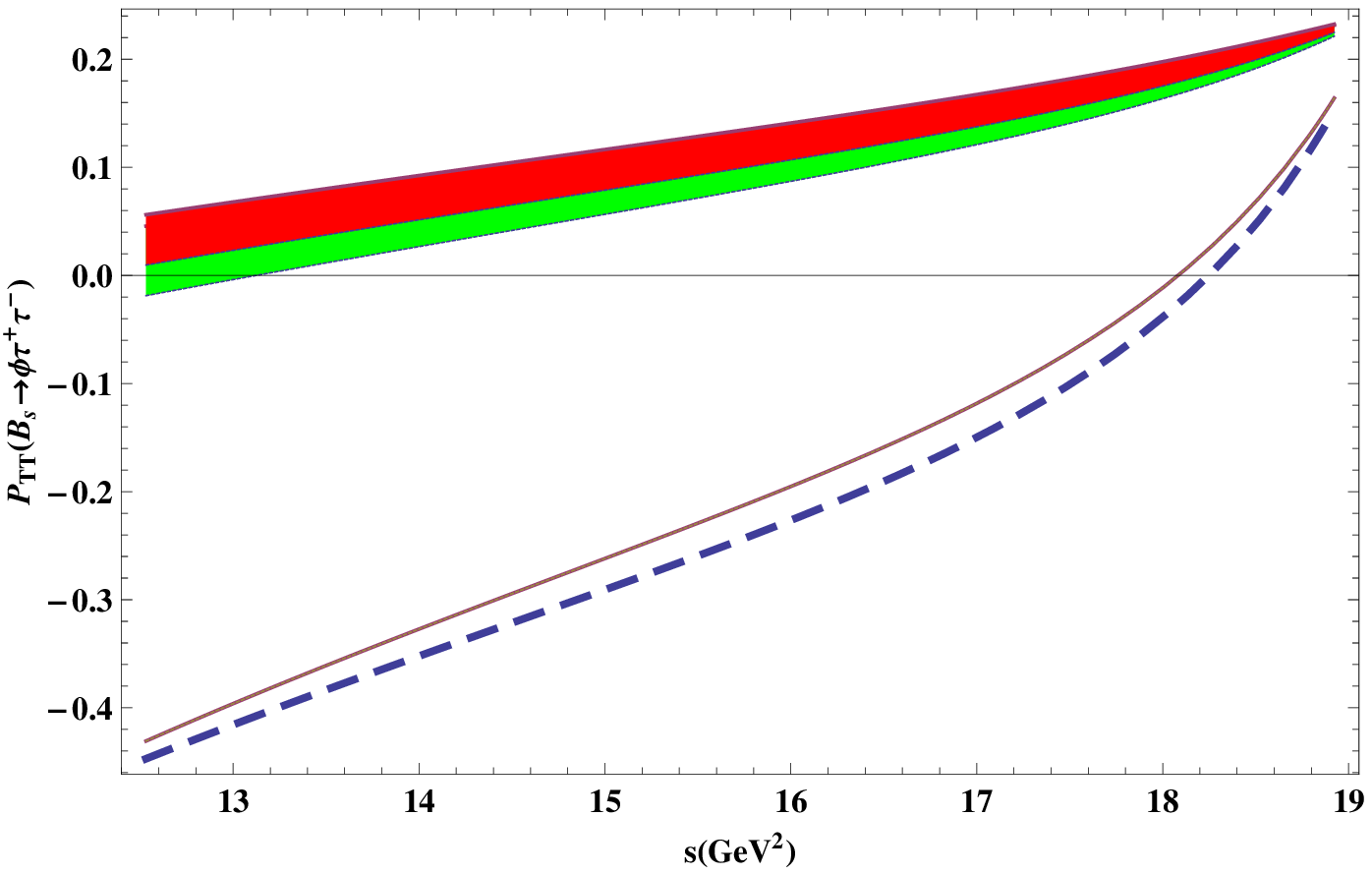,width=0.4\linewidth,clip=c} \put (-100,190){(c)} & %
\epsfig{file=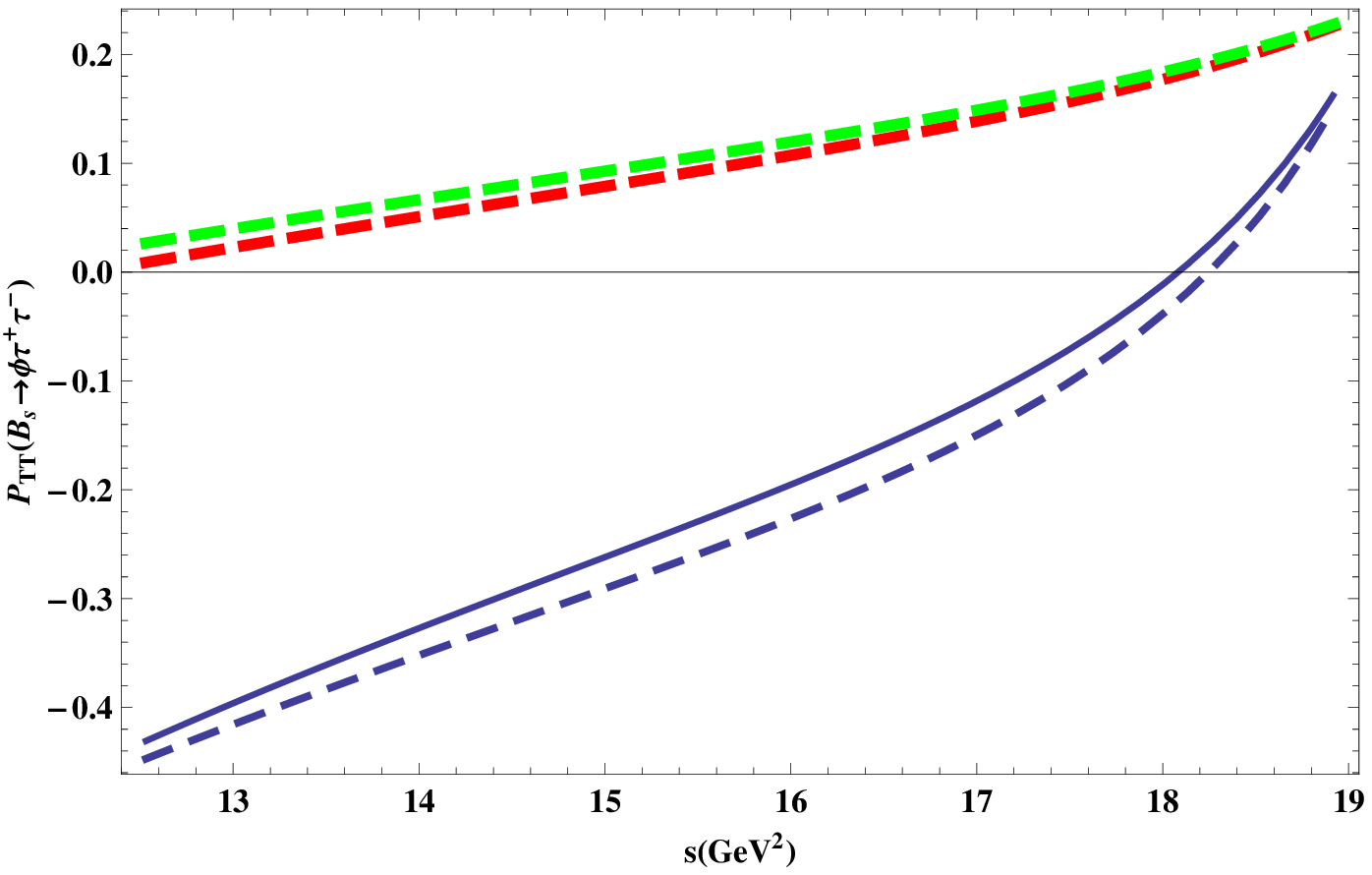,width=0.4\linewidth,clip=d} \put (-100,190){(d)}%
\end{tabular}%
\caption{$P_{TT}$ for the $B_{s} \to \protect\phi l^+l^-$ ($l=\protect\mu,
\protect\tau$) decays as functions of $q^2$. The legends are same as in
Fig.2. }
\label{PTT}
\end{figure}

The longitudinal $(f_{L})$ and the transverse $(f_{T}$) helicity fractions
of final state $\phi $ meson are depicted in Figs. 13 and 14. In Figure 13 one
can see that the values of longitudinal helicity fractions shifts
significantly for some of the NP scenarios when we have taus' as the final state
particles. The maximum deviation comes in the UED model and the reason is
the significant modification of the Wilson coefficients $C_{7}$ and $C_{9}$
in this model compared to their SM values. Similar effects can also be seen in case of the transverse helicity
fractions.

\begin{figure}[tbp]
\begin{tabular}{cc}
\epsfig{file=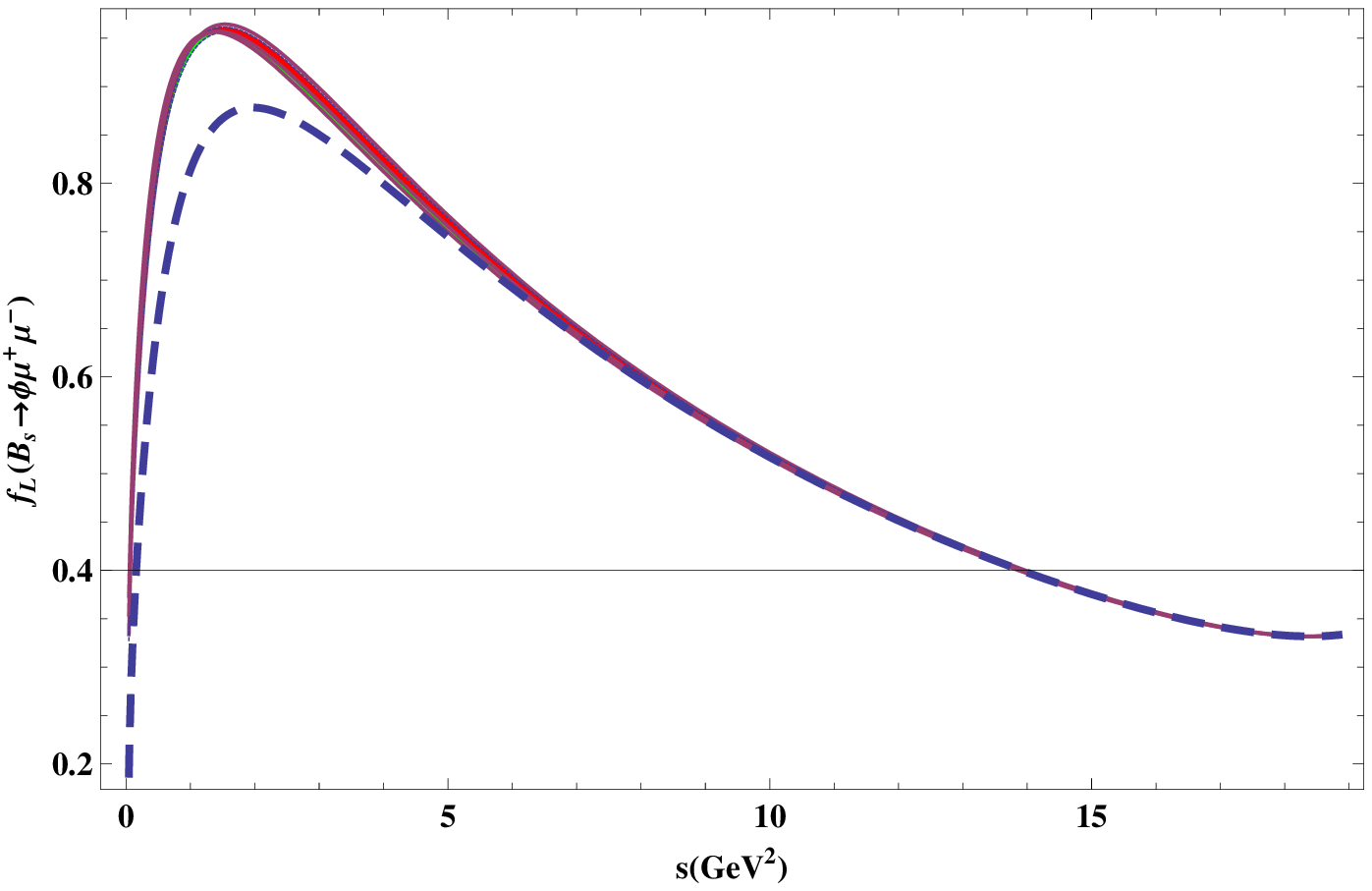,width=0.4\linewidth,clip=a} \put (-100,190){(a)} & %
\epsfig{file=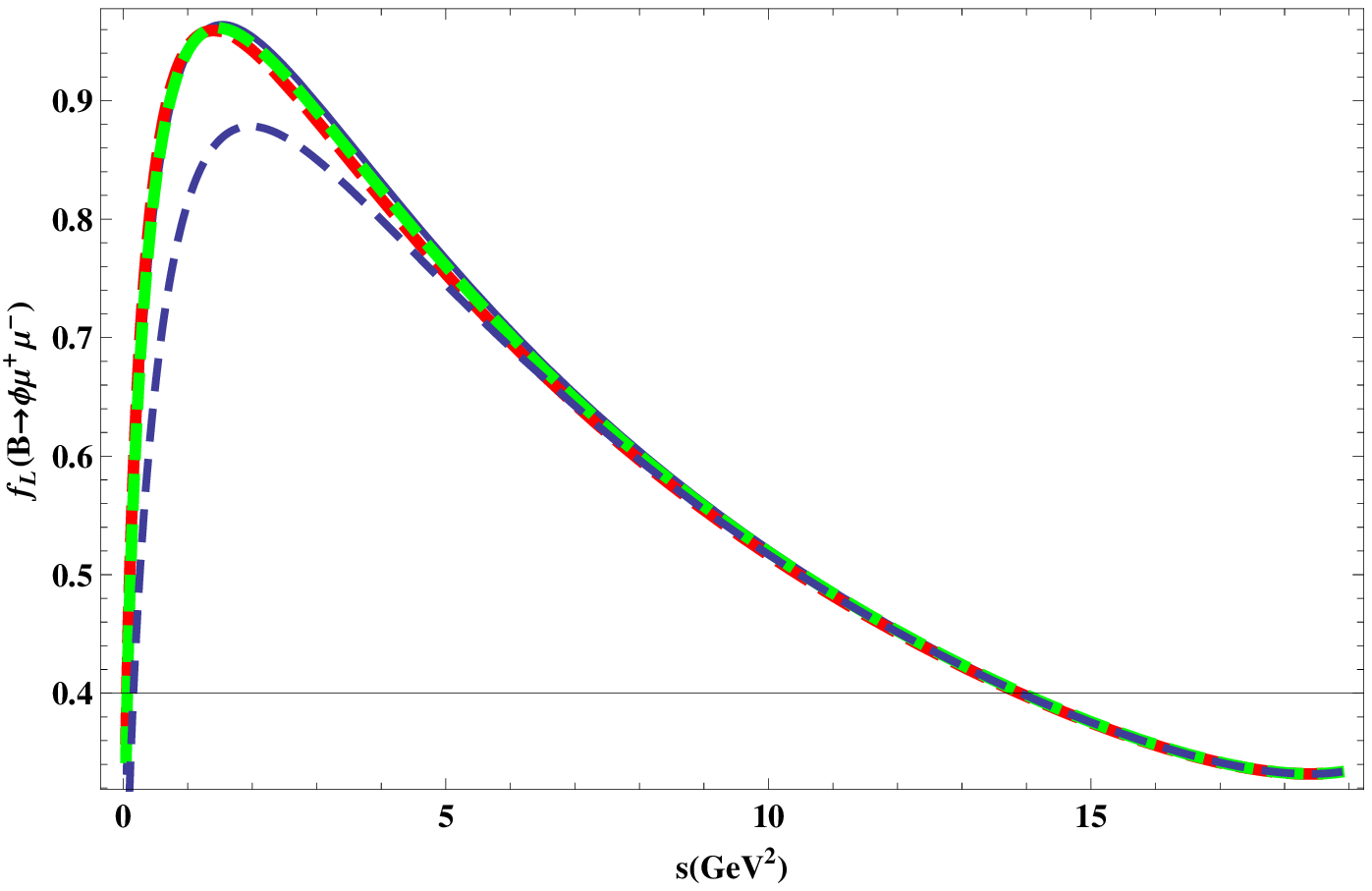,width=0.4\linewidth,clip=b} \put (-100,190){(b)} \\
\epsfig{file=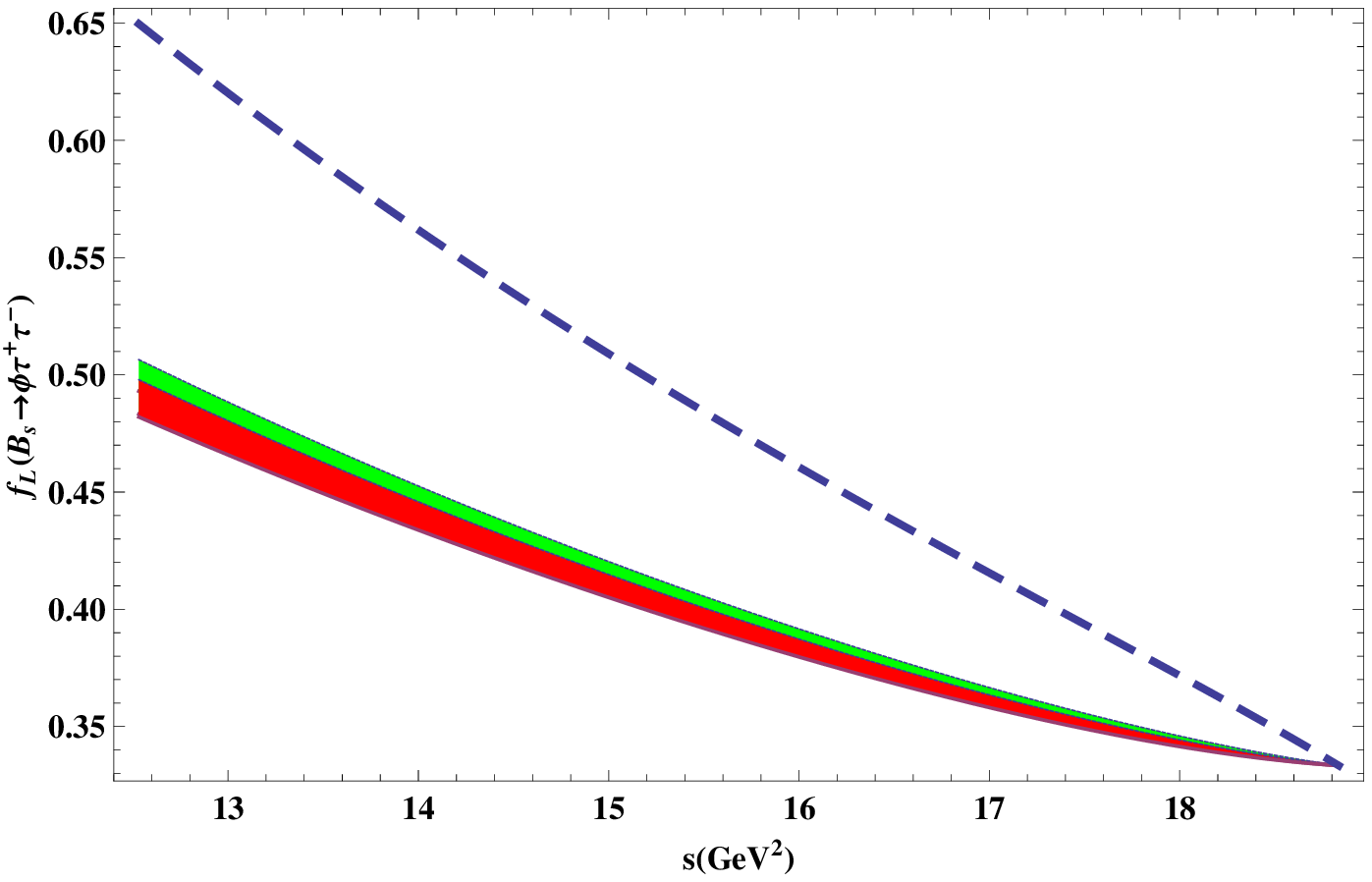,width=0.4\linewidth,clip=c} \put (-100,190){(c)} & %
\epsfig{file=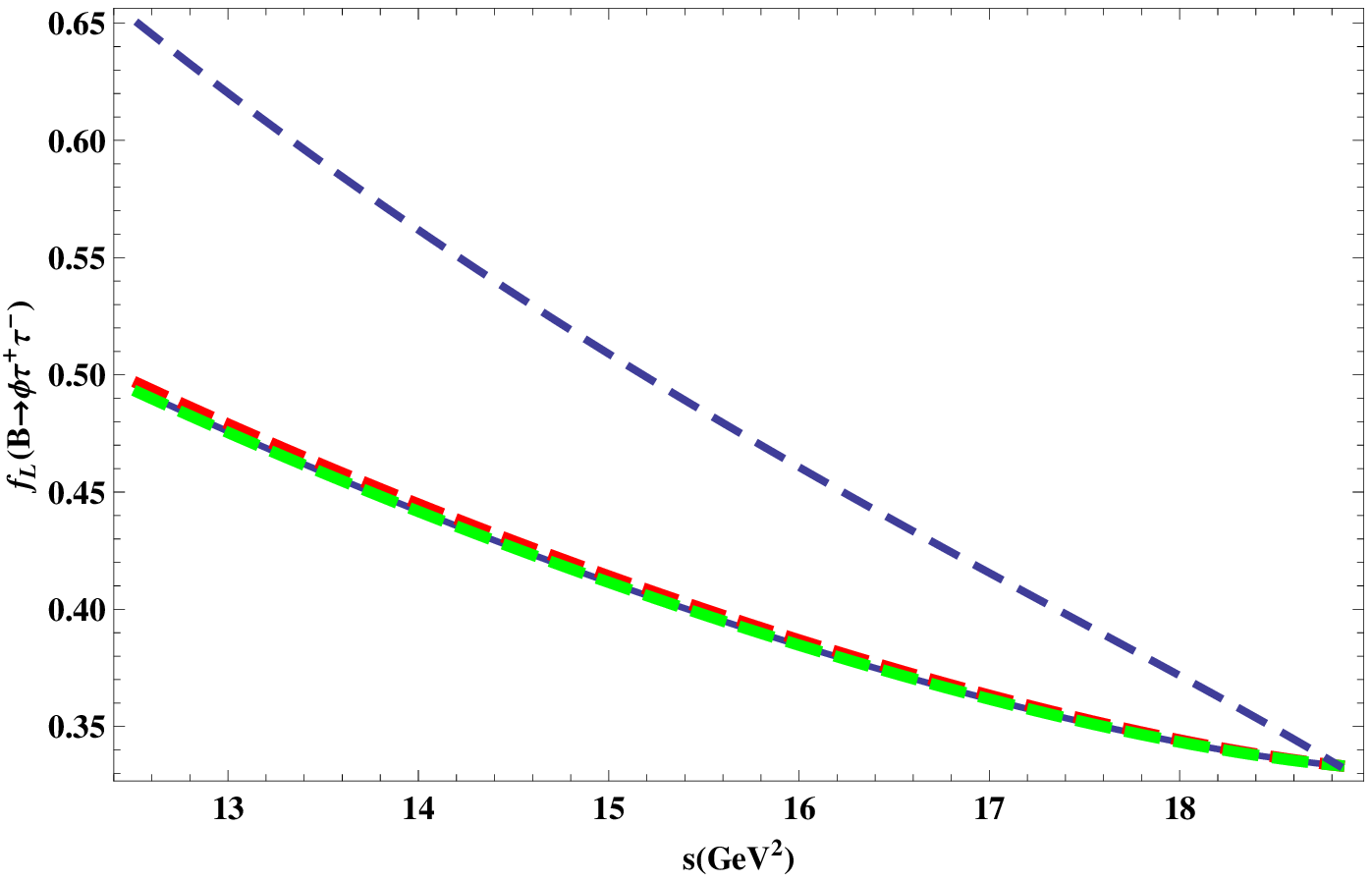,width=0.4\linewidth,clip=d} \put (-100,190){(d)}%
\end{tabular}%
\caption{The longitudinal helicity fractions for the $B_{s}\rightarrow
\protect\phi l^{+}l^{-}$ ($l=\protect\mu ,\protect\tau $) decays as
functions of $q^{2}$. The legends are same as in Fig.2.}
\label{LHF}
\end{figure}

\begin{figure}[tbp]
\begin{tabular}{cc}
\epsfig{file=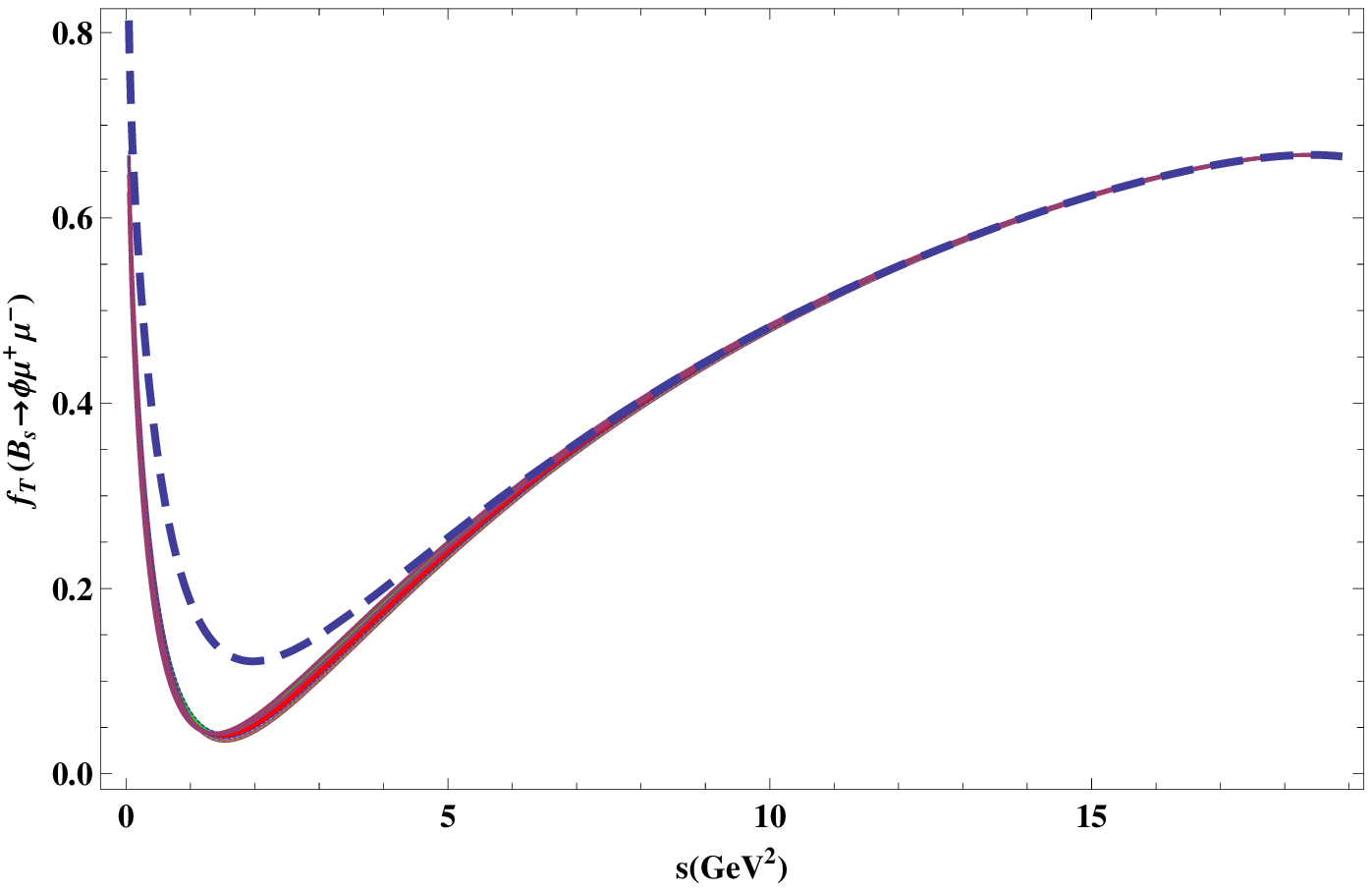,width=0.4\linewidth,clip=a} \put (-100,190){(a)} & %
\epsfig{file=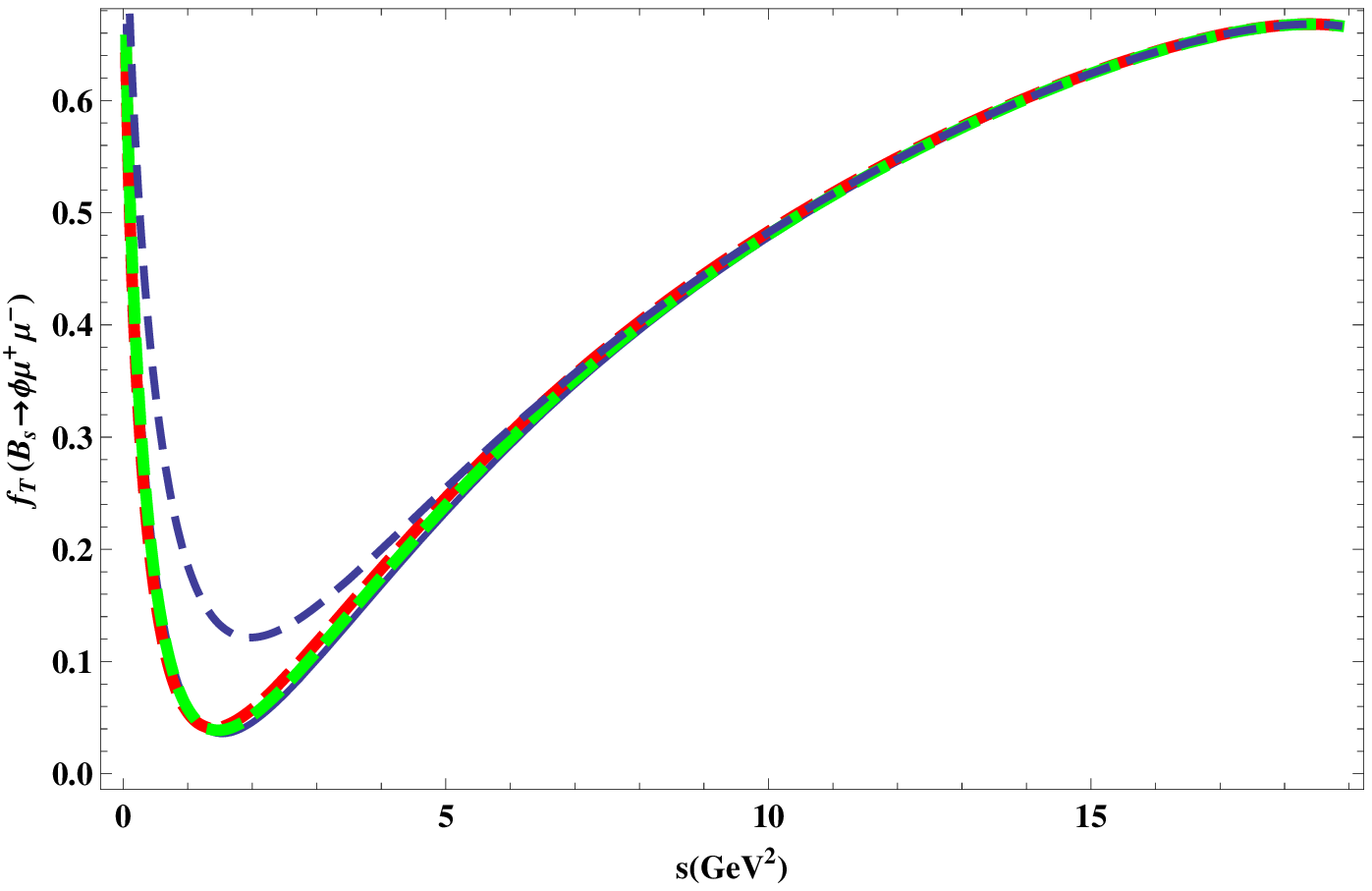,width=0.4\linewidth,clip=b} \put (-100,190){(b)} \\
\epsfig{file=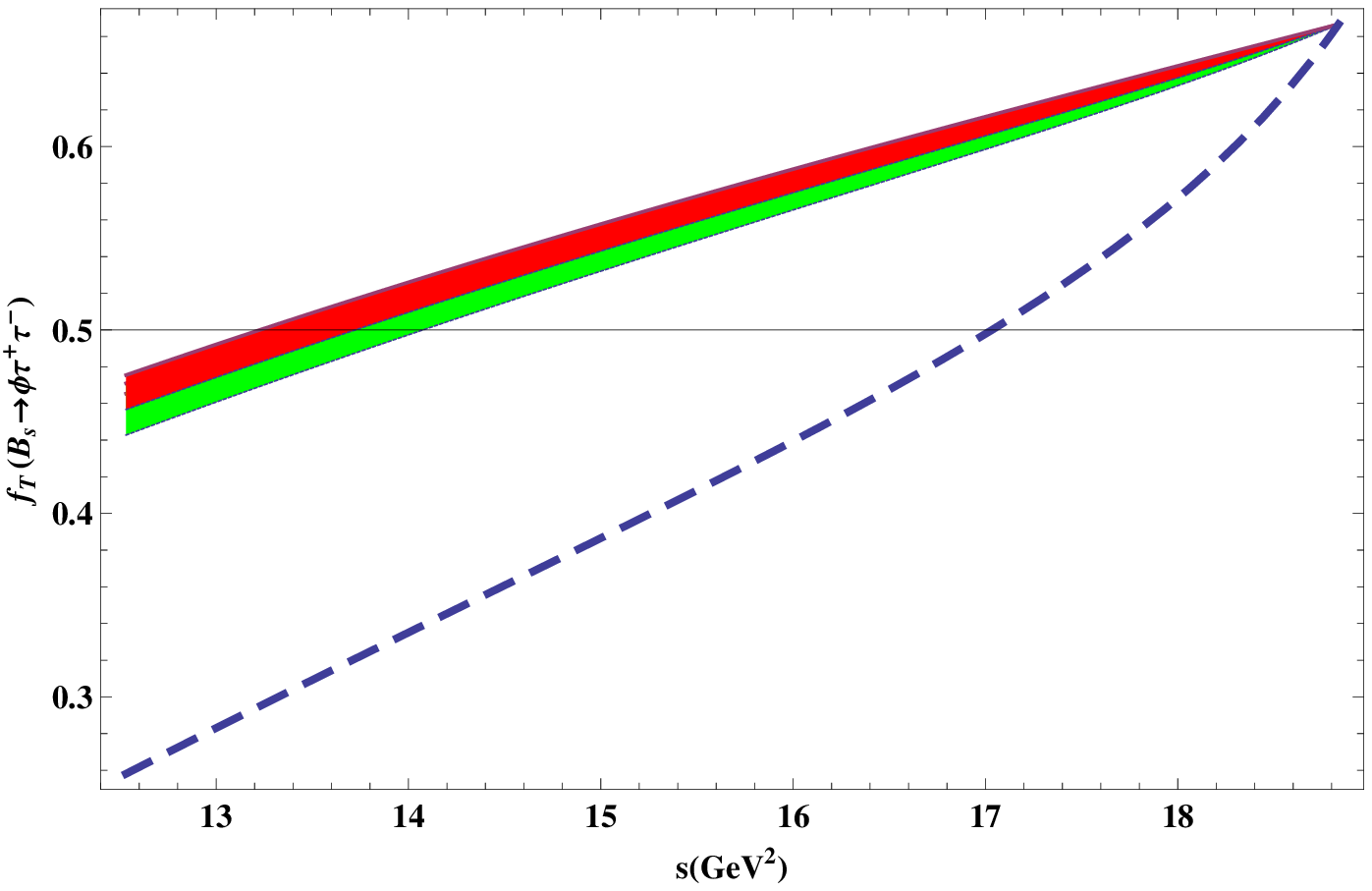,width=0.4\linewidth,clip=c} \put (-100,190){(c)} & %
\epsfig{file=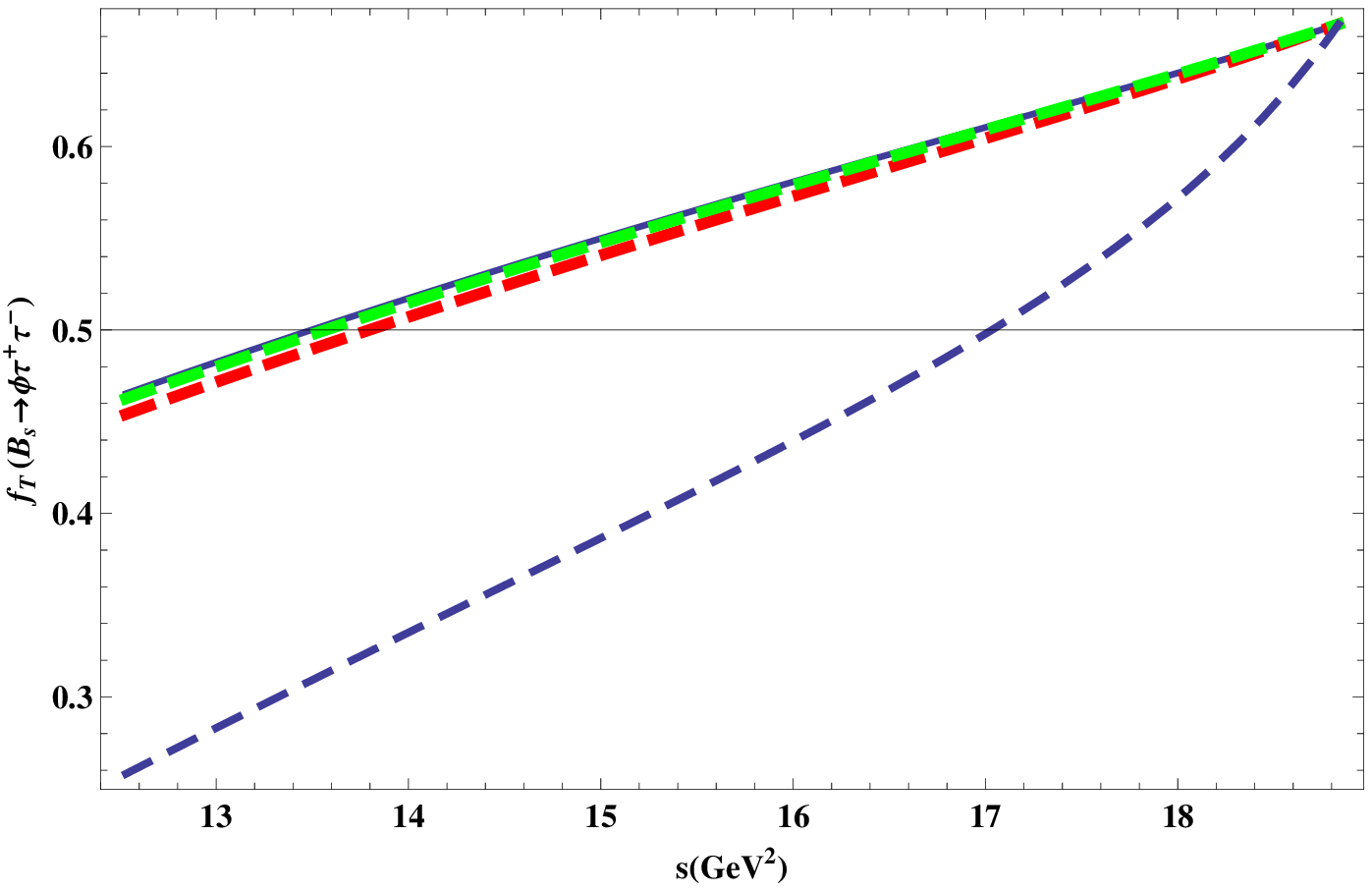,width=0.4\linewidth,clip=d} \put (-100,190){(d)}%
\end{tabular}%
\caption{The transverse helicity fractions for the $B_{s}\rightarrow \protect%
\phi l^{+}l^{-}$ ($l=\protect\mu ,\protect\tau $) decays as functions of $%
q^{2}$. The legends are same as in Fig.2.}
\label{THF}
\end{figure}

Just to summarize, in the present study we have observed
the sizeable difference between the predictions of various physical observables in the SM
and two different beyond the SM scenarios, namely, $Z^{\prime}$ and UED models.
Keeping in view that in certain physical observables the NP effects are obscured by the
uncertainties arising due to form factors but in different lepton polarization asymmetries
their effects are still considerable. We hope that the experimental study of
this channel will be a valuable source which provide an indirect way to dig out the new physics effects in an indirect way.

\section*{Acknowledgments}

The authors would like to thank Prof. Riazuddin and Prof.
Fayyazuddin for their valuable guidance and helpful discussions. The
author M. J. A would like to thank the support by Quaid-i-Azam
University through the University Research Fund. MAP would like to
acknowledge the grant (2012/13047-2) from FAPESP.


\begin{thebibliography}{999}
\bibitem{1} S. L. Glashow, J. Iliopoulos, and L. Maiani, \emph{Weak
Interactions with Lepton-Hadron Symmetry, Phys. Rev.} \textbf{D 2} (1970)
1285.

\bibitem{2} N. Cabibbo, \emph{Unitary Symmetry and Leptonic Decays, Phys.
Rev. Lett.} \textbf{10} (1963) 531.

\bibitem{3} M. Kobayashi and K. Maskawa, \emph{Symmetry breaking of chiral $%
U(3)\bigotimes U(3)$ and $X \to \eta \pi \pi$decay amplitude, Prog. Theor.
Phys.} \textbf{49} (1973) 652.

\bibitem{4a} H.~Y.~Cheng, C.~K.~Chua and A.~Soni, \emph{CP-violating
asymmetries in B0 decays to K+ K- K0(S(L)) and K0(S) K0(S) K0(S(L)), Phys.\
Rev.\ D} \textbf{72}, 094003 (2005) [arXiv:hep-ph/0506268].

\bibitem{4b} G.~Buchalla, G.~Hiller, Y.~Nir and G.~Raz, \emph{The Pattern of
CP asymmetries in $b\rightarrow s$ transitions, JHEP} \textbf{0509}, 074
(2005) [arXiv:hep-ph/0503151].

\bibitem{4c} E.~Lunghi and A.~Soni, \emph{Hints for the scale of new
CP-violating physics from B-CP anomalies, JHEP} \textbf{0908}, 051 (2009)
[arXiv:0903.5059 [hep-ph]].

\bibitem{5a} E.~Barberio \textit{et al.} [Heavy Flavor Averaging Group],
\emph{Averages of $b-$hadron and $c-$hadron Properties at the End of 2007,}
arXiv:0808.1297 [hep-ex].

\bibitem{5b} M.~Bona \textit{et al.} [UTfit Collaboration], \emph{First
Evidence of New Physics in $b \longleftrightarrow s$ Transitions, PMC Phys.\
A} \textbf{3}, 6 (2009) [arXiv:0803.0659 [hep-ph]].

\bibitem{6} S. Baek, C. W. Chiang and D. London, Phys. Lett. B 675 (2009) 59.

\bibitem{7} T. Feldmann and J. Matias, J. High Energy Phys. 0301 (2003) 074
[hep-ph/0212158]; B. Aubert et al. [BABAR Collaboration], Phys. Rev. Lett.
102 (2009) 091803 [arXiv:0807.4119 [hep-ex]].

\bibitem{8} Ken Kiers,Tal Knighton, David London, Matthew Russell, Alejandro
Szynkman and Kari Webster, \emph{Using $t \rightarrow b\bar{b}c$ to Search
for New Physics}, arXiv: 1107.0754 [hep-ph].

\bibitem{9} G.~Burdman, \emph{Short distance coefficients and the vanishing
of the lepton asymmetry in $B \to$ V $\ell^+$ lepton, Phys.\ Rev.\ D}
\textbf{57}, 4254 (1998) [arXiv:hep-ph/9710550].

\bibitem{10} A.~Ali, P.~Ball, L.~T.~Handoko and G.~Hiller, \emph{A
Comparative study of the decays $B \to$ ($K$, $K^{*)} \ell^+ \ell^-$ in
standard model and supersymmetric theories, Phys.\ Rev.\ D} \textbf{61},
074024 (2000) [arXiv:hep-ph/9910221].

\bibitem{11} M.~Beneke, T.~Feldmann and D.~Seidel, \emph{Systematic approach
to exclusive $B \to V \ell{+} \ell{-}$, $V \gamma$ decays, Nucl.\ Phys.\ B}
\textbf{612}, 25 (2001) [arXiv:hep-ph/0106067].

\bibitem{CERN} R. Aaij, LHCb Collaboration, \emph{Differential branching
fraction and angular analysis of the decay }$B^{0}\rightarrow \bar{K}^{0\ast
}\mu ^{+}\mu ^{-}$\emph{, Phys. Rev. Lett.} \textbf{108} (2012) 181806.

\bibitem{9a} A.Ishikawa et al., \emph{Measurement of forward-backward
asymmetry and Wilson coefficients in $B\to K^{\ast}l^+l^-$, Phys. Rev. Lett.}
\textbf{96} (2006) 251801 [hep-ex/0603018].

\bibitem{10a} J. T. Wei et al. [BELLE Collaboration], \emph{Measurement of
the differential branching fraction and forward-backward asymmetry for $B\to
K^{(*)}\ell^{+}\ell^{-}$ Phys. Rev. Lett.} \textbf{103} (2009) 171801
[arXiv:0904.0770 [hep-ex]].

\bibitem{11a} B.~Aubert \textit{et al.} [BABAR Collaboration], \emph{%
Measurements of branching fractions, rate asymmetries, and angular
distributions in the rare decays $B \to K \ell^{+} \ell^{-}$ and $B \to
K^{*} \ell^{+} \ell^{-}$, Phys.\ Rev.\ D} \textbf{73}, 092001 (2006)
[arXiv:hep-ex/0604007].

\bibitem{17} A.~S.~Cornell, N.~Gaur and S.~K.~Singh, \emph{FB asymmetries in
the $B \to K^{*} \ell{+} \ell{-}$ decay: A Model independent approach,}
arXiv:hep-ph/0505136.

\bibitem{18} A.~Ali, T.~Mannel and T.~Morozumi, \emph{Forward backward
asymmetry of dilepton angular distribution in the decay $b\to s \ell{+} \ell{%
-}$, Phys.\ Lett.\ B} \textbf{273}, 505 (1991).

\bibitem{19} A.~J.~Buras and M.~Munz, \emph{Effective Hamiltonian for $B \to
X_{s} e^{+} e^{-}$ beyond leading logarithms in the NDR and HV schemes,
Phys.\ Rev.\ D} \textbf{52}, 186 (1995) [arXiv:hep-ph/9501281].

\bibitem{20} M.~Misiak, \emph{The $b\to s e^{+} e^{-}$ and $b \to s \gamma$
decays with next-to-leading logarithmic QCD corrections, Nucl.\ Phys.\ B}
\textbf{393}, 23 (1993) [Erratum-ibid.\ B \textbf{439}, 461 (1995)].

\bibitem{21} F.~Kruger and E.~Lunghi, \emph{Looking for novel CP violating
effects in $\bar{B} \to K^{*} \ell^+$ lepton, Phys.\ Rev.\ D} \textbf{63},
014013 (2001) [arXiv:hep-ph/0008210].

\bibitem{22} A.~Ali, E.~Lunghi, C.~Greub and G.~Hiller, \emph{Improved model
independent analysis of semileptonic and radiative rare $B$ decays, Phys.\
Rev.\ D} \textbf{66}, 034002 (2002) [arXiv:hep-ph/0112300].

\bibitem{23} A.~Ghinculov, T.~Hurth, G.~Isidori and Y.~P.~Yao, \emph{New
NNLL results on the decay $B \to X_{s} l^{+}l^{-}$, Eur.\ Phys.\ J.}\ C
\textbf{33}, S288 (2004) [arXiv:hep-ph/0310187].

\bibitem{24} C.~Bobeth, T.~Ewerth, F.~Kruger and J.~Urban, \emph{Analysis of
neutral Higgs boson contributions to the decays $\bar{B}$( $s^{)} \to
\ell^{+} \ell^{-}$ and $\bar{B} \to K \ell^{+} \ell^{-}$, Phys.\ Rev.\ D}
\textbf{64}, 074014 (2001) [arXiv:hep-ph/0104284].

\bibitem{25} P.~H.~Chankowski and L.~Slawianowska, \emph{Effects of the
scalar FCNC in $b \to s l^{+} l^{-}$ transitions and supersymmetry, Eur.\
Phys.\ J.\ C} \textbf{33}, 123 (2004) [arXiv:hep-ph/0308032].

\bibitem{26} G.~Hiller and F.~Kruger, \emph{More model independent analysis
of $b \to s$ processes, Phys.\ Rev.\ D} \textbf{69}, 074020 (2004)
[arXiv:hep-ph/0310219].

\bibitem{27} A.~K.~Alok and S.~U.~Sankar, \emph{New physics upper bound on
the branching ratio of $B_s \to \ell^{+} \ell^{-}$, Phys.\ Lett.\ B} \textbf{%
620}, 61 (2005) [arXiv:hep-ph/0502120].

\bibitem{28} A.~K.~Alok, A.~Dighe and S.~U.~Sankar, \emph{Tension between
scalar/pseudoscalar new physics contribution to $B_{(s)}\to \mu^{+} \mu^{-}$
and $B \to K \mu^{+} \mu^{-}$ Mod.\ Phys.\ Lett.\ A} \textbf{25}, 1099
(2010) [arXiv:0803.3511 [hep-ph]].

\bibitem{29} A.~K.~Alok, A.~Dighe and S.~U.~Sankar, \emph{Probing extended
Higgs sector through rare $b \to s \mu^{+} \mu^{-}$ transitions, Phys.\
Rev.\ D} \textbf{78}, 034020 (2008) [arXiv:0805.0354 [hep-ph]].

\bibitem{30} A.~Hovhannisyan, W.~S.~Hou and N.~Mahajan, \emph{$B\to
K^{*}\ell^{+} \ell^{-}$ Forward-backward Asymmetry and New Physics, Phys.\
Rev.\ D} \textbf{77}, 014016 (2008) [arXiv:hep-ph/0701046].

\bibitem{31} F.~Kruger and J.~Matias, \emph{Probing new physics via the
transverse amplitudes of $B_{0} \to K_{0}^{\ast} (\to K^{-} \pi^{+})
\ell^{+} \ell^{-}$ at large recoil, Phys.\ Rev.\ D} \textbf{71}, 094009
(2005) [arXiv:hep-ph/0502060].

\bibitem{32} E.~Lunghi and J.~Matias, \emph{Huge right-handed current
effects in $B \to K^{\ast} (\to K \pi) \ell^{+} \ell^{-}$ in supersymmetry,
JHEP} \textbf{0704}, 058 (2007) [arXiv:hep-ph/0612166].

\bibitem{33} U.~Egede, T.~Hurth, J.~Matias, M.~Ramon and W.~Reece, \emph{New
observables in the decay mode $\bar{B}_{d} \to \bar{K}^{\ast}_0 \ell^{+}
\ell^{-}$, JHEP} \textbf{0811}, 032 (2008) [arXiv:0807.2589 [hep-ph]].

\bibitem{34} T.~M.~Aliev, V.~Bashiry and M.~Savci, \emph{Double lepton
polarization asymmetries in the $B\to K \ell^{+} \ell^{-}$ decay beyond the
standard model, Eur.\ Phys.\ J.}\ C \textbf{35}, 197 (2004)
[arXiv:hep-ph/0311294].

\bibitem{35} T.~M.~Aliev, V.~Bashiry and M.~Savci, \emph{Polarized lepton
pair forward backward asymmetries in $B \to K^{*} \ell^{+} \ell^{-}$ decay
beyond the standard model, JHEP} \textbf{0405}, 037 (2004)
[arXiv:hep-ph/0403282].

\bibitem{36} T.~M.~Aliev, M.~K.~Cakmak, A.~Ozpineci and M.~Savci, \emph{New
physics effects to the lepton polarizations in the $B \to K \ell^{+}
\ell^{-} $ decay, Phys.\ Rev.\ D} \textbf{64}, 055007 (2001)
[arXiv:hep-ph/0103039].

\bibitem{37} W.~Bensalem, D.~London, N.~Sinha and R.~Sinha, \emph{Lepton
polarization and forward backward asymmetries in $b \to s \tau^{+} \tau^{-}$%
, Phys.\ Rev.\ D} \textbf{67}, 034007 (2003) [arXiv:hep-ph/0209228].

\bibitem{38} A. K. Kumar, A. Dighe, D. Ghosh, D. London, J. Matias, M.
Nagashima and A. Sznkman, \emph{New Physics Contributions to the
Forward-Backward Asymmetry in $B \to K^{*} \mu^{+} \mu^{-}$ JHEP} \textbf{02}
(2010) 053.

\bibitem{39} CDF Collaboration, \emph{Measurement of forward-backward
asymmetry in $B\to K^{\ast} \mu^{+} \mu^{-}$ and first observation of $%
B_{s}^{0}\to \phi \mu^{+}\mu^{-}$, Phys. Rev. Lett.} \textbf{106}

\bibitem{RF1} R. Fleischer and R. Knegjens, \emph{Effective Lifetimes of $%
B_{s}$ Decay and their Constraints on the $B_{s}^{0}-\bar{B_{s}^{0}}$ Mixing
Parameters}:arXiv:1109.5115 [hep-ph].

\bibitem{RF2} F. Azfar et al. [CDF Collaboration], Public Note 10206 (2010).

\bibitem{RF3} V. M. Abazov et al. [D0 Collaboration], arXiv: 1109.3166
[hep-ex].

\bibitem{RF4} R. Van Kooten, Talk at Lepton-Photon 2011, Mumbai, India,
22-27 August 2011 [http://www.tifr.res.in/lp11].

\bibitem{RF5} V. M. Abazov et al. [D0 Collaboration], arXiv: 1106.6308
[hep-ex]

\bibitem{RF6} G. Raven, Talk at Lepton-Photon 2011, Mumbai, India, 22-27
August 2011 [http://www.tifr.res.in/lp11].

\bibitem{40} U.~O.~Yilmaz, \emph{Analysis of $B_s \to \phi \ell^{+} \ell^{-}
$ decay with new physics effects, Eur.\ Phys.\ J.\ C} \textbf{58}, 555
(2008) [arXiv:0806.0269 [hep-ph]].

\bibitem{41} G.~Erkol and G.~Turan, \emph{The Exclusive $B \to \phi \ell^{+}
\ell^{-}$ decay in the two Higgs doublet models, Eur.\ Phys.\ J.\ C} \textbf{%
25}, 575 (2002) [arXiv:hep-ph/0203038].

\bibitem{42} Q.~Chang and Y.~H.~Gao, \emph{Probe a family non-universal $%
Z^{\prime }$ boson effects in $\bar{B}_s \to \phi \mu^+ \mu^-$ decay, Nucl.\
Phys.\ B} \textbf{845}, 179 (2011) [arXiv:1101.1272 [hep-ph]].

\bibitem{43} R.~Mohanta and A.~K.~Giri, \emph{Study of FCNC mediated rare $%
B_s$ decays in a single universal extra dimension scenario, Phys.\ Rev.\ D}
\textbf{75}, 035008 (2007) [arXiv:hep-ph/0611068 ].

\bibitem{44} Ying Li and Juan Hua, arXiv: 1105.3031 [hep-ph].

\bibitem{45} P.~Ball and V.~M.~Braun, \emph{Exclusive semileptonic and rare
B meson decays in QCD, Phys.\ Rev.\ D} \textbf{58}, 094016 (1998)
[arXiv:hep-ph/9805422].

\bibitem{46} P.~Ball and R.~Zwicky, \emph{$B_{d,s}\to \rho, \omega, K^{*},
\phi$ decay form-factors from light-cone sum rules revisited, Phys.\ Rev.\ D}
\textbf{71}, 014029 (2005) [arXiv:hep-ph/0412079].

\bibitem{47} Y.~L.~Wu, M.~Zhong and Y.~B.~Zuo, \emph{$B_{s}, D_{s} \to \pi,
K, \eta, \rho, K^{*}, \omega, \phi$ Transition Form Factors and Decay Rates
with Extraction of the CKM parameters |V(ub)|, |V(cs)|,|V(cd)|, Int.\ J.\
Mod.\ Phys.\ A} \textbf{21}, 6125 (2006) [arXiv:hep-ph/0604007].

\bibitem{48} W.~Wang, R.~H.~Li and C.~D.~Lu, \emph{Radiative charmless $%
B_s\to V \gamma$ and $B_s\to A \gamma$ decays in pQCD approach,}
arXiv:0711.0432 [hep-ph].

\bibitem{49} D.~Melikhov and B.~Stech, \emph{Weak form-factors for heavy
meson decays: An Update, Phys.\ Rev.\ D} \textbf{62}, 014006 (2000)
[arXiv:hep-ph/0001113].

\bibitem{50} A.~Deandrea and A.~D.~Polosa, \emph{The Exclusive $B_s \to \phi
\mu^{+} \mu^{-}$ process in a constituent quark model, Phys.\ Rev.\ D}
\textbf{64}, 074012 (2001) [arXiv:hep-ph/0105058].

\bibitem{51} C.~Q.~Geng and C.~C.~Liu, \emph{Study of $B_s \to$ ($\eta$, $%
\eta^\prime$, $\phi^{)} \ell \bar{\ell}$ decays,'' J.\ Phys.\ G} \textbf{29}%
, 1103 (2003) [arXiv:hep-ph/0303246].

\bibitem{latt} C. M. Bouchard \textit{et al} \emph{Form factors for $B$ and
$B_s$ semileptonic decays with NRQCD/HISQ quarks}, arXiv:1210.6992 [hep-lat].

\bibitem{52} G.~Buchalla, A.~J.~Buras and M.~E.~Lautenbacher, \emph{Weak
decays beyond leading logarithms, Rev.\ Mod.\ Phys.} \textbf{68}, 1125
(1996) [arXiv:hep-ph/9512380].

\bibitem{53} A.~J.~Buras and M.~Munz, \emph{Effective Hamiltonian for $B\to
X_s e^{+} e^{-}$ beyond leading logarithms in the NDR and HV schemes, Phys.\
Rev.\ D} \textbf{52}, 186 (1995) [arXiv:hep-ph/9501281].

\bibitem{54} C.~S.~Lim, T.~Morozumi and A.~I.~Sanda, \emph{A Prediction for
d gamma (b ---> s Lepton anti-Lepton) / d q**2 Including the Long Distance
Effects, Phys.\ Lett.\ B} \textbf{218}, 343 (1989).

\bibitem{55} A.~Ali, T.~Mannel and T.~Morozumi, \emph{Forward backward
asymmetry of dilepton angular distribution in the decay $b \to s l^+ l^-$
Phys.\ Lett.\ B} \textbf{273}, 505 (1991).

\bibitem{56} F.~Kruger and L.~M.~Sehgal, \emph{Lepton polarization in the
decays $B \to X_{s} \mu^{+}\mu^{-}$ and $B \to X_{s}\tau^{+} \tau^{-}$,
Phys.\ Lett.\ B} \textbf{380}, 199 (1996) [arXiv:hep-ph/9603237].

\bibitem{57} B.~Grinstein, M.~J.~Savage and M.~B.~Wise, \emph{$B \to X_s e^+
e^-$ in the Six Quark Model, Nucl.\ Phys.\ B} \textbf{319}, 271 (1989).

\bibitem{58} G.~Cella, G.~Ricciardi and A.~Vicere, \emph{QCD corrections to
the $\bar{B} \to X_s e^+ e^-$ decay, Phys.\ Lett.\ B} \textbf{258}, 212
(1991).

\bibitem{59} C.~Bobeth, M.~Misiak and J.~Urban, \emph{Photonic penguins at
two loops and m(t) dependence of BR[B ---> X(s) lepton+ lepton-], Nucl.\
Phys.\ B} \textbf{574}, 291 (2000) [arXiv:hep-ph/9910220].

\bibitem{60} H.~H.~Asatrian, H.~M.~Asatrian, C.~Greub and M.~Walker, \emph{%
Two loop virtual corrections to B ---> X(s) lepton+ lepton- in the standard
model, Phys.\ Lett.\ B} \textbf{507}, 162 (2001) [arXiv:hep-ph/0103087].

\bibitem{61} M.~Misiak, \emph{The $b \to s e^{+} e^{-}$ and $b \to s \gamma $
decays with next-to-leading logarithmic QCD corrections, Nucl.\ Phys.\ B}
\textbf{393}, 23 (1993) [Erratum-ibid.\ B \textbf{439}, 461 (1995)].

\bibitem{62} D.~Melikhov, N.~Nikitin and S.~Simula, \emph{Lepton asymmetries
in exclusive $b \to s l^{+} l^{-}$ decays as a test of the standard model,
Phys.\ Lett.\ B} \textbf{430}, 332 (1998) [arXiv:hep-ph/9803343].

\bibitem{63} J.~M.~Soares, \emph{CP violation in radiative b decays, Nucl.\
Phys.\ B} \textbf{367}, 575 (1991).

\bibitem{64} G.~M.~Asatrian and A.~Ioannisian, \emph{CP violation in the
decay $b \to s \gamma$ in the left-right symmetric model, Phys.\ Rev.\ D}
\textbf{54}, 5642 (1996) [arXiv:hep-ph/9603318].

\bibitem{65} J.~M.~Soares, \emph{The contribution of the J / Psi resonance
to the radiative B decays, Phys.\ Rev.\ D }\textbf{53}, 241 (1996)
[arXiv:hep-ph/9503285].

\bibitem{66} C.~H.~Chen and C.~Q.~Geng, \emph{Baryonic rare decays of $%
\Lambda_{b} \to \Lambda l^{+} \l ^{-}$, Phys. Rev. D} \textbf{64}, 074001
(2001) [arXiv:hep-ph/0106193].

\bibitem{67} T.~Huber, T.~Hurth and E.~Lunghi, \emph{The Role of Collinear
Photons in the Rare Decay $\bar{B} \to X_{s} \ell^{+} \ell^{-}$,}
arXiv:0807.1940 [hep-ph].

\bibitem{68} D.~Melikhov, N.~Nikitin and S.~Simula, \emph{Lepton asymmetries
in exclusive $b \to s l^{+} l^{-}$ decays as a test of the standard model,
Phys.\ Lett.\ B} \textbf{430}, 332 (1998) [arXiv:hep-ph/9803343].

\bibitem{69} J.~M.~Soares, \emph{CP violation in radiative b decays, Nucl.\
Phys.\ B} \textbf{367}, 575 (1991).

\bibitem{70} G.~M.~Asatrian and A.~Ioannisian, \emph{CP violation in the
decay $b\rightarrow s\gamma $ in the left-right symmetric
model,\textquotedblright\ Phys.\ Rev.\ D} \textbf{54}, 5642 (1996)
[arXiv:hep-ph/9603318].

\bibitem{ACD1} T. Appelquist, H. C. Cheng and B. A. Dobrescu, \emph{Bounds
on universal extra dimensions, Phys. Rev. D}\textbf{64} (2001) 035002.

\bibitem{ACD2} A. J. Buras, M. Spranger and A.Weiler, \emph{The Impact of
Universal Extra Dimensions on the Unitarity Triangle and Rare K and B
Decays, Nucl. Phys. B}\textbf{660 }(2003)\textbf{\ }225

\bibitem{ACD3} A. J. Buras, A. Poschenrieder, M. Spranger and A.Weiler,
\emph{The Impact of Universal Extra Dimensions on }$B\rightarrow Xs\gamma $%
\emph{, }$B\rightarrow X_{s}g$\emph{, }$B\rightarrow Xs\mu ^{+}\mu ^{-}$%
\emph{, }$K_{L}\rightarrow \pi ^{0}e^{+}e^{-}$\emph{, and }$\varepsilon
/\varepsilon ^{\prime }$\emph{, Nucl. Phys. B}\textbf{678 }(2004)\textbf{\ }%
455.

\bibitem{ACD4} K. Agashe, N. G. Deshpande and G. H. Wu, \emph{Universal
Extra Dimensions and }$b\rightarrow s\gamma $\emph{, Phys. Lett. B}\textbf{%
514} (2001)\textbf{\ }309.

\bibitem{76} P.~Langacker and M.~Plumacher, ``Flavor changing effects in
theories with a heavy $Z^\prime$ boson with
Phys.\ Rev.\ D \textbf{62}, 013006 (2000) [arXiv:hep-ph/0001204].

\bibitem{77} V.~Barger, L.~Everett, J.~Jiang, P.~Langacker, T.~Liu and
C.~Wagner, \emph{\textquotedblleft Family Non-universal U(1)-prime Gauge
Symmetries and }$b->s$\emph{\ Phys.\ Rev.\ D }\textbf{80}, 055008 (2009)
[arXiv:0902.4507 [hep-ph]].

\bibitem{jam} M.Jamil Aslam , Cai-Dian Lu and Yu-Ming Wang, \emph{$B \to
K^{\ast}_{0} \ell^{+} \ell^{-}$ decays in supersymmetric theories, Phys. Rev.%
} \textbf{D 79}, 074007 (2009) [arXiv:0902.0432].

\bibitem{jama} M.Jamil Aslam, Yu-Ming Wang and Cai-Dian Lu, \emph{\
Exclusive semileptonic decays of $\Lambda_{b}\to \Lambda \ell^+ \ell^-$ in
supersymmetric theories, Phys. Rev.} \textbf{D 78}, 114032 (2008)
[arXiv:0808.2113].

\bibitem{Aliev} T. M. Aliev and M. Savci, \emph{$\Lambda _{b}\rightarrow
\Lambda \ell ^{+}\ell ^{-}$ decay in universal extra dimensions, Eur. Phys.
J.} \textbf{C 50},91 (2007) [arXiv: hep-ph/0606225].

\bibitem{Colangelo} P. Colangelo \emph{et al.}, Phys. Rev. D 74, (2006)
115006 [hep-ph/0610044]; A. Siddique \emph{et al.}, arXiv:0803.0192.

\bibitem{paracha} M. A. Paracha \emph{et al.}, arXiv:1101.2323 (2011).

\bibitem{DPLBashiry} V. Bashiry and K. Azizi, \emph{Systematic analysis of
the }$B_{s}\rightarrow f_{0}\ell ^{+}\ell ^{-}$\emph{\ in the universal
extra dimension, }arXiv: 1112.5243 [hep-ph] .

\bibitem{13IJ} Ishtiaq Ahmed and M. Jamil Aslam, in prepration.

\bibitem{Haisch} U. Haisch, A. Weiler, \emph{Bound on minimal universal
extra dimensions from $\bar{B}\to X_{s}\gamma$, Phys. Rev. D} \textbf{76},
034014 (2007) .

\bibitem{Gogoladze} I. Gogoladze and C. Macesanu, \emph{Precision
electroweak constraints on Universal Extra Dimensions revisited Phys. Rev. D}
\textbf{74}, 093012 (2006).

\bibitem{Feng} J. A. R. Cembranos, J. L. Feng and L. E. Strigari, \emph{%
Exotic Collider Signals from the Complete Phase Diagram of Minimal Universal
Extra Dimensions Phys. Rev. D} \textbf{75}, 036004 (2007) .

\bibitem{CDFNEW} T. Aaltonen et al. [CDF Collaboration], arXiv: 1103.2482
[hep-ex].

\bibitem{CDF} T. Aaltonen et al. [CDF Collaboration], \textit{Phys. Rev.
Lett.} \textbf{100}, 161803 (2008); P. Q. Hung and M. Sher, \textit{Phys.
Rev. D} \textbf{77}, 037302 (2008).

\bibitem{Zp1} V. Barger, C. W. Chiang, P. Langacker and H. S. Lee, \emph{%
Solution to the }$B\rightarrow \pi K$\emph{\ Puzzle in a Flavor-Changing }$%
Z^{\prime }$\emph{\ Model , Phys. Lett. B} \textbf{598} (2004) 218
[hep-ph/0406126].

\bibitem{Zp2} Q. Chang, X. Q. Li and Y. D. Yang, \emph{Constraints on the
nonuniversal }$Z^{\prime }$\emph{\ couplings from }$B\rightarrow \pi K$\emph{%
, }$\pi K^{\ast }$\emph{\ and }$\rho K$\emph{\ Decays, JHEP} 0905 (2009)
056, arXiv:0903.0275 [hep-ph].

\bibitem{Zp3} V. Barger, L. Everett, J. Jiang, P. Langacker, T. Liu and C.
Wagner, \emph{Family Non-universal }$U^{\prime }\left( 1\right) $\emph{\
Gauge Symmetries and }$b\rightarrow s$\emph{\ Transitions, Phys. Rev. D}
\textbf{80} (2009) 055008 arXiv:0902.4507 [hep-ph].

\bibitem{Zp4} V. Barger, L. Everett, J. Jiang, P. Langacker, T. Liu and C.
Wagner, $b\rightarrow s$\emph{\ Transitions in Family-dependent }$U^{\prime
}\left( 1\right) $\emph{\ Models, JHEP} 0912 (2009) 048, arXiv:0906.3745
[hep-ph].

\bibitem{Zp5} Q. Chang, X. Q. Li and Y. D. Yang, \emph{Family Non-universal }%
$Z^{\prime }$\emph{\ effects on }$\bar{B}_{q}-B_{q}$\emph{\ mixing, }$%
B\rightarrow X_{s}\mu ^{+}\mu ^{-}$\emph{\ and }$B_{s}\rightarrow \mu
^{+}\mu ^{-}$\emph{\ Decays, JHEP} 1002 (2010) 082, arXiv:0907.4408 [hep-ph].

\end{thebibliography}
\end{document}